\documentclass{aa}  

\usepackage{graphicx}
\usepackage{txfonts}
\usepackage{natbib}
\usepackage{amsmath,amssymb}
\usepackage{ccaption}

\def\lsim{\!\!\!\phantom{\le}\smash{\buildrel{}\over
 {\lower2.5dd\hbox{$\buildrel{\lower2dd\hbox{$\displaystyle<$}}\over
                                 \sim$}}}\,\,}
\def\gsim{\!\!\!\phantom{\ge}\smash{\buildrel{}\over
{\lower2.5dd\hbox{$\buildrel{\lower2dd\hbox{$\displaystyle>$}}\over
                               \sim$}}}\,\,}

\def\twCO{$^{12}$CO }
\def\thCO{$^{13}$CO }

\begin{document}

   \title{Molecular gas and dust properties of galaxies from the \\ Great Observatories All-sky LIRG Survey}

\titlerunning{Molecular gas and dust properties in (U)LIRGs from GOALS}
\authorrunning{Herrero-Illana et al.}

\author{R. Herrero-Illana\inst{1,2}\thanks{\email{rherrero@eso.org}},
G.~C.~Privon\inst{3,4}, 
A.~S.~Evans\inst{5,6},
T.~D\'iaz-Santos\inst{7},
M.~\'A.~P\'erez-Torres\inst{2,8},
V.~U\inst{9},
A.~Alberdi\inst{2},
K.~Iwasawa\inst{10},
L.~Armus\inst{11},
S.~Aalto\inst{12},
J.~Mazzarella\inst{13},
J.~Chu\inst{14},
D.~B.~Sanders\inst{15},
L.~Barcos-Mu\~noz\inst{5,6},
V.~Charmandaris\inst{16},
S.~T.~Linden\inst{5},
I.~Yoon\inst{6},
D.~T.~Frayer\inst{17}, 
H.~Inami\inst{18},
D.-C.~Kim\inst{6},
H.~J.~Borish\inst{5},
J.~Conway\inst{12},
E.~J~Murphy\inst{6,13},
Y.~Song\inst{5}, 
S.~Stierwalt\inst{11},
\and
J.~Surace\inst{11}.
}

\institute{
$^{1}$European Southern Observatory (ESO), Alonso de C\'ordova 3107, Vitacura, Casilla 19001, Santiago de Chile, Chile\\
$^2$Instituto de Astrof\'isica de Andaluc\'ia - CSIC, PO Box 3004, 18008, Granada, Spain\\
$^3$ Instituto de Astrof\'isica, Facultad de F\'isica, Pontificia Universidad Cat\'olica de Chile, Casilla 306, Santiago 22, Chile\\
$^{4}$Department of Astronomy, University of Florida, 211 Bryant Space Sciences Center, Gainesville, FL 32607, USA \\
$^{5}$Department of Astronomy, University of Virginia, Charlottesville, VA 22903, USA \\
$^6$National Radio Astronomy Observatory, Charlottesville, VA 22903, USA\\
$^7$N\'ucleo de Astronom\'ia de la Facultad de Ingenier\'ia, Universidad Diego Portales, Av. Ej\'ercito Libertador 441, Santiago, Chile\\
$^8$Visiting Scientist: Facultad de Ciencias, Univ. de Zaragoza, Spain\\
$^{9}$Department of Physics and Astronomy, 4129 Frederick Reines Hall, University of California, Irvine, CA 92697, USA \\
$^{10}$ICREA and Institut de Ci\`encies del Cosmos (ICC), Universitat de Barcelona (IEEC-UB), Mart\'i i Franqu\`es 1, 08028, Barcelona, Spain\\
$^{11}$Spitzer Science Center, California Institute of Technology, Pasadena, CA 91106, USA\\
$^{12}$Chalmers University of Technology, Department of Earth and Space Sciences, Onsala Space Observatory, 43992 Onsala, Sweden \\
$^{13}$Infrared Processing and Analysis Center, California Institute of Technology, Pasadena, CA 91125, USA \\
$^{14}$Gemini North Observatory, 670 N. A`ohoku Place, Hilo, HI 96720, USA \\
$^{15}$University of Hawaii, Institute for Astronomy, 2680 Woodlawn Dr., Honolulu, HI 96822, USA \\
$^{15}$Joint ALMA Observatory, Alonso de C\'ordova 3107, Vitacura, Santiago, Chile \\
$^{16}$Institute for Astronomy, Astrophysics, Space Applications \& Remote Sensing, National Observatory of Athens, GR-15236, Penteli, Greece \\
$^{17}$Green Bank Observatory, Green Bank WV,  24944, USA \\
$^{18}$National Optical Astronomy Observatory, 950 North Cherry Avenue, Tucson, AZ 85719, USA \\
}

   \date{Received Month DD, YYYY; accepted Month DD, YYYY}
   
 
  \abstract{
  We present IRAM-30\,m Telescope \twCO and \thCO observations of a sample of 55 luminous and ultraluminous infrared galaxies (LIRGs and ULIRGs) in the local universe. This sample is a subset of the Great Observatory All-Sky LIRG Survey (GOALS), for which we use ancillary multi-wavelength data to better understand their interstellar medium and star formation properties. Fifty-three (96\%) of the galaxies are detected in
$^{12}$CO, and 29 (52\%) are also detected in $^{13}$CO above a 3$\sigma$ level. The median full width at zero intensity (FWZI) velocity of the CO line emission is 661\,km s$^{-1}$, and $\sim$54\% of the galaxies show a multi-peak CO profile. \emph{Herschel} photometric data is used to construct the far-IR spectral energy distribution of each galaxy, which are fit with a modified blackbody model that allows us to derive dust temperatures and masses, and infrared luminosities.  We make the assumption that the gas-to-dust mass ratio of (U)LIRGs is comparable to local spiral galaxies with a similar stellar mass (i.e., gas/dust of mergers is comparable to their progenitors) 
to derive a CO-to-H$_2$ conversion factor of $\langle\alpha\rangle=1.8^{+1.3}_{-0.8}\,M_\odot$\,(K km\,s$^{-1}$\,pc$^{2}$)$^{-1}$; such a value is comparable
to that derived for (U)LIRGs based on dynamical mass arguments.
We derive gas depletion times of $400-600$\,Myr for the (U)LIRGs, compared to the 1.3\,Gyr for local spiral galaxies.
Finally, we re-examine the relationship between the \twCO/\thCO ratio and dust temperature, confirming a transition to elevated ratios in warmer systems.
}


   \maketitle
%

\section{Introduction}
Luminous and ultraluminous infrared galaxies (LIRGs: $L_\mathrm{IR}(8- 1\,000\,\mu\mathrm{m})>10^{11}L_\odot$; ULIRGs; $L_\mathrm{IR}>10^{12}L_\odot$) 
are known to host powerful starbursts, making them ideal for studying 
the properties of extreme star-forming environments. Since the discovery of (U)LIRGs as an important  galaxy population by the Infrared Astronomical Satellite (\emph{IRAS}), which found that 30\%-50\% of the total bolometric luminosity of galaxies in the local universe is emitted at infrared (IR) and sub-millimeter wavelengths \citep{soifer86}, CO observations of (U)LIRGs have been carried out to characterize the molecular gas responsible 
for the ongoing star formation, as well as to trace galaxy kinematics 
\citep[e.g.,][]{tinney90,sanders91,aalto95, solomon97,downes98,bryant99,yao03, narayanan05, papadopoulos12b}. 
Much of the interpretation of the data is reliant on the CO luminosity-to-molecular gas mass conversion factor, 
$\alpha = M({\rm H}_2) / L'_{\rm CO}$; the commonly used ``Milky Way'' conversion factor, $\alpha_\mathrm{MW}\simeq4\,M_\odot$\,(K km\,s$^{-1}$\,pc$^{2}$)$^{-1}$, has been called into question 
for its use with (U)LIRGs, for which alternative values have been proposed \citep[e.g.,][]{downes98, bolatto13}.
The choice directly impacts upon estimates of the total gas mass, the star formation efficiency, and the gas depletion timescale, which can differ by almost an order of magnitude. 
In addition to CO, observations of high density and optically thin gas tracers have provided the opportunity to better assess the physical conditions of
the interstellar medium (ISM), in part by determining the physical state of the gas
more directly involved in the formation of future stars.

The advent of broadband receivers on millimeter-wave telescopes has enabled the detection of multiple lines at once, probing a range of optical depths, critical densities, and temperatures. In our present study, we make use of the 30\,m Telescope of the Institut de Radioastronomie Millim/'etrique (IRAM-30\,m) to detect multiple millimeter-wave lines from a sample of 55 nearby (U)LIRGs selected from the Great Observatories All-Sky LIRG Survey \citep[GOALS,][]{armus09}. We focus here primarily on the $^{12}$CO(1$\to$0) and
$^{13}$CO(1$\to$0) data, which is combined with 
new {\it Herschel Space Observatory} observations in order to estimate
the conversion factor $\alpha$ {from an assumed gas-to-dust mass ratio}, the star formation efficiency, and gas depletion timescales, and to examine the relation
between the $^{12}$CO(1$\to$0)-to-$^{13}$CO(1$\to$0) ratio
and the dust temperature.
A complementary study of dense gas tracers HCN(1$\to$0) and HCO$^+$(1$\to$0), observed as part of this IRAM campaign, 
was presented in \citet{privon15}. 

This paper is organized as follows: we describe our sample in Sect.~\ref{sec:sample}. Observations and data reduction are summarized in Sect.~\ref{sec:data}. The main results are presented in Sect.~\ref{sec:results}. In Sect.~\ref{sec:discussion} we determine the CO-to-H$_2$ conversion factor (\ref{sec:g2d}), derive the star formation properties of the sample (\ref{sec:sfprop}), and reexamine the \twCO/\thCO relationship with dust temperature (\ref{sec:12co13coratio}). Finally, we summarize our results in Sect.~\ref{sec:summary}.
Throughout this study, we adopt a Wilkinson Microwave Anisotropy Probe (WMAP) cosmology of $H_{\rm 0} = 69.3$ km s$^{-1}$ Mpc$^{-1}$, $\Omega_\Lambda = 0.714$ and $\Omega_m = 0.286$ \citep{hinshaw13}.

\section{Sample} \label{sec:sample}

Our sources are selected from the GOALS sample, which consists of all 180 luminous and 22 ultraluminous nearby IR galaxy systems in the \emph{IRAS} Revised Bright Galaxy Sample ($f_{60\,\mu\mathrm{m}}>5.24\,\mathrm{Jy}$: \citealt{sanders03}). The GOALS sample has been observed with the \emph{Hubble Space Telescope} \citep[\emph{HST;}][]{haan11,kim13}, the \emph{Spitzer Space Telescope} \citep{diaz-santos10,diaz-santos11,petric11,stierwalt13,stierwalt14}, the {\it Herschel Space Observatory} \citep{chu17}, the Galaxy Evolution Explorer \emph{(GALEX)} \citep{howell10}, and the \emph{Chandra X-ray Observatory} \citep{iwasawa11}, making it the most complete multiwavelength sample of nearby, IR-bright galaxies. In addition, ground-based H\,{\sc i}, radio, and near-IR spectroscopic data complement the space-telescope-based dataset. GOALS contains a sufficient number of LIRGs to cover the full range of galaxy-galaxy interactions and merger stages, including isolated sources. The sample of 55 (U)LIRGs observed for our
present CO survey consists of GOALS
sources accessible from Pico Veleta that had not been previously observed by the IRAM-30\,m Telescope. {Unfortunately, due to the lack of a public archive for the IRAM-30\,m Telescope, previously observed data for GOALS galaxies cannot be obtained in its raw format, so a systematic data reduction could not be guaranteed. For this reason, we restricted the current study to the aforementioned sample of 55 (U)LIRGs.}

\begin{figure}\centering
\includegraphics[width=.9\columnwidth]{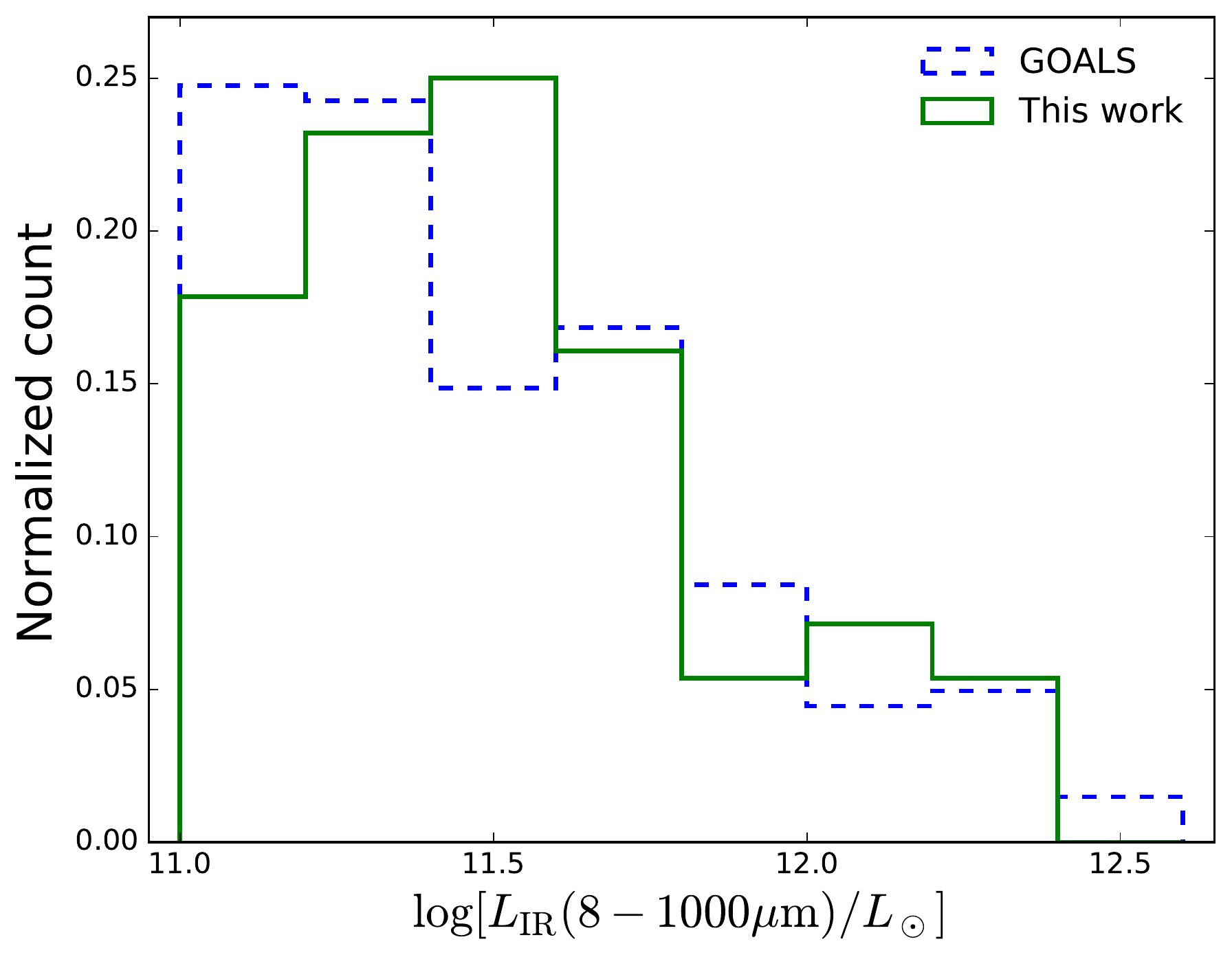}
\caption{Normalized histogram comparing $L_\mathrm{IR}$ from the complete set of 202 GOALS galaxies and our observed sample. All $L_\mathrm{IR}$ were obtained from \citet{armus09}.}
\label{fig:histgoals}
\end{figure}

In the observed sample, 48 of the galaxies are LIRGs or components of LIRGs, and seven are ULIRGs or component of ULIRGs. {In the cases where each component of a system was observed independently, these were treated as individual sources throughout the analysis}. The list of galaxies is presented in Table~\ref{table:intro}, along with $L_\mathrm{IR}$ from \citet{armus09}. As shown in the histogram of Fig.~\ref{fig:histgoals}, our observed sources are representative of the complete GOALS sample in terms of luminosity.

To compare our (U)LIRG sample with local spiral galaxies, we have used published observations from the HERA CO-Line Extragalactic Survey (HERACLES) \citep{leroy09} and the Five College Radio Astronomy Observatory (FCRAO) \citep{young95}, for which CO and \emph{Herschel} data are available. {These include a total of 22 sources.}

{For comparison, our (U)LIRG sample has a median redshift (see Sect.~\ref{sec:results} for details) of $0.0248\pm0.0163$ and a median $\log(L_\mathrm{IR}/L_\odot)$ of $(11.45\pm0.33)$ according to \citet{armus09}, where the errors indicate the standard deviation of the sample. On the other hand, the local comparison sample (both HERACLES and FCRAO) has a median redshift of $0.0023\pm0.0018$ and a median $\log(L_\mathrm{IR}/L_\odot)$ of $10.28\pm0.51$.}

\begin{table*}
\caption{\label{table:intro}Source list}
\centering
\begin{tabular}{lrrc}
\hline\hline
{Source name} & R.A. & Dec.& $\log(L_\mathrm{IR})$ \\
 & (J2000) & (J2000) & ($L_\odot$) \\
\hline
NGC 0034 & 00 11 06.55 & -12 06 27.90 & 11.49 \\
Arp 256N & 00 18 50.17 & -10 21 44.78 & 11.48\tablefootmark{$\dagger$} \\
Arp 256S & 00 18 50.90 & -10 22 36.19 & 11.48\tablefootmark{$\dagger$} \\
IC 1623 & 01 07 47.53 & -17 30 25.88 & 11.71 \\
MCG -03-04-014 & 01 10 08.96 & -16 51 09.79 & 11.65 \\
IRAS F01364-1042 & 01 38 52.92 & -10 27 11.41 & 11.85 \\
IC 0214 & 02 14 05.47 & 05 10 25.21 & 11.43 \\
UGC 01845 & 02 24 07.99 & 47 58 10.81 & 11.12 \\
NGC 0958 & 02 30 42.85 & -02 56 20.51 & 11.20 \\
ESO 550-IG025 & 04 21 20.00 & -18 48 39.38 & 11.51 \\
UGC 03094 & 04 35 33.83 & 19 10 18.19 & 11.41 \\
NGC 1797 & 05 07 44.85 & -08 01 08.69 & 11.04 \\
IRAS F05189-2524 & 05 21 01.47 & -25 21 45.40 & 12.16 \\
IRAS F05187-1017 & 05 21 06.54 & -10 14 46.79 & 11.30 \\
IRAS F06076-2139 & 06 09 45.81 & -21 40 23.70 & 11.65 \\
NGC 2341 & 07 09 12.00 & 20 36 10.01 & 11.17 \\
NGC 2342 & 07 09 18.06 & 20 38 10.39 & 11.31 \\
IRAS 07251-0248 & 07 27 37.55 & -02 54 54.11 & 12.39 \\
IRAS F09111-1007 W & 09 13 36.40 & -10 19 30.00 & 12.06\tablefootmark{$\dagger$} \\
IRAS F09111-1007 E & 09 13 38.80 & -10 19 20.32 & 12.06\tablefootmark{$\dagger$} \\
UGC 05101 & 09 35 51.60 & 61 21 11.81 & 12.01 \\
2MASX J11210825-0259399\tablefootmark{1} & 11 21 08.28 & -02 59 39.01 & 11.43\tablefootmark{$\dagger$} \\
CGCG 011-076 & 11 21 12.22 & -02 59 02.18 & 11.43\tablefootmark{$\dagger$} \\
IRAS F12224-0624 & 12 25 03.89 & -06 40 51.71 & 11.36 \\
CGCG 043-099 & 13 01 50.80 & 04 19 59.99 & 11.68 \\
ESO 507-G070 & 13 02 52.35 & -23 55 17.69 & 11.56 \\
NGC 5104 & 13 21 23.10 & 00 20 32.89 & 11.27 \\
IC 4280 & 13 32 53.30 & -24 12 25.88 & 11.15 \\
NGC 5258 & 13 39 57.25 & 00 49 47.60 & 11.62\tablefootmark{$\dagger$} \\
UGC 08739 & 13 49 13.91 & 35 15 26.21 & 11.15 \\
NGC 5331 & 13 52 16.20 & 02 06 05.62 & 11.66 \\
CGCG 247-020 & 14 19 43.21 & 49 14 11.90 & 11.39 \\
IRAS F14348-1447 & 14 37 38.34 & -15 00 22.79 & 12.39 \\
CGCG 049-057 & 15 13 13.09 & 07 13 32.02 & 11.35 \\
NGC 5936 & 15 30 00.86 & 12 59 22.20 & 11.14 \\
IRAS F16164-0746 & 16 19 11.79 & -07 54 02.81 & 11.62 \\
CGCG 052-037 & 16 30 56.50 & 04 04 58.51 & 11.45 \\
IRAS F16399-0937 & 16 42 40.21 & -09 43 14.41 & 11.63 \\
NGC 6285 & 16 58 23.99 & 58 57 21.31 & 11.37\tablefootmark{$\dagger$} \\
NGC 6286 & 16 58 31.55 & 58 56 12.19 & 11.37\tablefootmark{$\dagger$} \\
IRAS F17138-1017 & 17 16 35.76 & -10 20 39.80 & 11.49 \\
UGC 11041 & 17 54 51.83 & 34 46 34.50 & 11.11 \\
CGCG 141-034 & 17 56 56.63 & 24 01 01.31 & 11.20 \\
IRAS 18090+0130 & 18 11 38.41 & 01 31 40.12 & 11.65 \\
NGC 6701 & 18 43 12.50 & 60 39 11.20 & 11.12 \\
NGC 6786 & 19 10 54.00 & 73 24 35.71 & 11.49\tablefootmark{$\dagger$} \\
UGC 11415 & 19 11 04.40 & 73 25 32.02 & 11.49\tablefootmark{$\dagger$} \\
ESO 593-IG008 & 19 14 31.15 & -21 19 06.31 & 11.93 \\
NGC 6907 & 20 25 06.60 & -24 48 32.11 & 11.11 \\
IRAS 21101+5810 & 21 11 30.40 & 58 23 03.19 & 11.81 \\
ESO 602-G025 & 22 31 25.49 & -19 02 04.31 & 11.34 \\
UGC 12150 & 22 41 12.18 & 34 14 57.01 & 11.35 \\
IRAS F22491-1808 & 22 51 49.36 & -17 52 24.82 & 12.20 \\
CGCG 453-062 & 23 04 56.55 & 19 33 07.09 & 11.38 \\
2MASX\,J23181352+0633267\tablefootmark{2} & 23 18 13.52 & 06 33 26.50 & 11.12\tablefootmark{$\dagger$} \\
\hline
\end{tabular}
\tablefoot{Coordinates refer to the pointing position of our observations.}
\tablefoottext{1}{Southwest component of CGCG 011-076.} 
\tablefoottext{2}{Southwest component of NGC 7591.}
\tablefoottext{$\dagger$}{Component of a system. In these cases, the shown $L_\mathrm{IR}$ refers to the whole system.}
\end{table*}

\section{Observations and data reduction} \label{sec:data}

The observations\footnote{Based on observations carried out with the IRAM-30\,m Telescope. IRAM is supported by INSU/CNRS (France), MPG (Germany) and IGN (Spain).} presented in this study were carried out with the Eight Mixer Receiver (EMIR) multiband millimeter-wave receiver \citep{carter12} at the IRAM-30\,m Telescope on Pico Veleta, Spain, in five observing periods: June 2010,  September 2011, December 2011, October 2012 (PI: K. Iwasawa), and March 2014 (PI: R. Herrero-Illana). The final March 2014 run was obtained during the director's discretionary time in order to verify the $^{13}$CO line strength of three sources (IRAS\,F05189-2524, IRAS\,22491-1808, and IRAS\,07251-0248). 
We tuned the receiver to two frequency windows: one centered
at the redshifted frequencies of HCN$(1\to0)$ and HCO$^+(1\to0)$ \citep[{rest frequencies of 88.632 and 89.189\,GHz, respectively;} presented in][]{privon15}, and the other at the frequencies of $^{12}$CO$(1\to0)$ and $^{13}$CO$(1\to0)$ {(rest frequencies of 115.271 and 110.201\,GHz, respectively)}. {The EMIR receiver has a total bandwidth of 8\,GHz.} All the observations were performed in wobbler switching mode.
For the majority of the galaxies, the peak of their \emph{Spitzer} IRAC channel 4 (i.e., 8$\mu$m) emission was used as the pointing center. A complete journal of observations is presented in Table~\ref{table:measured}, along with the measured line intensities.

\begin{table*}
\caption{\label{table:measured}Measured parameters.}
\centering
\begin{tabular}{lcccccc}
\hline\hline
{Source name} & {Date} & {$t_\mathrm{int}$} & {$T_\mathrm{sys}$}  & {$S_\mathrm{{^{12}}CO}\Delta v$} & {$S_\mathrm{{^{13}}CO}\Delta v$} & FWZI \\
{} & {} & {(min)} & {(K)}  & {(Jy\,km\,s$^{-1}$)} & {(Jy\,km\,s$^{-1}$)} & (km\,s$^{-1}$) \\  
{(1)} & {(2)} & {(3)} & {(4)} & {(5)} & {(6)} & (7) \\
\hline
NGC 0034 & 2011-09 & 10.5 & 278, 182 & $126.0 \pm 3.6$ & $< 5.4$           & $763.5$  \\
Arp 256N & 2011-12 & 21.0 & 188, 138 & $15.1 \pm 1.3$ & $4.6 \pm 1.3$ & $313.1$  \\
Arp 256S & 2011-12 & 16.0 & 197, 142 & $45.8 \pm 1.6$ & $< 4.8$ & $495.7$  \\
IC 1623 & 2011-12 & 21.5 & 200, 143 & $469.2 \pm 1.9$ & $10.9 \pm 1.3$ & $699.5$  \\
MCG -03-04-014 & 2011-12 & 16.0 & 175, 133 & $96.0 \pm 1.2$ & $<5.4$ & $578.3$  \\
IRAS F01364-1042 & 2011-09 & 21.0 & 171, 135 & $37.4 \pm 1.4$ & $ <4.2$ & $763.3$  \\
IC 0214 & 2011-12 & 16.0 & 183, 177 & $62.8 \pm 1.7$ & $5.9 \pm 1.1$ & $549.5$  \\
UGC 01845 & 2012-10 & 10.0 & 204, 140 & $208.0 \pm 2.5$ & $11.3 \pm 1.9$ & $670.7$  \\
NGC 0958 & 2012-10 & 15.5 & 285, {176} & $144.1 \pm 4.2$ & {12.1}\,$\pm\,${2.1} & $854.3$  \\
ESO 550-IG025 & 2011-09 & 16.0 & 222, 157 & $70.3 \pm 2.7$ & $<4.2$ & $751.6$  \\
UGC 03094 & 2012-10 & 10.0 & 197, 136 & $130.9 \pm 2.8$ & $12.7 \pm 1.4$ & $858.6$  \\
NGC 1797 & 2012-10 & 15.5 & 219, 148 & $126.4 \pm 2.2$ & $7.3 \pm 1.1$ & $541.4$  \\
IRAS F05189-2524 & 2014-03 & 128.0 & 236, 173 & $31.0 \pm 0.5$ & $<1.5$ & $423.8$  \\
IRAS F05187-1017 & 2014-03 & 51.0 & 166, 117 & $62.0 \pm 0.6$ & $3.2 \pm 0.6$ & $600.9$  \\
IRAS F06076-2139 & 2011-09 & 16.0 & 200, 145 & $37.6 \pm 1.7$ & $<$\,{3.3} & $539.7$  \\
NGC 2341 & 2014-03 & 25.5 & 235, 150 & $105.0 \pm 2.3$ & $6.8 \pm 1.2$ & $568.3$  \\
NGC 2342 & 2012-10 & 15.5 & 169, 117 & $111.5 \pm 1.7$ & $7.2 \pm 1.2$ & $568.6$  \\
IRAS 07251-0248 & 2014-03 & 21.5 & 136, 106 & $18.0 \pm 1.4$ & $<2.1$ & $763.8$  \\
IRAS F09111-1007 W & 2011-09 & 21.5 & 184, $\cdots^\dagger$ & $38.5 \pm 1.1$ & $\cdots$ & $493.5$  \\
IRAS F09111-1007 E & 2011-09 & 21.5 & 189, $\cdots^\dagger$ & $19.2 \pm 1.9$ & $\cdots$ & $630.9$  \\
UGC 05101 & 2010-06 & 53.0 & 159, 122 & $86.6 \pm 1.5$ & $<2.4$ & $973.1$  \\
2MASX J11210825-0259399  & 2014-03 & 30.5 & 203, 134 & $<$\,{3.7}$^\ddagger$ & $<$\,{3.7}$^\ddagger$ & $\cdots$ \\
CGCG 011-076 & 2012-10 & 10.0 & 201, 179 & $114.3 \pm 2.6$ & $8.3 \pm 2.6$ & $728.8$  \\
IRAS F12224-0624 & 2014-03 & 20.5 & 190, 126 & $19.9 \pm 0.9$ & $<2.7$ & $312.8$  \\
CGCG 043-099 & 2011-09 & 26.5 & 194, 141 & $72.8 \pm 1.3$ & $4.9 \pm 1.3$ & $701.5$  \\
ESO 507-G070 & 2011-12 & 10.5 & 227, 135 & $126.2 \pm 3.6$ & $<4.5$ & $882.0$  \\
NGC 5104 & 2012-10 & 10.0 & 186, 151 & $131.6 \pm 2.9$ & $9.1 \pm 2.2$ & $879.6$  \\
IC 4280 & 2012-10 & 10.0 & 252, 178 & $119.7 \pm 3.5$ & $8.4 \pm 1.8$ & $593.6$  \\
NGC 5258 & 2011-12 & 10.5 & 243, 145 & $128.6 \pm 3.2$ & $9.9 \pm 1.9$ & $675.5$  \\
UGC 08739 & 2012-10 & 15.5 & 232, 166 & $208.4 \pm 2.7$ & $14.2 \pm 1.3$ & $774.8$  \\
NGC 5331 & 2011-09 & 16.0 & 176, 134 & $119.7 \pm 2.5$ & $8.6 \pm 1.7$ & $1155.3$ \\
CGCG 247-020 & 2012-10 & 20.5 & 309, 231 & $63.4 \pm 1.9$ & $<4.2$ & $390.8$  \\
IRAS F14348-1447 & 2011-09 & 10.5 & 173, 147 & $53.0 \pm 2.8$ & $<$\,{6.3} & $788.4$  \\
CGCG 049-057 & 2012-10 & 61.5 & 253, 179 & $120.1 \pm 1.1$ & $7.2 \pm 1.1$ & $488.8$  \\
NGC 5936 & 2012-10 & 10.0 & 176, 122 & $155.8 \pm 1.7$ & $11.1 \pm 1.2$ & $308.8$  \\
IRAS F16164-0746 & 2011-12 & 10.5 & 240, 152 & $104.9 \pm 3.2$ & $<5.7$ & $702.0$  \\
CGCG 052-037 & 2012-10 & 20.0 & 182, 125 & $100.5 \pm 1.3$ & $5.2 \pm 1.3$ & $676.5$  \\
IRAS F16399-0937 & 2011-12 & 16.0 & 243, 154 & $99.0 \pm 2.5$ & $4.7 \pm 1.2$ & $652.1$  \\
NGC 6285 & 2011-12 & 10.5 & 170, 109 & $65.7 \pm 1.9$ & $<3.9$ & $698.9$  \\
NGC 6286 & 2011-12 & 10.5 & 216, 116 & $270.4 \pm 3.0$ & $19.6 \pm 1.4$ & $957.3$  \\
IRAS F17138-1017 & 2011-12 & 10.5 & 245, 141 & $174.7 \pm 2.8$ & $5.3 \pm 1.1$ & $542.6$  \\
UGC 11041 & 2012-10 & 10.0 & 196, 148 & $161.5 \pm 2.3$ & $11.3 \pm 1.7$ & $542.1$  \\
CGCG 141-034 & 2012-10 & 10.0 & 171, 118 & $53.3 \pm 2.6$ & $5.4 \pm 1.3$ & $699.6$  \\
IRAS 18090+0130 & 2011-09 & 16.0 & 186, 137 & $113.1 \pm 2.0$ & $4.8 \pm 1.3$ & $722.5$  \\
NGC 6701 & 2012-10 & 10.5 & 211, 155 & $202.3 \pm 2.1$ & $14.0 \pm 1.3$ & $308.8$  \\
NGC 6786 & 2011-09 & 10.5 & 193, 134 & $77.2 \pm 2.2$ & $7.8 \pm 1.1$ & $479.9$  \\
UGC 11415 & 2011-09 & 10.5 & 192, 134 & $58.8 \pm 2.2$ & $<3.3$ & $506.6$  \\
ESO 593-IG008 & 2014-03 & 20.5 & 190, 147 & $112.5 \pm 2.5$ & $<5.1$ & $1172.0$ \\
NGC 6907 & 2012-10 & 15.5 & 367, 245 & $278.4 \pm 4.4$ & $<11.1$ & $924.1$  \\
IRAS 21101+5810 & 2011-09 & 10.5 & 165, 127 & $42.8 \pm 1.9$ & $<3.9$ & $648.5$  \\
ESO 602-G025 & 2014-03 & 25.5 & 251, 162 & $134.4 \pm 2.0$ & $8.2 \pm 1.3$ & $729.0$  \\
UGC 12150 & 2012-10 & 15.5 & 247, 156 & $77.3 \pm 2.7$ & $<4.2$ & $778.4$  \\
IRAS F22491-1808 & 2014-03 & 133.0 & 142, 117 & $19.5 \pm 0.6$ & $<1.8$ & $520.2$  \\
CGCG 453-062 & 2012-10 & 15.5 & 293, 177 & $60.4 \pm 3.0$ & $<3.6$ & $598.7$  \\
2MASX J23181352+0633267 & 2014-03 & 13.0 & 219, 147 & $<$\,{7.5}$^\ddagger$ & $<$\,{7.5}$^\ddagger$ & $\cdots$ \\
\hline
\end{tabular}
\tablefoot{(1) Source name; (2) Observing date; (3) Integration time; (4) System temperatures for $^{12}$CO and $^{13}$CO, respectively; (5) Line intensity for $^{12}$CO; (6) Line intensity for $^{13}$CO. In those cases where the detection was not significant, the $3\sigma$ upper limit is quoted. {(7) Full width at zero intensity.} \\
$^\dagger$ {\thCO line was not observed for these sources due to a non-optimal receiver setup.}\\
{$^\ddagger$ To obtain the upper limits for the \twCO and \thCO non-detections, we have assumed a FWZI corresponding to the median of the sample.}}
\end{table*}

\begin{table*}
\caption{\label{table:derived}
Derived parameters}
\centering
\begin{tabular}{lcccc}
\hline\hline
{Source name}  & {$L'_\mathrm{{^{12}}CO}$} & {$L'_\mathrm{{^{13}}CO}$}& {$z$} & {$M_{\mathrm{H}_2}$} \\
{}   & {($10^8$\,K\,km\,s$^{-1}$\,pc$^2$)} & {($10^8$\,K\,km\,s$^{-1}$\,pc$^2$)} &{} & {($10^9M_\odot$)}  \\
{(1)} & {(2)} & {(3)} & {(4)} & {(5)} \\
\hline
NGC 0034 & $21.52 \pm 2.24$ & $<0.96$ & $0.0194$ & $3.83 \pm 0.40$ \\
Arp 256N & $5.08 \pm 0.67$ & $1.74 \pm 0.46$ & $0.0271$ & $0.90 \pm 0.12$ \\
Arp 256S & $15.35 \pm 1.63$ & $<1.68$ & $0.0271$ & $2.73 \pm 0.29$ \\
IC 1623 & $84.94 \pm 8.50$ & $2.23 \pm 0.31$ & $0.0200$ & $15.12 \pm 1.51$ \\
MCG -03-04-014 & $54.06 \pm 5.45$ & $<2.97$ & $0.0351$ & $9.62 \pm 0.97$ \\
IRAS F01364-1042 & $40.12 \pm 4.27$ & $<4.38$ & $0.0482$ & $7.14 \pm 0.76$ \\
IC 0214 & $26.11 \pm 2.71$ & $2.78 \pm 0.53$ & $0.0302$ & $4.65 \pm 0.48$ \\
UGC 01845 & $23.25 \pm 2.34$ & $1.42 \pm 0.25$ & $0.0157$ & $4.14 \pm 0.42$ \\
NGC 0958 & $24.34 \pm 2.54$ & {2.30}$\,\pm\,${0.41} & $0.0193$ & $4.33 \pm 0.45$ \\
ESO 550-IG025 & $33.10 \pm 3.55$ & $<1.98$ & $0.0321$ & $5.89 \pm 0.63$ \\
UGC 03094 & $35.66 \pm 3.65$ & $3.88 \pm 0.52$ & $0.0245$ & $6.35 \pm 0.65$ \\
NGC 1797 & $12.83 \pm 1.30$ & $0.83 \pm 0.14$ & $0.0150$ & $2.28 \pm 0.23$ \\
IRAS F05189-2524 & $26.08 \pm 2.64$ & $<1.26$ & $0.0428$ & $4.64 \pm 0.47$ \\
IRAS F05187-1017 & $23.24 \pm 2.33$ & $1.35 \pm 0.25$ & $0.0289$ & $4.14 \pm 0.42$ \\
IRAS F06076-2139 & $24.27 \pm 2.67$ & $<2.22$ & $0.0375$ & $4.32 \pm 0.48$ \\
NGC 2341 & $13.90 \pm 1.42$ & $1.01 \pm 0.18$ & $0.0171$ & $2.47 \pm 0.25$ \\
NGC 2342 & $15.59 \pm 1.58$ & $1.13 \pm 0.19$ & $0.0176$ & $2.77 \pm 0.28$ \\
IRAS 07251-0248 & $64.84 \pm 8.22$ & $<7.65$ & $0.0877$ & $11.54 \pm 1.46$ \\
IRAS F09111-1007 W & $52.43 \pm 5.46$ & $\cdots$ & $0.0543$ & $9.33 \pm 0.97$ \\
IRAS F09111-1007 E & $26.52 \pm 3.71$ & $\cdots$ & $0.0547$ & $4.72 \pm 0.66$ \\
UGC 05101 & $61.37 \pm 6.23$ & $<1.68$ & $0.0393$ & $10.92 \pm 1.11$ \\
2MASX J11210825-0259399 & $\cdots$ & $\cdots$ & $\cdots$ & $\cdots$ \\
CGCG 011-076 & $31.94 \pm 3.28$ & $2.61 \pm 0.76$ & $0.0248$ & $5.69 \pm 0.58$ \\
IRAS F12224-0624 & $6.33 \pm 0.69$ & $<0.84$ & $0.0264$ & $1.13 \pm 0.12$ \\
CGCG 043-099 & $46.61 \pm 4.74$ & $3.51 \pm 0.90$ & $0.0374$ & $8.30 \pm 0.84$ \\
ESO 507-G070 & $26.23 \pm 2.73$ & $<0.93$ & $0.0214$ & $4.67 \pm 0.49$ \\
NGC 5104 & $20.65 \pm 2.11$ & $1.60 \pm 0.37$ & $0.0186$ & $3.68 \pm 0.38$ \\
IC 4280 & $14.35 \pm 1.50$ & $1.13 \pm 0.23$ & $0.0163$ & $2.55 \pm 0.27$ \\
NGC 5258 & $31.08 \pm 3.20$ & $2.69 \pm 0.52$ & $0.0231$ & $5.53 \pm 0.57$ \\
UGC 08739 & $26.61 \pm 2.68$ & $2.05 \pm 0.25$ & $0.0168$ & $4.74 \pm 0.48$ \\
NGC 5331 & $59.89 \pm 6.12$ & $4.85 \pm 0.95$ & $0.0331$ & $10.66 \pm 1.09$ \\
CGCG 247-020 & $19.31 \pm 2.02$ & $<1.29$ & $0.0258$ & $3.44 \pm 0.36$ \\
IRAS F14348-1447 & $169.52 \pm 19.23$ & $<20.64$ & $0.0826$ & $30.17 \pm 3.42$ \\
CGCG 049-057 & $9.21 \pm 0.92$ & $0.63 \pm 0.10$ & $0.0130$ & $1.64 \pm 0.16$ \\
NGC 5936 & $12.55 \pm 1.26$ & $1.00 \pm 0.13$ & $0.0133$ & $2.23 \pm 0.22$ \\
IRAS F16164-0746 & $26.72 \pm 2.80$ & $<1.50$ & $0.0237$ & $4.76 \pm 0.50$ \\
CGCG 052-037 & $27.45 \pm 2.77$ & $1.61 \pm 0.38$ & $0.0245$ & $4.89 \pm 0.49$ \\
IRAS F16399-0937 & $32.98 \pm 3.40$ & $1.77 \pm 0.44$ & $0.0270$ & $5.87 \pm 0.61$ \\
NGC 6285 & $10.91 \pm 1.14$ & $<0.63$ & $0.0191$ & $1.94 \pm 0.20$ \\
NGC 6286 & $43.05 \pm 4.33$ & $3.51 \pm 0.38$ & $0.0187$ & $7.66 \pm 0.77$ \\
IRAS F17138-1017 & $23.72 \pm 2.40$ & $0.81 \pm 0.17$ & $0.0173$ & $4.22 \pm 0.43$ \\
UGC 11041 & $19.33 \pm 1.95$ & $1.52 \pm 0.24$ & $0.0163$ & $3.44 \pm 0.35$ \\
CGCG 141-034 & $9.80 \pm 1.09$ & $1.12 \pm 0.26$ & $0.0201$ & $1.74 \pm 0.19$ \\
IRAS 18090+0130 & $42.98 \pm 4.36$ & $2.06 \pm 0.54$ & $0.0289$ & $7.65 \pm 0.78$ \\
NGC 6701 & $15.84 \pm 1.59$ & $1.24 \pm 0.15$ & $0.0132$ & $2.82 \pm 0.28$ \\
NGC 6786 & $22.15 \pm 2.30$ & $2.50 \pm 0.39$ & $0.0251$ & $3.94 \pm 0.41$ \\
UGC 11415 & $17.17 \pm 1.84$ & $<0.99$ & $0.0253$ & $3.06 \pm 0.33$ \\
ESO 593-IG008 & $123.62 \pm 12.67$ & $<5.82$ & $0.0488$ & $22.00 \pm 2.26$ \\
NGC 6907 & $14.17 \pm 1.43$ & $<0.57$ & $0.0106$ & $2.52 \pm 0.26$ \\
IRAS 21101+5810 & $30.17 \pm 3.30$ & $<2.67$ & $0.0392$ & $5.37 \pm 0.59$ \\
ESO 602-G025 & $38.77 \pm 3.92$ & $2.66 \pm 0.45$ & $0.0252$ & $6.90 \pm 0.70$ \\
UGC 12150 & $16.27 \pm 1.72$ & $<0.87$ & $0.0215$ & $2.90 \pm 0.31$ \\
IRAS F22491-1808 & $55.12 \pm 5.74$ & $<4.86$ & $0.0777$ & $9.81 \pm 1.02$ \\
CGCG 453-062 & $17.01 \pm 1.90$ & $<1.02$ & $0.0249$ & $3.03 \pm 0.34$ \\
2MASX J23181352+0633267 & $\cdots$ & $\cdots$ & $\cdots$ & $\cdots$ \\
\hline
\end{tabular}
\tablefoot{(1) Source name; (2) \twCO line luminosity; (3) \thCO line luminosity; (4) Redshift obtained from the $^{12}$CO line; (5) Molecular gas mass, derived using $\alpha=1.8\,M_\odot$\,(K\,km\,s$^{-1}$\,pc$^2$)$^{-1}$ (see Sect.~\ref{sec:g2d} for details). In those cases where the detection was not significant, the $3\sigma$ upper limit is quoted.}
\end{table*}

\begin{figure*}\centering
\includegraphics[width=1\textwidth]{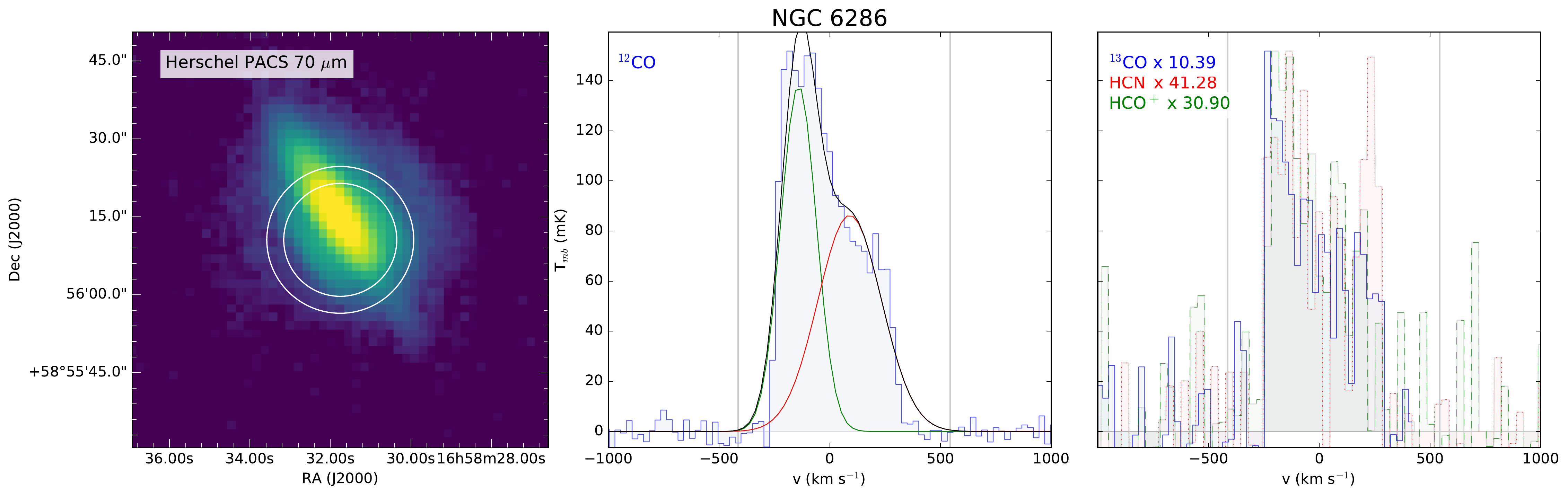}
\caption{Spectral line profiles for NGC\,6286. The left panel shows the \emph{Herschel} PACS 70\,$\mu$m map \citep[see][]{chu17} with a logarithmic stretch to highlight the full extent of the infrared emission relative to the IRAM-30\,m Telescope. Overlaid circles represent the pointing position and beam sizes (FWHM) corresponding to the \twCO and \thCO (inner), and HCN and HCO$^+$ (outer) IRAM observations. The middle panel shows the {continuum-subtracted} CO spectra, fitted with Gaussian components. {Velocity frame is defined according to the redshifts quoted in Table~\ref{table:derived}.} Gray vertical lines correspond to the velocity integration limits of the FWZI. The right panel shows the {continuum-subtracted spectra} of \thCO (blue, solid line), HCN (red, dotted line), and HCO$^+$ (green, dashed line) in the same velocity range as the CO spectrum. The spectra have been normalized to the peak of the CO line, and the normalization factor is shown for each line. The complete data set with all the observations, as well as a table with the fitted Gaussian components of the \twCO spectra is shown in Appendix~\ref{app:lineprof}.}
\label{fig:lineprof}
\end{figure*}

The data were reduced using the Continuum and Line Analysis Single-dish Software (CLASS)\footnote{\url{http://iram.fr/IRAMFR/GILDAS/}} package. For each scan, gain elevation calibration, platforming correction (when the Fast Fourier Transform Spectrometer (FTS) backend was used), 
and order one baseline subtraction were performed. Then, the scans for each galaxy were averaged together and the average spectrum was 
smoothed to $\sim25$\,km\,s$^{-1}$ resolution. A final linear baseline was subtracted over the channels that did not contain line emission.
We obtained line fluxes by integrating all channels 
within the line profile. To account for systematic uncertainties, we added a standard 10\% of error in quadrature to the rms of the fluxes, which is the uncertainty for IRAM-30\,m data {as estimated by the observatory}\footnote{{\url{http://www.iram.fr/GENERAL/calls/s17/30mCapabilities.pdf}}}. Figure~\ref{fig:lineprof} contains an example of a resultant spectrum, and the spectra of the remaining galaxies 
are shown in Appendix \ref{app:lineprof}.

\section{Results}\label{sec:results}

In Tables~\ref{table:measured} and \ref{table:derived} we present the measured and derived properties from the IRAM observations, respectively.
From the 55 observed sources we detect \twCO in 53 ($\sim96\%$) and \thCO in 29 ($\sim52\%$), above a $3\sigma$ level. {The only two sources with no \twCO detection (2MASX J11210825-0259399 and 2MASX J23181352+0633267) are subcomponents of LIRGs and, given the non-detections, were excluded from the analysis.}
Line intensities and luminosities for \thCO were obtained by integrating over the velocity range corresponding to the \twCO detections.
We find that $\sim54\%$ of the sources show a multiple peak profile (see Figs.~\ref{fig:lineprof} and \ref{figapp:lineprof}). For that reason, the full width at half maximum (FWHM) cannot be clearly determined, so we calculate the full width at zero intensity (FWZI), {which was obtained as the width where the sum of the fitted Gaussians of each source (see Fig~\ref{fig:lineprof}) is above 0.5\,mK}. The FWZI of our sample has a median value of $661$\,km\,s$^{-1}$.

We derived the integrated CO luminosities, measured in K\,km\,s$^{-1}$\,pc$^2$, via
\begin{equation}
L'_\mathrm{CO} = 3.25\times10^7 (S_\mathrm{CO}\Delta v) \nu_\mathrm{obs}^{-2}D_L^2(1+z)^{-3}
\label{eq:lpco}
\end{equation}
\citep{solomon92}, where $S_\mathrm{CO}\Delta v$ is the velocity integrated flux (Jy\,km\,s$^{-1}$), $\nu_\mathrm{obs}$ is the observed frequency (GHz), $D_L$ is the luminosity distance (Mpc), and $z$ is the redshift of each source, {which was obtained from a visual determination of the center of the spectral profile.}  We use a point source sensitivity of $S/T_A^*=6.1\,$Jy/K. Considering that $\nu_\mathrm{rest}=\nu_\mathrm{obs}(1+z)$ and substituting 
$\nu_\mathrm{rest}(^{12}\mathrm{CO})=115.271$\,GHz and 
$\nu_\mathrm{rest}(^{13}\mathrm{CO})=110.201$\,GHz, we can rewrite Eq.~(\ref{eq:lpco}) in the useful form 

\begin{equation}
L'_\mathrm{^nCO} = A_{^n\mathrm{CO}}\times10^3\left(\frac{S_\mathrm{CO}\Delta v}{\mathrm{Jy\,km\,s}^{-1}}\right)\left(\frac{D_L}{\mathrm{Mpc}}\right)^2(1+z)^{-1},
\end{equation}
where $A_\mathrm{^{12}CO} = 2.45$ and $A_\mathrm{^{13}CO} = 2.68$.

{Three sources in our sample have available archive CO observations \citep{sanders91} with the National Radio Astronomy Observatory (NRAO) 12\,m telescope. {Their CO flux for IRAS\,22491-1808 (($55.0\pm11.0)\times10^8$\,K\,km\,s$^{-1}$\,pc$^2$) is compatible with our results. On the other hand, their IRAS\,F05189-2524  and IRAS\,F14348-1447 measurements (($47.9\pm9.6)\times10^8$ and $(123.0\pm24.6)\times10^8$\,K\,km\,s$^{-1}$\,pc$^2$, respectively), although significantly higher, most likely due to extended emission and the larger FWHM of the NRAO 12\,m telescope {(55\,arcsec)}, are also compatible with our result within $2\sigma$.} To take the extended emission into account, we have performed aperture photometry as detailed in Sect.~\ref{sec:g2d}.}

We obtained the molecular gas mass ($M_\mathrm{H_2}$) from the integrated CO intensities assuming a constant CO-to-H$_2$ conversion factor, $\alpha=1.8\,M_\odot$\,(K\,km\,s$^{-1}$\,pc$^2$)$^{-1}$. The derivation of this factor is {also} explained in detail in Sect.~\ref{sec:g2d}.

\section{Discussion}\label{sec:discussion}

The relationship between the molecular gas content and the star formation properties (rate, efficiency, or depletion time) has been thoroughly studied \citep[e.g.,][]{leroy08,daddi10,genzel10}. Our study allows us to constrain several of these correlations with a large and uniformly observed sample. However, to that end we first need to investigate whether the conversion between CO and H$_2$, $\alpha$, is identical for local spiral galaxies and (U)LIRGs, as is often assumed, and determine its value.

\subsection{Gas-to-dust ratio and determination of $\alpha$}\label{sec:g2d}

There are extensive discussions in the literature over the proper value of $\alpha$ to use when 
deriving molecular gas masses of (U)LIRGs \citep[e.g.,][]{solomon97,downes98,yao03,papadopoulos12a,bolatto13}. Prior to 1997, the standard was to make use of the value of $\alpha$ derived for the Milky Way, $\alpha_\mathrm{MW}\simeq4\,M_\odot$\,(K km\,s$^{-1}$\,pc$^{2}$)$^{-1}$, with 
the reasoning that the molecular gas was bound in physically distinct, virialized clouds, in which increases (decreases) in
the temperature of the gas, $T$, were offset by decreases (increases) in the gas density, $\rho$; that is, 
\begin{equation}M_{\rm cloud} \propto (\rho^{1/2} / T) L'_{\rm CO} 
= ({\rm constant}) \times L'_{\rm CO}\end{equation} 
\citep[e.g.,][]{scoville87}.
However, arguments based on estimates of dynamical masses in (U)LIRGs led to the hypothesis that $\alpha_\mathrm{MW}$ was too high for this type of object. That is, the dynamical masses of (U)LIRGs as traced by CO kinematics 
were lower than the gas masses derived using $\alpha_\mathrm{MW}$ -- and therefore the value of $\alpha$ in these extreme starburst galaxies should be at least proportionally lower \citep[e.g.,][]{downes98}. It was speculated that a primary difference between normal spiral galaxies
and (U)LIRGs is that the ``inter-cloud'' medium regions of the latter population likely contain significant amounts of molecular gas \citep{solomon97}; this means that the molecular gas distribution is more uniform within the starburst region, invalidating the Milky Way assumption of discrete molecular clouds.

To address this issue, in this paper we
assume that (i) the relative mass fraction of atomic and molecular hydrogen is similar among our galaxies ($M_{\rm HI}/M_{\rm H_2} = \gamma$), 
and (ii) any observed offset is driven by the physics setting the value of $\alpha$ for each population. Therefore, we assume that
\begin{equation}\label{eq:fractions}
\begin{split}{M_{\rm gas} \over M_{\rm dust}} = {(1+\gamma)\alpha_\mathrm{n} L'_{\rm CO} \over M_{\rm dust}} ({\rm normal})  = {(1+\gamma)\alpha_\mathrm{L} L'_{\rm CO} 
\over M_{\rm dust}}({\rm LIRG})
\end{split}
,\end{equation}
where n and L subscripts are used for normal and LIRGs, respectively \citep[see][for an extensive discussion of dust-based conversion factor determinations]{bolatto13}.
To proceed, we used for comparison the Key Insights on Nearby Galaxies: a Far-Infrared Survey with Herschel (KINGFISH)\footnote{\url{https://www.ast.cam.ac.uk/research/kingfish}} sample of local spiral galaxies observed with 
{\it Herschel} by \citet{dale12} that have pre-existing CO measurements {in two surveys: the FCRAO extragalactic CO survey \citep{young95}, with $^{12}$CO(1-0) measurements, and the HERACLES survey from \citet{leroy09}, with $^{12}$CO(2-1) observations, for which we have used a conversion factor of $R_{21}=0.8$ for the conversion between $^{12}$CO(2-1) and $^{12}$CO(1-0) as suggested by the authors \citep{leroy09}.} Both sets of data encompass most of the galaxies' emission. In particular, the FCRAO sample was observed with a 14\,m dish (i.e., FWHM as large as $\simeq45\,$arcsec), while the HERACLES galaxies were spatially mapped with the Heterodyne Receiver Array (HERA) multi-pixel receiver on the IRAM-30\,m telescope.

\begin{figure}\centering
\includegraphics[width=.95\columnwidth]{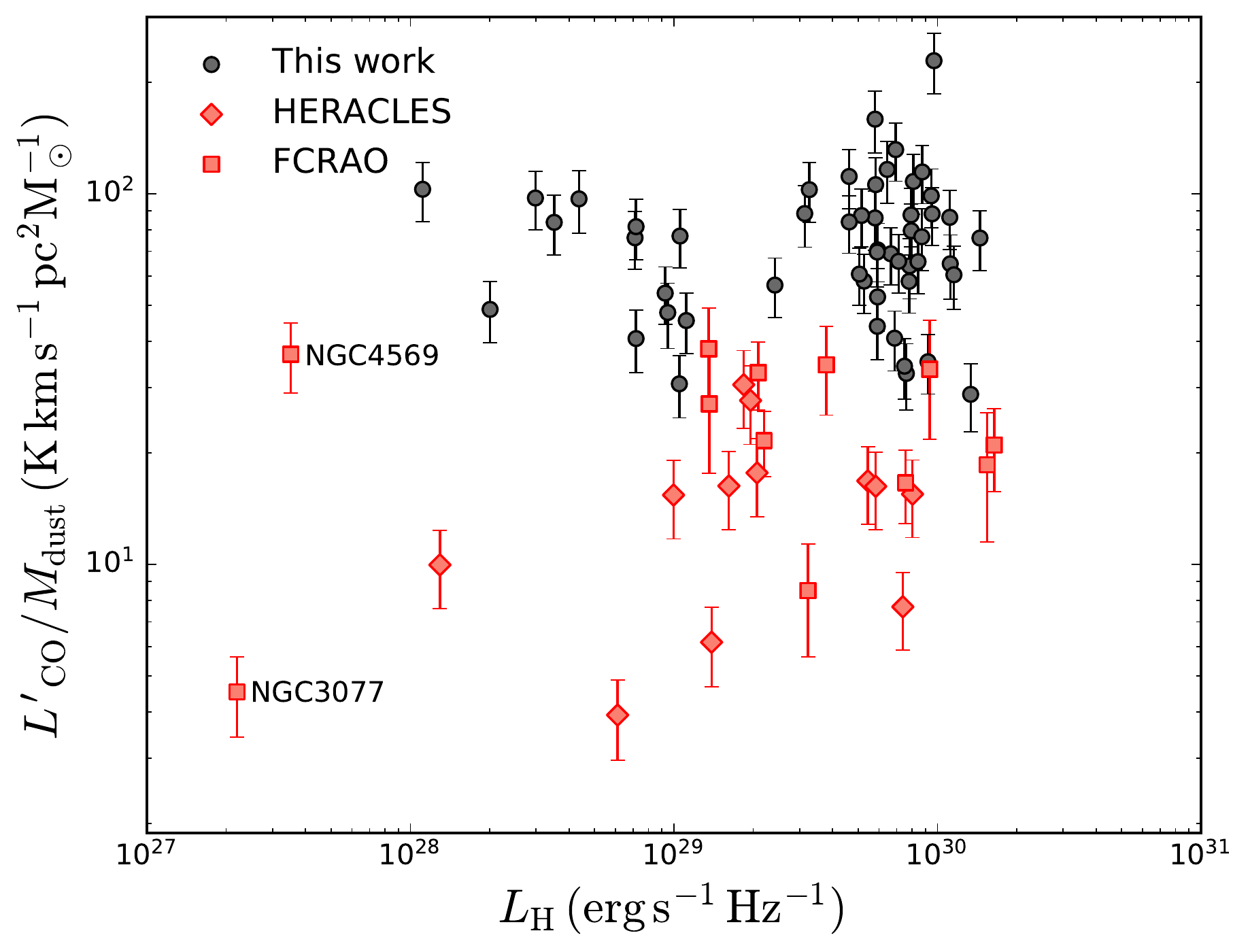}
\caption{Ratio between the CO luminosity and the dust mass as a function of the rest-frame H-band luminosity, tracer of stellar mass. There is no obvious trend in our sample (in black) or in the local comparison samples (in red). {Given the low H-band luminosity of NGC\,3077 and NGC\,4569, these were excluded from the comparison. Details are given in the main text.}}
\label{fig:hband}
\end{figure}

There is a correlation between the gas-to-dust ratio and both the stellar mass \citep{cortese16} and metallicity \citep[e.g.,][]{remy-ruyer14}. These two correlations are likely linked through the mass-metallicity relation \citep[][]{tremonti04}.
To check whether we expect a variation of gas-to-dust ratios among our samples, we compared $L'_\mathrm{CO}/M_\mathrm{dust}$ with the Two Micron All-Sky Survey (2MASS) H-band luminosity, which is a first order tracer of the stellar mass of galaxies. As shown in Fig.~\ref{fig:hband}, we do not find any trend, either in our sample or in the comparison FCRAO or HERACLES sources, so no stellar mass correction was applied to our derived gas-to-dust ratios.
{The reason why a clear dependency is not seen in the figure is likely to be in the mass-metallicity relation itself: while we expect a difference in metallicity of $\sim0.1$\,dex for a stellar mass variation of two orders of magnitude (the range of GOALS galaxies is $9.5 < \log(M_\star) < 11.5$) \citep{torrey18}, there is a large scatter between the two magnitudes, of approximately $\sim0.2-0.3$\,dex.
This scatter is the cause of why we do not see variations with the gas-to-dust ratio over the traced range of stellar masses.}
From Fig. \ref{fig:hband} it is also noticeable that two sources from the FCRAO comparison sample (NGC\,3077 and NGC\,4569) have significantly lower H-band luminosities than any other source in this study. These were excluded from the analysis below, as they are not representative of the global characteristics of both populations.

To obtain the dust properties of the galaxies, both for our sources and for the local comparison samples, we constructed far infrared (FIR) spectral energy distributions (SEDs) using the new {\it Herschel} photometric catalog by \citep{chu17} between 70 and 350$\mu$m in order to perform a model fitting. Although the angular resolution of the IRAM-30\,m telescope ($\sim22''$) is generally sufficient to encompass most of the FIR extension of each galaxy, this is not always the case, as shown in the left panels of the molecular gas observations of Figs.~\ref{fig:lineprof} and \ref{figapp:lineprof}, where we have plotted the IRAM beam sizes over the \emph{Herschel} $70\,\mu$m images. In order to compare gas (derived from CO) and dust (derived from FIR) properties, it is essential to consider the same apertures for both datasets. To ensure this, we convolved our \emph{Herschel} images to the angular resolution of our IRAM-30\,m data using Gaussian smoothing, and performed aperture photometry centered in the IRAM-30\,m pointing position. We did not smooth our \emph{Herschel} Spectral and Photometric Imaging Receiver (SPIRE) $350\,\mu$m images, which already have a resolution ($\sim25''$) comparable with the \twCO measurements. We excluded our $500\,\mu$m images due to their coarser resolution than the IRAM-30\,m beam.

We then fit each SED to an optically thin modified blackbody to derive the dust temperatures, $T_\mathrm{dust}$, and masses, $M_\mathrm{dust}$ for each galaxy (see Appendix~\ref{app:mbbsedfit}). 
{To calculate the dust masses, \citet{scoville16} argue for the use of a common mass-weighted dust temperature for all sources ($T_\mathrm{dust}=25$\;K), instead of that obtained from their individual SED fits, which is more representative from localized warm, luminous regions.
In this paper we follow this strategy, so we note that any comparison between samples must take this difference into account. For our sample, {the use of} a fixed $T_\mathrm{dust}=25$\;K yields an average $M_\mathrm{dust}$ 1.53 {times} higher than when a SED fitted temperature is used.
}

By integrating below the fitted blackbody we directly obtain the luminosity between 42.5 and 122.5\,$\mu$m, $L_\mathrm{FIR}$. 
Additionally, we also calculate the total infrared (TIR) luminosity (between 3 and 1100\,$\mu$m, $L_\mathrm{TIR}$), making use of the prescription provided by \citet{dale02}:
\begin{equation}\label{eq:ltir}
\begin{split}
L_\mathrm{TIR} = 1.559\,\nu L_\nu(24\,\mu\mathrm{m}) + 0.7686\,\nu L_\nu(70\,\mu\mathrm{m}) + \\
+ 1.347\,\nu L_\nu(160\,\mu\mathrm{m}),
\end{split}
\end{equation}
using the \emph{Spitzer} Multiband Imaging Photometer (MIPS) $24\,\mu\mathrm{m}$ data from Mazzarella et al. (in prep.) for our galaxies (after applying the same Gaussian smoothing described above), and the local sample data from \citet{dale07,dale09} and \citet{engelbracht08}. We note that the TIR luminosities are obtained from galaxy-integrated photometry, and thus are not matched to the IRAM beam. All fitted parameters (plus $L_\mathrm{TIR}$) are presented in Table~\ref{table:sedfit}. For completeness, we provide as well the the galaxy-integrated parameters obtained by using the total flux for each filter (including SPIRE $500\,\mu$m in this case). Unless specifically mentioned otherwise, throughout this paper we always refer to or use the parameters obtained from the SED fitting to the aperture-matched photometry. For details on the modified blackbody SED fitting, we refer the reader to Appendix~\ref{app:mbbsedfit}.

\begin{table*}
\caption{\label{table:sedfit}Derived dust parameters and luminosities for our (U)LIRG sample.}
\centering\small
\begin{tabular}{lccccccccccc}
\hline\hline
{} & \multicolumn{3}{c}{Overall SED fit}  &  {} & \multicolumn{5}{c}{Within IRAM-30\,m beam}  \\
\cline{2-4}\cline{6-10} \\
    {Source name} & {$T_\mathrm{dust}$} & {$M_\mathrm{dust}^\mathrm{fit}$} & {$\log(L_\mathrm{FIR})$} & {\,\,\,} & {$T_\mathrm{dust}$} & {$M_\mathrm{dust}^\mathrm{fit}$} & {$M_{\mathrm{dust}}^\mathrm{25K}$} & {$\log(L_\mathrm{FIR})$} & {$\log(L_\mathrm{TIR})$} \\
{} & {(K)} & {($10^7\,M_\odot$)} & {($L_\odot$)}  & {\,\,\,} &{(K)} & {($10^7\,M_\odot$)} & {($10^7\,M_\odot$)} & {($L_\odot$)} & {($L_\odot$)} \\
\hline
NGC 0034 & $34.87 \pm 0.59$ & $4.11_{-0.40}^{+0.42}$ & $11.15_{-0.08}^{+0.07}$ &  & $36.32 \pm 1.08$ & $1.35_{-0.22}^{+0.23}$ & $3.17_{-0.50}^{+0.55}$ & $10.77_{-0.13}^{+0.12}$ & $11.02$ \\
Arp 256N & $25.65 \pm 2.14$ & $4.18_{-2.13}^{+2.87}$ & $10.29_{-0.50}^{+0.37}$  &  & $26.04 \pm 0.51$ & $1.25_{-0.17}^{+0.19}$ & $1.38_{-0.19}^{+0.21}$ & $9.81_{-0.12}^{+0.09}$ & $10.12$ \\
Arp 256S & $31.50 \pm 0.64$ & $6.11_{-0.83}^{+0.89}$ & $11.04_{-0.11}^{+0.09}$&  & $33.22 \pm 0.87$ & $1.88_{-0.29}^{+0.31}$ & $3.68_{-0.56}^{+0.60}$ & $10.67_{-0.12}^{+0.11}$ & $10.99$ \\
IC 1623 & $30.60 \pm 0.58$ & $13.72_{-1.25}^{+1.31}$ & $11.32_{-0.09}^{+0.07}$ &  & $32.11 \pm 0.85$ & $3.71_{-0.58}^{+0.63}$ & $6.73_{-1.05}^{+1.14}$ & $10.89_{-0.13}^{+0.11}$ & $11.21$ \\
MCG -03-04-014 & $30.14 \pm 0.56$ & $15.29_{-1.78}^{+1.89}$ & $11.32_{-0.10}^{+0.07}$ &  & $30.68 \pm 0.74$ & $6.26_{-0.94}^{+1.01}$ & $10.31_{-1.55}^{+1.67}$ & $10.98_{-0.12}^{+0.10}$ & $11.27$ \\
IRAS F01364-1042 & $34.92 \pm 0.82$ & $10.50_{-1.42}^{+1.52}$ & $11.54_{-0.11}^{+0.09}$ &  & $35.89 \pm 1.06$ & $4.14_{-0.66}^{+0.72}$ & $9.63_{-1.54}^{+1.67}$ & $11.20_{-0.13}^{+0.12}$ & $11.37$ \\
IC 0214 & $28.09 \pm 0.85$ & $11.72_{-2.27}^{+2.51}$ & $11.00_{-0.16}^{+0.13}$ &  & $29.84 \pm 0.69$ & $3.70_{-0.54}^{+0.58}$ & $5.71_{-0.84}^{+0.90}$ & $10.67_{-0.12}^{+0.10}$ & $10.96$ \\
UGC 01845 & $29.54 \pm 0.41$ & $5.40_{-0.50}^{+0.52}$ & $10.82_{-0.07}^{+0.06}$ &  & $29.92 \pm 0.72$ & $2.09_{-0.31}^{+0.34}$ & $3.23_{-0.48}^{+0.52}$ & $10.44_{-0.12}^{+0.10}$ & $10.73$ \\
NGC 0958 & $22.96 \pm 0.78$ & $30.26_{-6.27}^{+7.08}$ & $10.81_{-0.18}^{+0.15}$ &  & $23.86 \pm 0.45$ & $4.51_{-0.65}^{+0.70}$ & $3.99_{-0.57}^{+0.62}$ & $10.11_{-0.12}^{+0.09}$ & $10.50$ \\
ESO 550-IG025 & $28.79 \pm 0.35$ & $9.23_{-0.81}^{+0.84}$ & $10.97_{-0.07}^{+0.05}$ &  & $28.19 \pm 0.64$ & $5.18_{-0.78}^{+0.84}$ & $7.00_{-1.05}^{+1.14}$ & $10.66_{-0.13}^{+0.10}$ & $10.96$ \\
UGC 03094 & $26.39 \pm 0.59$ & $18.25_{-2.62}^{+2.82}$ & $11.02_{-0.12}^{+0.09}$ &  & $27.23 \pm 0.60$ & $5.43_{-0.81}^{+0.88}$ & $6.74_{-1.01}^{+1.09}$ & $10.58_{-0.12}^{+0.10}$ & $10.91$ \\
NGC 1797 & $29.05 \pm 0.56$ & $4.23_{-0.52}^{+0.55}$ & $10.67_{-0.10}^{+0.08}$ &  & $30.79 \pm 0.76$ & $1.25_{-0.19}^{+0.20}$ & $2.06_{-0.31}^{+0.34}$ & $10.30_{-0.12}^{+0.11}$ & $10.62$ \\
IRAS F05189-2524 & $37.28 \pm 0.76$ & $10.49_{-1.20}^{+1.27}$ & $11.70_{-0.09}^{+0.08}$ &  & $37.63 \pm 1.19$ & $4.31_{-0.71}^{+0.77}$ & $11.01_{-1.82}^{+1.98}$ & $11.34_{-0.13}^{+0.12}$ & $11.69$ \\
IRAS F05187-1017 & $28.98 \pm 0.61$ & $10.10_{-1.31}^{+1.40}$ & $11.03_{-0.11}^{+0.09}$ &  & $29.46 \pm 0.71$ & $4.10_{-0.63}^{+0.68}$ & $6.14_{-0.94}^{+1.01}$ & $10.68_{-0.13}^{+0.10}$ & $10.91$ \\
IRAS F06076-2139 & $30.95 \pm 0.83$ & $12.17_{-1.97}^{+2.14}$ & $11.29_{-0.14}^{+0.11}$ &  & $32.09 \pm 0.85$ & $4.60_{-0.72}^{+0.78}$ & $8.39_{-1.31}^{+1.42}$ & $10.96_{-0.13}^{+0.11}$ & $11.22$ \\
NGC 2341 & $28.58 \pm 0.52$ & $5.50_{-0.63}^{+0.67}$ & $10.73_{-0.09}^{+0.08}$ &  & $30.20 \pm 0.72$ & $1.43_{-0.21}^{+0.23}$ & $2.26_{-0.34}^{+0.36}$ & $10.30_{-0.12}^{+0.10}$ & $10.59$ \\
NGC 2342 & $26.77 \pm 0.64$ & $10.50_{-1.49}^{+1.61}$ & $10.82_{-0.12}^{+0.10}$ &  & $28.23 \pm 0.64$ & $1.78_{-0.26}^{+0.28}$ & $2.40_{-0.36}^{+0.38}$ & $10.21_{-0.12}^{+0.10}$ & $10.55$ \\
IRAS 07251-0248 & $33.71 \pm 0.81$ & $45.47_{-6.38}^{+6.86}$ & $12.05_{-0.12}^{+0.09}$ &  & $33.28 \pm 0.95$ & $22.53_{-3.70}^{+4.03}$ & $45.51_{-7.47}^{+8.14}$ & $11.71_{-0.15}^{+0.11}$ & $11.96$ \\
IRAS F09111-1007 W & $30.30 \pm 0.55$ & $35.48_{-4.49}^{+4.77}$ & $11.68_{-0.11}^{+0.07}$ &  & $29.77 \pm 0.72$ & $15.29_{-2.34}^{+2.54}$ & $23.66_{-3.63}^{+3.92}$ & $11.27_{-0.13}^{+0.09}$ & $11.52$ \\
IRAS F09111-1007 E & $27.79 \pm 0.78$ & $16.79_{-3.22}^{+3.55}$ & $11.11_{-0.17}^{+0.11}$ &  & $26.55 \pm 0.55$ & $8.09_{-1.18}^{+1.28}$ & $9.47_{-1.39}^{+1.49}$ & $10.66_{-0.14}^{+0.08}$ & $11.01$ \\
UGC 05101 & $29.49 \pm 0.42$ & $47.76_{-4.29}^{+4.49}$ & $11.74_{-0.08}^{+0.05}$ &  & $29.96 \pm 0.74$ & $17.41_{-2.63}^{+2.85}$ & $27.21_{-4.11}^{+4.45}$ & $11.35_{-0.13}^{+0.10}$ & $11.61$ \\
2MASX J11210825-0259399  & $22.94 \pm 2.60$ & $1.65_{-0.90}^{+1.32}$ & $9.54_{-0.61}^{+0.45}$ &  & $25.06 \pm 0.49$ & $0.53_{-0.08}^{+0.08}$ & $0.53_{-0.08}^{+0.08}$ & $9.32_{-0.12}^{+0.09}$ & $9.67$ \\
CGCG 011-076 & $27.01 \pm 0.52$ & $13.98_{-1.71}^{+1.82}$ & $10.97_{-0.11}^{+0.08}$ &  & $28.23 \pm 0.65$ & $3.81_{-0.57}^{+0.62}$ & $5.16_{-0.78}^{+0.84}$ & $10.53_{-0.12}^{+0.10}$ & $10.88$ \\
IRAS F12224-0624 & $33.46 \pm 0.67$ & $46.70_{-5.70}^{+6.05}$ & $12.05_{-0.10}^{+0.08}$ &  & $32.09 \pm 0.85$ & $2.06_{-0.32}^{+0.35}$ & $3.74_{-0.58}^{+0.63}$ & $10.62_{-0.13}^{+0.11}$ & $10.82$ \\
CGCG 043-099 & $28.89 \pm 0.54$ & $21.17_{-2.57}^{+2.73}$ & $11.33_{-0.10}^{+0.07}$ &  & $30.16 \pm 0.73$ & $5.41_{-0.82}^{+0.88}$ & $8.58_{-1.29}^{+1.40}$ & $10.86_{-0.13}^{+0.10}$ & $11.12$ \\
ESO 507-G070 & $33.06 \pm 0.61$ & $7.91_{-0.89}^{+0.94}$ & $11.29_{-0.09}^{+0.08}$ &  & $33.67 \pm 0.93$ & $2.25_{-0.35}^{+0.38}$ & $4.53_{-0.71}^{+0.77}$ & $10.79_{-0.13}^{+0.11}$ & $11.00$ \\
NGC 5104 & $27.07 \pm 0.43$ & $12.25_{-1.28}^{+1.35}$ & $10.92_{-0.09}^{+0.07}$ &  & $27.75 \pm 0.61$ & $3.14_{-0.46}^{+0.50}$ & $4.08_{-0.60}^{+0.65}$ & $10.40_{-0.12}^{+0.10}$ & $10.72$ \\
IC 4280 & $25.62 \pm 0.47$ & $12.86_{-1.56}^{+1.66}$ & $10.78_{-0.10}^{+0.08}$ &  & $26.82 \pm 0.57$ & $2.47_{-0.36}^{+0.39}$ & $2.95_{-0.43}^{+0.46}$ & $10.20_{-0.12}^{+0.10}$ & $10.53$ \\
NGC 5258 & $26.14 \pm 0.50$ & $18.35_{-2.15}^{+2.29}$ & $10.99_{-0.10}^{+0.08}$ &  & $27.59 \pm 0.59$ & $3.52_{-0.51}^{+0.55}$ & $4.51_{-0.66}^{+0.71}$ & $10.43_{-0.12}^{+0.09}$ & $10.75$ \\
UGC 08739 & $24.07 \pm 0.56$ & $20.62_{-3.03}^{+3.29}$ & $10.79_{-0.13}^{+0.10}$ &  & $24.95 \pm 0.49$ & $4.38_{-0.63}^{+0.68}$ & $4.36_{-0.63}^{+0.67}$ & $10.23_{-0.12}^{+0.09}$ & $10.57$ \\
NGC 5331 & $29.33 \pm 0.87$ & $16.55_{-3.90}^{+4.34}$ & $11.27_{-0.18}^{+0.14}$ &  & $28.15 \pm 0.62$ & $7.87_{-1.15}^{+1.24}$ & $10.60_{-1.55}^{+1.67}$ & $10.83_{-0.12}^{+0.09}$ & $11.12$ \\
CGCG 247-020 & $32.02 \pm 0.72$ & $5.11_{-0.70}^{+0.75}$ & $11.01_{-0.11}^{+0.09}$ &  & $32.52 \pm 0.84$ & $1.88_{-0.29}^{+0.31}$ & $3.51_{-0.54}^{+0.58}$ & $10.62_{-0.12}^{+0.11}$ & $10.91$ \\
IRAS F14348-1447 & $34.02 \pm 0.61$ & $26.38_{-2.72}^{+2.87}$ & $11.85_{-0.09}^{+0.07}$ &  & $32.71 \pm 0.90$ & $26.18_{-4.21}^{+4.57}$ & $50.76_{-8.15}^{+8.86}$ & $11.74_{-0.14}^{+0.10}$ & $11.96$ \\
CGCG 049-057 & $30.98 \pm 0.52$ & $7.50_{-0.80}^{+0.84}$ & $11.10_{-0.08}^{+0.07}$ &  & $31.59 \pm 0.82$ & $2.02_{-0.31}^{+0.34}$ & $3.53_{-0.55}^{+0.59}$ & $10.58_{-0.12}^{+0.11}$ & $10.79$ \\
NGC 5936 & $26.55 \pm 0.50$ & $8.95_{-1.09}^{+1.17}$ & $10.73_{-0.10}^{+0.08}$ &  & $28.43 \pm 0.64$ & $1.44_{-0.21}^{+0.23}$ & $1.97_{-0.29}^{+0.31}$ & $10.14_{-0.12}^{+0.10}$ & $10.48$ \\
IRAS F16164-0746 & $31.91 \pm 0.62$ & $12.02_{-1.47}^{+1.56}$ & $11.37_{-0.10}^{+0.08}$ &  & $32.05 \pm 0.85$ & $3.02_{-0.47}^{+0.51}$ & $5.46_{-0.86}^{+0.93}$ & $10.79_{-0.13}^{+0.11}$ & $11.01$ \\
CGCG 052-037 & $28.57 \pm 0.60$ & $13.65_{-1.86}^{+1.99}$ & $11.12_{-0.11}^{+0.09}$ &  & $29.95 \pm 0.71$ & $3.57_{-0.53}^{+0.57}$ & $5.54_{-0.83}^{+0.89}$ & $10.67_{-0.12}^{+0.10}$ & $10.97$ \\
IRAS F16399-0937 & $29.74 \pm 1.13$ & $12.66_{-3.17}^{+3.60}$ & $11.20_{-0.21}^{+0.17}$ &  & $29.69 \pm 0.76$ & $4.31_{-0.69}^{+0.75}$ & $6.56_{-1.05}^{+1.14}$ & $10.73_{-0.13}^{+0.11}$ & $11.06$ \\
NGC 6285 & $26.13 \pm 0.76$ & $3.19_{-0.57}^{+0.63}$ & $10.24_{-0.15}^{+0.12}$ &  & $28.41 \pm 0.62$ & $0.83_{-0.12}^{+0.13}$ & $1.14_{-0.16}^{+0.18}$ & $9.89_{-0.12}^{+0.10}$ & $10.17$ \\
NGC 6286 & $25.12 \pm 0.31$ & $23.83_{-2.01}^{+2.10}$ & $10.99_{-0.07}^{+0.05}$ &  & $26.17 \pm 0.53$ & $6.25_{-0.90}^{+0.97}$ & $7.02_{-1.01}^{+1.09}$ & $10.53_{-0.12}^{+0.09}$ & $10.84$ \\
IRAS F17138-1017 & $31.49 \pm 0.60$ & $7.82_{-0.93}^{+0.98}$ & $11.16_{-0.09}^{+0.08}$ &  & $31.33 \pm 0.81$ & $2.46_{-0.38}^{+0.41}$ & $4.22_{-0.65}^{+0.71}$ & $10.64_{-0.12}^{+0.11}$ & $10.94$ \\
UGC 11041 & $26.01 \pm 0.47$ & $9.67_{-1.11}^{+1.18}$ & $10.70_{-0.10}^{+0.08}$ &  & $27.56 \pm 0.60$ & $2.30_{-0.34}^{+0.36}$ & $2.94_{-0.43}^{+0.46}$ & $10.25_{-0.12}^{+0.10}$ & $10.57$ \\
CGCG 141-034 & $30.20 \pm 0.57$ & $5.69_{-0.70}^{+0.74}$ & $10.90_{-0.10}^{+0.08}$ &  & $30.04 \pm 0.73$ & $2.01_{-0.30}^{+0.33}$ & $3.14_{-0.47}^{+0.51}$ & $10.43_{-0.12}^{+0.10}$ & $10.70$ \\
IRAS 18090+0130 & $28.06 \pm 0.92$ & $4.78_{-1.08}^{+1.21}$ & $10.61_{-0.19}^{+0.15}$ &  & $27.49 \pm 0.62$ & $7.39_{-1.12}^{+1.22}$ & $9.39_{-1.43}^{+1.54}$ & $10.74_{-0.13}^{+0.10}$ & $11.06$ \\
NGC 6701 & $27.13 \pm 0.46$ & $8.04_{-0.88}^{+0.93}$ & $10.75_{-0.09}^{+0.07}$ &  & $28.87 \pm 0.67$ & $1.50_{-0.22}^{+0.24}$ & $2.13_{-0.32}^{+0.34}$ & $10.20_{-0.12}^{+0.10}$ & $10.52$ \\
NGC 6786 & $27.52 \pm 0.47$ & $9.49_{-1.07}^{+1.14}$ & $10.86_{-0.10}^{+0.07}$ &  & $28.52 \pm 0.64$ & $2.78_{-0.41}^{+0.44}$ & $3.86_{-0.56}^{+0.61}$ & $10.43_{-0.12}^{+0.10}$ & $10.74$ \\
UGC 11415 & $29.03 \pm 0.58$ & $4.76_{-0.59}^{+0.63}$ & $10.71_{-0.10}^{+0.08}$ &  & $30.18 \pm 0.73$ & $1.59_{-0.24}^{+0.26}$ & $2.51_{-0.38}^{+0.41}$ & $10.34_{-0.12}^{+0.10}$ & $10.72$ \\
ESO 593-IG008 & $28.57 \pm 0.54$ & $46.04_{-5.57}^{+5.93}$ & $11.63_{-0.11}^{+0.07}$ &  & $28.74 \pm 0.67$ & $16.27_{-2.46}^{+2.66}$ & $23.10_{-3.50}^{+3.78}$ & $11.20_{-0.13}^{+0.09}$ & $11.50$ \\
NGC 6907 & $24.47 \pm 0.51$ & $14.87_{-1.94}^{+2.09}$ & $10.71_{-0.11}^{+0.09}$ &  & $26.75 \pm 0.57$ & $1.24_{-0.18}^{+0.20}$ & $1.47_{-0.22}^{+0.23}$ & $9.90_{-0.12}^{+0.10}$ & $10.25$ \\
IRAS 21101+5810 & $32.71 \pm 0.77$ & $11.78_{-1.79}^{+1.92}$ & $11.42_{-0.12}^{+0.10}$ &  & $28.26 \pm 0.73$ & $6.31_{-1.06}^{+1.15}$ & $8.59_{-1.44}^{+1.57}$ & $10.75_{-0.14}^{+0.11}$ & $11.07$ \\
ESO 602-G025 & $27.22 \pm 0.51$ & $13.79_{-1.71}^{+1.82}$ & $10.99_{-0.10}^{+0.08}$ &  & $28.75 \pm 0.67$ & $3.92_{-0.59}^{+0.64}$ & $5.55_{-0.83}^{+0.90}$ & $10.60_{-0.12}^{+0.10}$ & $10.92$ \\
UGC 12150 & $28.18 \pm 0.36$ & $11.61_{-0.97}^{+1.02}$ & $11.01_{-0.07}^{+0.05}$ &  & $28.87 \pm 0.67$ & $3.99_{-0.60}^{+0.65}$ & $5.69_{-0.85}^{+0.92}$ & $10.62_{-0.12}^{+0.10}$ & $10.93$ \\
IRAS F22491-1808 & $37.05 \pm 0.83$ & $17.27_{-2.10}^{+2.23}$ & $11.88_{-0.10}^{+0.09}$ &  & $36.99 \pm 1.15$ & $7.91_{-1.31}^{+1.42}$ & $19.93_{-3.29}^{+3.58}$ & $11.54_{-0.13}^{+0.12}$ & $11.75$ \\
CGCG 453-062 & $28.28 \pm 0.62$ & $12.32_{-1.80}^{+1.94}$ & $11.05_{-0.12}^{+0.09}$ &  & $29.26 \pm 0.70$ & $3.87_{-0.59}^{+0.64}$ & $5.71_{-0.87}^{+0.94}$ & $10.64_{-0.12}^{+0.10}$ & $10.92$ \\
2MASX J23181352+0633267 & $25.38 \pm 0.77$ & $11.07_{-2.08}^{+2.30}$ & $10.69_{-0.16}^{+0.13}$ &  & $23.42 \pm 0.40$ & $0.38_{-0.05}^{+0.05}$ & $0.32_{-0.04}^{+0.05}$ & $8.98_{-0.11}^{+0.08}$ & $9.46$           \\
\hline
\end{tabular}
\tablefoot{While $L_\mathrm{FIR}$ is measured between 42.5 and 122.5\,$\mu$m, $L_\mathrm{TIR}$ is measured between 3 and 1100\,$\mu$m and is obtained from Eq.~\ref{eq:ltir}. We have assumed an uncertainty in $L_\mathrm{TIR}$ of 20\%. {We note that dust masses $M_\mathrm{dust}^\mathrm{fit}$ are derived evaluating the modified blackbody function using the dust temperature from the models, in contrast with $M_\mathrm{dust}^\mathrm{25K}$, obtained assuming $T_\mathrm{dust}=25$\;K, as suggested in \citet{scoville16}. The latter are used as the preferred method in this study}.}
\end{table*}

\begin{figure*}\centering
\includegraphics[width=.76\textwidth]{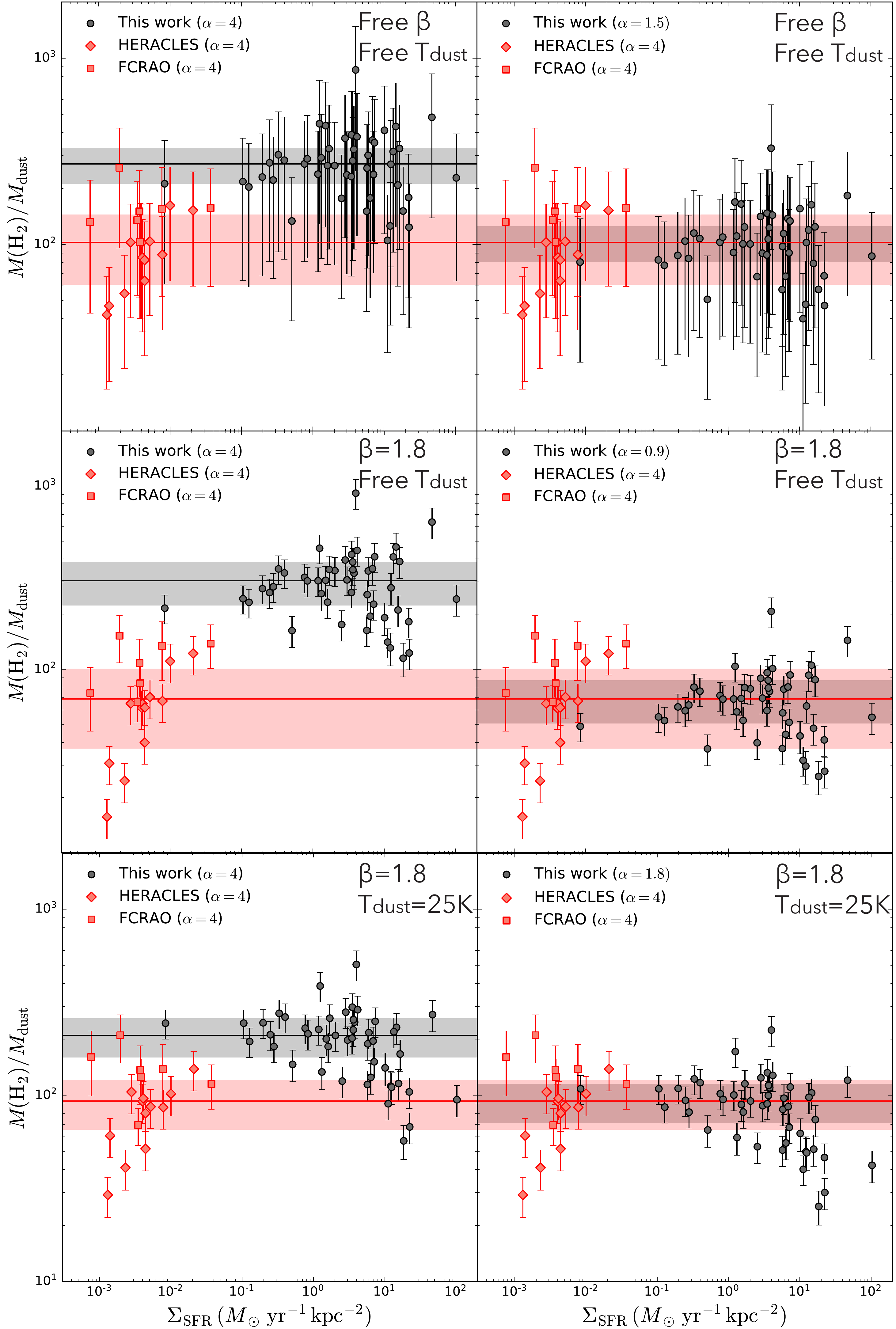}
\caption{{Gas-to-dust mass ratio for both the reference sample, consisting of HERACLES (red diamonds) and FCRAO (red squares) sources, and our (U)LIRG sample (black circles) {as a function of $\Sigma_\mathrm{SFR}$}. Horizontal lines show the median value for each sample, and shaded areas indicate their {median absolute deviations}. In the top panels, $M_\mathrm{dust}$ has been obtained leaving $\beta$ as a free parameter in the SED fit and using the fitted $T_\mathrm{dust}$ to derive dust mass. In the middle panels the emissivity index has been fixed to $\beta=1.8$, but $M_\mathrm{dust}$ is still obtained from the fitted dust temperatures. Finally, in the bottom panels, emissivity is fixed to 1.8, and dust masses are obtained assuming a fixed $25$\,K dust temperature (see main text for further details). Left panels show the results of using a Milky Way CO-to-H$_2$ conversion factor ($\alpha=4$). To make the gas-to-dust mass ratio of (U)LIRGs consistent with the reference sample, $\alpha$ has been modified to the value quoted in the right panels. 
For the remainder of this work, a fixed dust emissivity of $\beta=1.8$ is assumed, and we adopt a CO-to-H$_2$ factor of $\alpha=1.8$ for (U)LIRGs, as required to make the median gas-to-dust mass ratio of (U)LIRGs match that of the reference sample of (non-LIRG) spiral galaxies.}
}
\label{fig:gasdustlir}
\end{figure*}

In Fig.~\ref{fig:gasdustlir} we show the gas-to-dust mass ratio ($M_{\rm H_2} / M_{\rm dust}$) for both our (U)LIRGs and the local control sample (FCRAO + HERACLES) {as a function of the star formation rate (SFR) surface density ($\Sigma_\mathrm{SFR}$; obtained from \citet{leroy13} and \citet{diaz-santos17}), showing that the two populations have distinctly different $\Sigma_\mathrm{SFR}$ and thus physical conditions \citep[e.g.,][]{diaz-santos17}.} 
{The top panels of the figure show the gas-to-dust mass ratios obtained by deriving the dust masses based on the fitted 250\,$\mu$m continuum, using a mass absorption coefficient of $\kappa(250\,\mu\mathrm{m})=0.48\,\mathrm{m}^2\,\mathrm{kg}^{-1}$ (see Appendix~\ref{app:mbbsedfit} for details) and by leaving $T_\mathrm{dust}$ and $\beta$ as free parameters. The middle panels are the same as the top, but for a fixed emissivity index of $\beta=1.8$ \citep{planck11}. Finally, the bottom panels show the gas-to-dust mass ratios obtained by fixing $\beta$ and assuming a fixed 25\,K dust temperature for every source, as proposed by \citet{scoville16}. For the remainder of the discussion we have adopted the results from the latter method, while we occasionally refer to the results obtained when letting $T_\mathrm{dust}$ free.
}

If a Milky Way $\alpha$ is adopted for the reference sample of normal star-forming galaxies, we find that in order to match the offset between the median gas-to-dust ratio of (U)LIRGs relative to the reference sample, within their {median absolute deviations}, {it is necessary to adopt a value $\alpha_\mathrm{(U)LIRG}=1.8^{+1.3}_{-0.8}\,M_\odot$\,(K km\,s$^{-1}$\,pc$^{2}$)$^{-1}$, assuming a fixed $T_\mathrm{dust}$. If we instead use the dust temperatures from SED fitting to derive the dust masses, we require $\alpha_\mathrm{(U)LIRG}=0.9^{+0.9}_{-0.5}$ in order to match the gas-to-dust ratios}. 
Our derived value of $\alpha$ is compatible with the values derived by \citet{downes98} ($\alpha=0.8$), \citet{papadopoulos12a} ($\alpha=0.6\pm0.2$), and \citet{scoville15} for the (U)LIRG Arp\,220 ($\alpha\sim2$).

\subsubsection{Impact of varying ISM assumptions}
We stress that our derived value of $\alpha$ strongly depends on two assumptions: a constant gas-to-dust ratio and a constant relative fraction of atomic to molecular hydrogen gas, $\gamma$. Some studies have claimed evidence of the gas-to-dust ratio of (U)LIRGs being compatible with that of normal galaxies \citep[see, e.g.,][]{wilson08}.  
However, these ratios are calibrated based on a lower value of $\alpha$ for (U)LIRGs, falling in a circular argument.

{{In any case, even if either of the two ratios is proven to be different in LIRGs, the following relation derived from Eq.~\ref{eq:fractions} and the offset between samples seen in Fig.~\ref{fig:gasdustlir} should still hold:
}}
\begin{equation}
\alpha_\mathrm{L} = \left(0.45^{+0.3}_{-0.2}\right)\alpha_\mathrm{n}\left(\frac{1+\gamma_\mathrm{n}}{1+\gamma_\mathrm{L}}\right)\left(\frac{L'_\mathrm{CO}}{M_\mathrm{dust}}\right)_\mathrm{n}\left(\frac{L'_\mathrm{CO}}{M_\mathrm{dust}}\right)^{-1}_\mathrm{L}
.\end{equation}

{{We note that if the assumption is made that the gas-to-dust mass ratio and $\alpha$ are constant, but instead we vary the relative fraction of atomic to molecular gas between the two samples, the end result would be a decrease in $\gamma$ for the (U)LIRGs. In particular, we would get $(1+\gamma_\mathrm{n})/(1+\gamma_\mathrm{L})=2.2$. This equation yields unphysical negative $\gamma_\mathrm{L}$ values for any $\gamma_\mathrm{n}$ value below $1.2$.}
The observed difference in Fig.~\ref{fig:gasdustlir} between the two is unlikely thus to be ascribed solely to a different $\gamma$.
}

\subsection{Star formation efficiencies and gas depletion timescales} \label{sec:sfprop}

\begin{figure}\centering
\includegraphics[width=0.9\columnwidth]{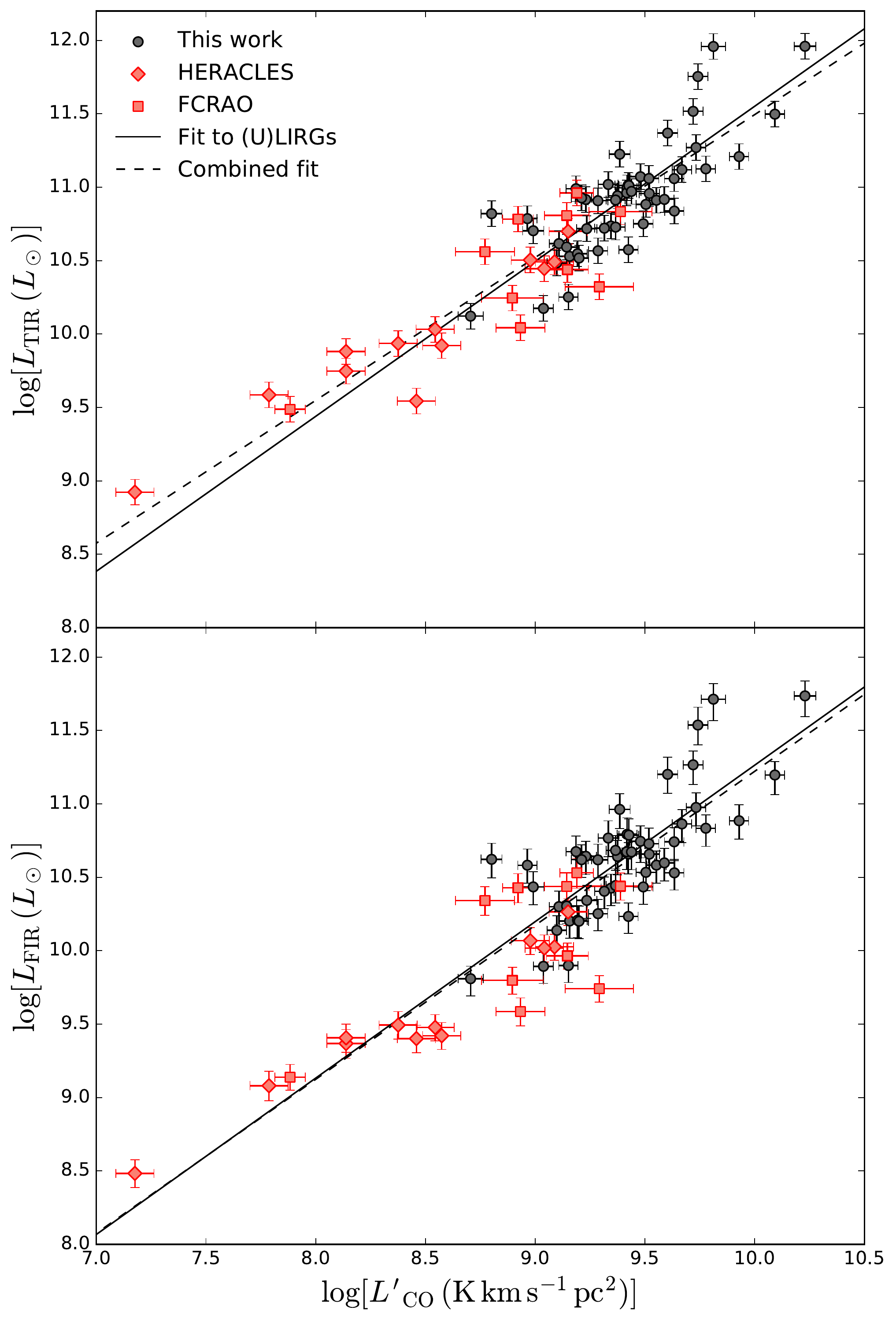}
\caption{Total and far IR luminosities versus the CO luminosity. Solid lines show linear fits to our (U)LIRG sample, yielding $\log(L_\mathrm{TIR}) = (1.05\pm0.12)\log(L'_\mathrm{CO}) + (1.02\pm1.12)$ and $\log(L_\mathrm{FIR}) = (1.07\pm0.13)\log(L'_\mathrm{CO}) + (0.55\pm1.26)$. Dashed lines are fits including the comparison samples, which also show linear slopes when using $L_\mathrm{TIR}$ ($0.97\pm0.06$) or $L_\mathrm{FIR}$ ($1.05\pm0.06$).
To avoid AGN contamination, only those (U)LIRGs with an AGN bolometric fraction $<0.20$ \citep[obtained from][]{diaz-santos17} were plotted and fitted.
}
\label{fig:lpco_vs_logirfir}
\end{figure}

Figure~\ref{fig:lpco_vs_logirfir} shows the FIR (derived from the modified blackbody SED fits) and the TIR (obtained from Eq.~\ref{eq:ltir}) luminosities as a function of the $L'_{\rm CO}$ for our sample of GOALS galaxies and for the local comparison samples.
 In optically thick environments, and 
in the absence of significant
dust heating by an active galactic nucleus (AGN), the plot of $L_{\rm TIR,FIR}$ versus $L'_{\rm CO}$ 
is essentially a comparison between the energy generated by the embedded starburst(s) versus the total reservoir of gas available to form new stars. 

For the purpose of fitting the data, we excluded (U)LIRGs where an AGN contributes $>20\%$ to the energy output, where we used the AGN bolometric fractions obtained from the combination of several mid-IR diagnostics \citep{diaz-santos17}. The reason behind this exclusion is to avoid contamination in the estimation of the star formation rates, efficiencies, and gas depletion times, as strong AGN can contribute significantly to the FIR luminosity. These sources were not excluded from the analysis in Sect.~\ref{sec:g2d} because there is not a known dependence of gas-to-dust mass ratio with AGN fraction, and our size sample does not allow us to study possible effects on it. No sources were removed from the local comparison sample, which is  not dominated by AGN by design.

Three sources were discarded based on the above {AGN} criterion: NGC\,0958, UGC\,05101, and IRAS\,F05189-2524.
Furthermore, to avoid the instability  of ordinary least-squares fits when both dependent and independent variables have an associated error, we followed the same approach as \citet{greve14} and used a Bayesian method to fit the data. Specifically, we used Josh Mayer's Python port\footnote{\url{https://github.com/jmeyers314/linmix/}} of the \texttt{LINMIX\_ERR} IDL package \citep{kelly07}.

The combined fit (our (U)LIRG sample and the local, normal, star-forming sources) to the data in Fig.~\ref{fig:lpco_vs_logirfir} has the functional form
\begin{equation}\label{eq:lcolir}
\begin{aligned}
\log(L_\mathrm{TIR}) &= (0.97\pm0.06)\log(L'_\mathrm{CO}) + (1.75\pm0.51) \\
\log(L_\mathrm{FIR}) &= (1.05\pm0.06)\log(L'_\mathrm{CO}) + (0.73\pm0.58)
\end{aligned}
,\end{equation}
with median Pearson correlation coefficients of $\rho=0.923$ and $\rho=0.915$, respectively, {and a robust statistical significance ($p$-values of $4.1\times10^{-28}$ and $5.0\times10^{-26}$, respectivey)}. We note that $L_\mathrm{TIR}$ is derived from integrated-galaxy photometry (see Eq. \ref{eq:ltir}), while $L_\mathrm{FIR}$ is obtained from integrating the fit to the aperture-matched SEDs. More luminous galaxies, with warmer dust, will have a larger contribution to the emission from shorter wavelengths (our fit uses $70<\lambda/\mu\mathrm{m}<350$), and thus $L_\mathrm{FIR}$ could be underestimated for those sources. We find, however, that there is no significant difference in the use of $L_\mathrm{TIR}$ or $L_\mathrm{FIR}$ for our sample. 
By comparison, \citet{greve14} derived a linear value for the slope of $(1.00\pm0.05)$ and $(0.99\pm0.04)$ for $L_\mathrm{(8-1000\,\mu\mathrm{m})}$ and $L_\mathrm{(50-300\,\mu\mathrm{m})}$, respectively. Their luminosity dynamic range was $10.29\leq \log(L_{(8-1000\mu\mathrm{m})})\leq12.56$. We also note that their sample included high-z galaxies which, if they had not been included, would have caused the slope to be sub-linear.
We only find a very slight steepening in the slope of the fit to the (U)LIRGs with respect to the local comparison sample, but not as pronounced as found by \citet{gao04}, where a clear super-linear slope is derived.

A way of interpreting the relation between infrared and CO luminosities is through the star formation efficiency, which is the ratio of the total energy from young massive stars per unit of star-forming molecular gas, and which is usually represented by $L_{\rm TIR,FIR}/L'_{\rm CO}$.
However, the use of $L'_{\rm CO}$ 
can be misleading, as it cannot be directly converted into a $M_{\rm H_2}$ estimate without the use of an $\alpha$ conversion factor, which we showed to vary between the samples in Sect.~\ref{sec:g2d}.

We can rework the above in terms of the gas depletion timescales, $t_{\rm dep} = M(\mathrm{H_2}) / \mathrm{SFR}$, where the molecular gas mass is derived using $\alpha=1.8\,M_\odot$\,(K\,km\,s$^{-1}$\,pc$^2$)$^{-1}$ for (U)LIRGs (see Sect.~\ref{sec:g2d}), and the Milky Way value for the local sample. The star formation rate (SFR) is obtained from the infrared and UV luminosity \citep[see][]{murphy11}
\begin{equation}
\frac{\textrm{SFR}}{M_\odot\,\mathrm{yr}^{-1}}=4.42\times10^{-44}\left(\frac{L_\mathrm{FUV}+0.88L_\mathrm{FIR}}{\textrm{erg\,s}^{-1}}\right),
\end{equation}
where we have used our aperture-matched, SED-derived {FIR} luminosity, while the UV luminosities (integrated luminosity within the GALEX far ultraviolet (FUV) filter; $\lambda_\mathrm{eff}=1528$\AA) were obtained from \citet{howell10} for the (U)LIRGs, and from the GALEX Ultraviolet Atlas of Nearby Galaxies \citep{gildepaz07} for the local comparison sample. Most of the UV emission is expected to be inside the IRAM-30\,m beam, thus any correction would be within the errors.

\begin{figure*}\centering
\includegraphics[width=0.7\textwidth]{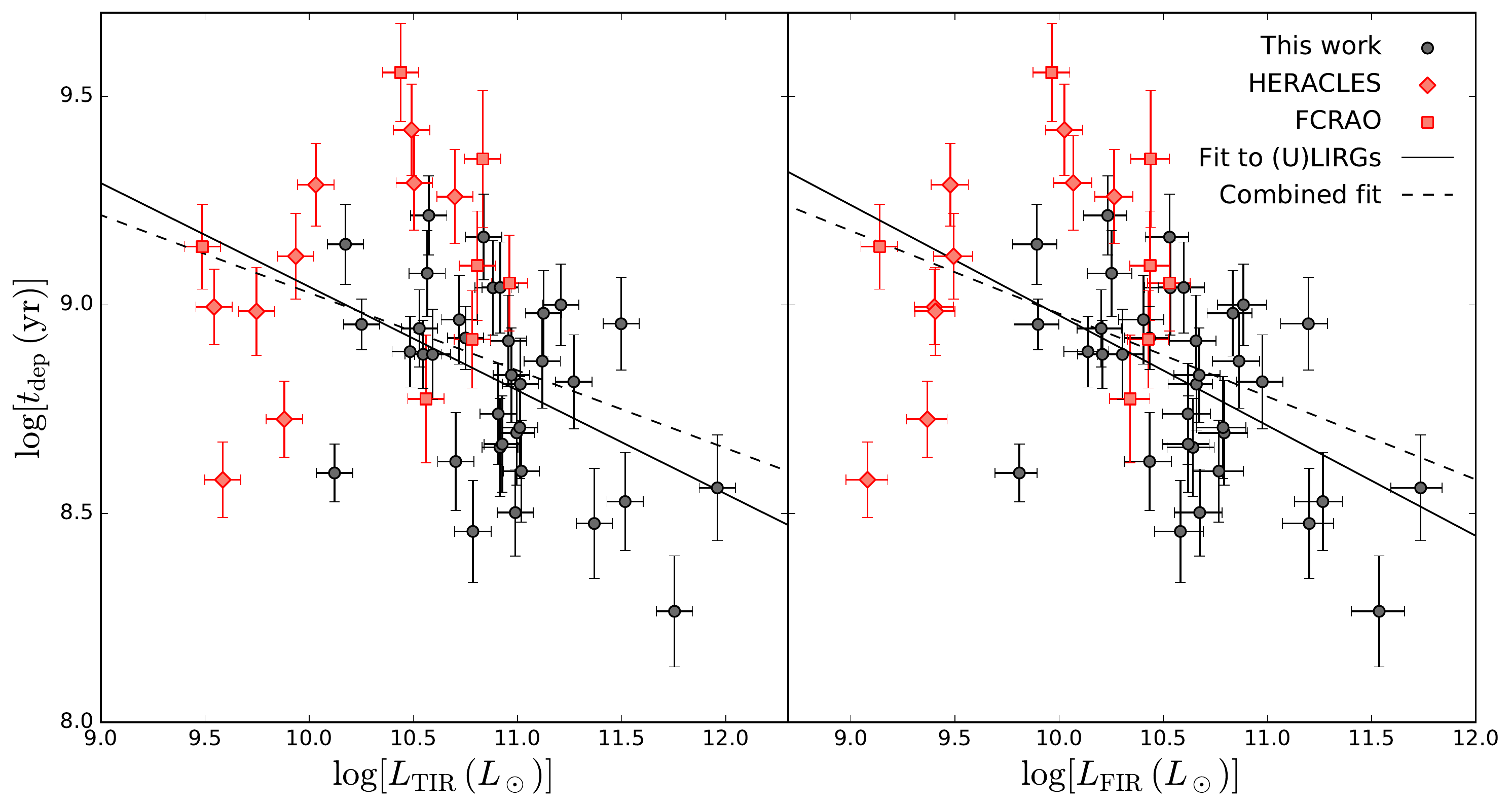}
\caption{Depletion time as a function of TIR and FIR luminosities. Solid lines are fits to the (U)LIRG sample, where $\log(t_{\rm dep}) = (-0.25\pm0.10)\log(L_\mathrm{TIR}) + (11.54\pm1.05)$ and $\log(t_{\rm dep}) = (-0.26\pm0.09)\log(L_\mathrm{FIR}) + (11.61\pm0.96)$. Dashed lines are a fit to the combined sample and yield $\log(t_{\rm dep}) = (-0.19\pm0.07)\log(L_\mathrm{TIR}) + (10.89\pm0.77)$ and $\log(t_{\rm dep}) = (-0.20\pm0.06)\log(L_\mathrm{FIR}) + (10.95\pm0.67)$. This trend implies that gas is consumed faster in (U)LIRGs and thus star formation is more efficient in these systems, compared to lower luminosity spiral galaxies.}
\label{fig:tdep}
\end{figure*}

Figure~\ref{fig:tdep} presents the variation of $t_\mathrm{dep}$ as a function of $L_\mathrm{TIR, FIR}$. 
Fits to the trends considering all three samples combined yield the correlations
\begin{equation} \label{eq:tdep}
\begin{aligned}
\log(t_{\rm dep}) &= (-0.19\pm0.07)\log(L_\mathrm{TIR}) + (10.89\pm0.77) \\
\log(t_{\rm dep}) &= (-0.20\pm0.06)\log(L_\mathrm{FIR}) + (10.95\pm0.67)
\end{aligned}
,\end{equation}
with median Pearson correlation coefficients of $\rho=-0.396$ and $\rho=-0.454$, respectively, {and p-values of $0.006$ and $0.001$, showing a moderately strong statistical significance.}

We obtain a median $t_\mathrm{dep}$ of 1.3\,Gyr for the local sample, compared to the 2.1\,Gyr from \citet{kennicutt98} or 2.35\,Gyr from \citet{bigiel11}. This timescale is obtained assuming there is not gas replenishing and that there is a constant star formation rate. We note that $t_\mathrm{dep}$ increases by a factor of two to three if interstellar gas recycling is considered \citep{ostriker75,kennicutt94}.
Nevertheless, the above correlation shows that $L_{\rm TIR} = 10^{11}$ L$_\odot$ LIRGs would deplete their available gas in $\sim630$\,Myr, whereas $L_{\rm TIR} = 10^{12}$ L$_\odot$ ULIRGs would deplete their gas in $\sim400$\,Myr, {a period more than three times shorter than for local normal spiral galaxies}. The compression of the molecular gas clouds due to large-scale shocks, which would enhance star formation, has been proposed to account for this difference \citep{jog92, barnes04,genzel10}. However, the quoted depletion times are obtained from the total molecular gas (as traced by \twCO), and not specifically from the dense gas phase of the molecular medium, which is expected to be more closely associated to regions where star formation is imminent.

{We note the strong dependence of depletion time (Eq.~\ref{eq:tdep}) with $\alpha_\mathrm{CO}$. Indeed, if dust masses were derived using the fitted $T_\mathrm{dust}$ instead of the fixed 25\,K, the relation would be $\log(t_{\rm dep}) = -0.36\log(L_\mathrm{TIR}) + 12.46$, implying an even larger difference (one order of magnitude) in the depletion time between (U)LIRGs and the comparison local sample. }

\begin{figure}\centering
\includegraphics[width=0.9\columnwidth]{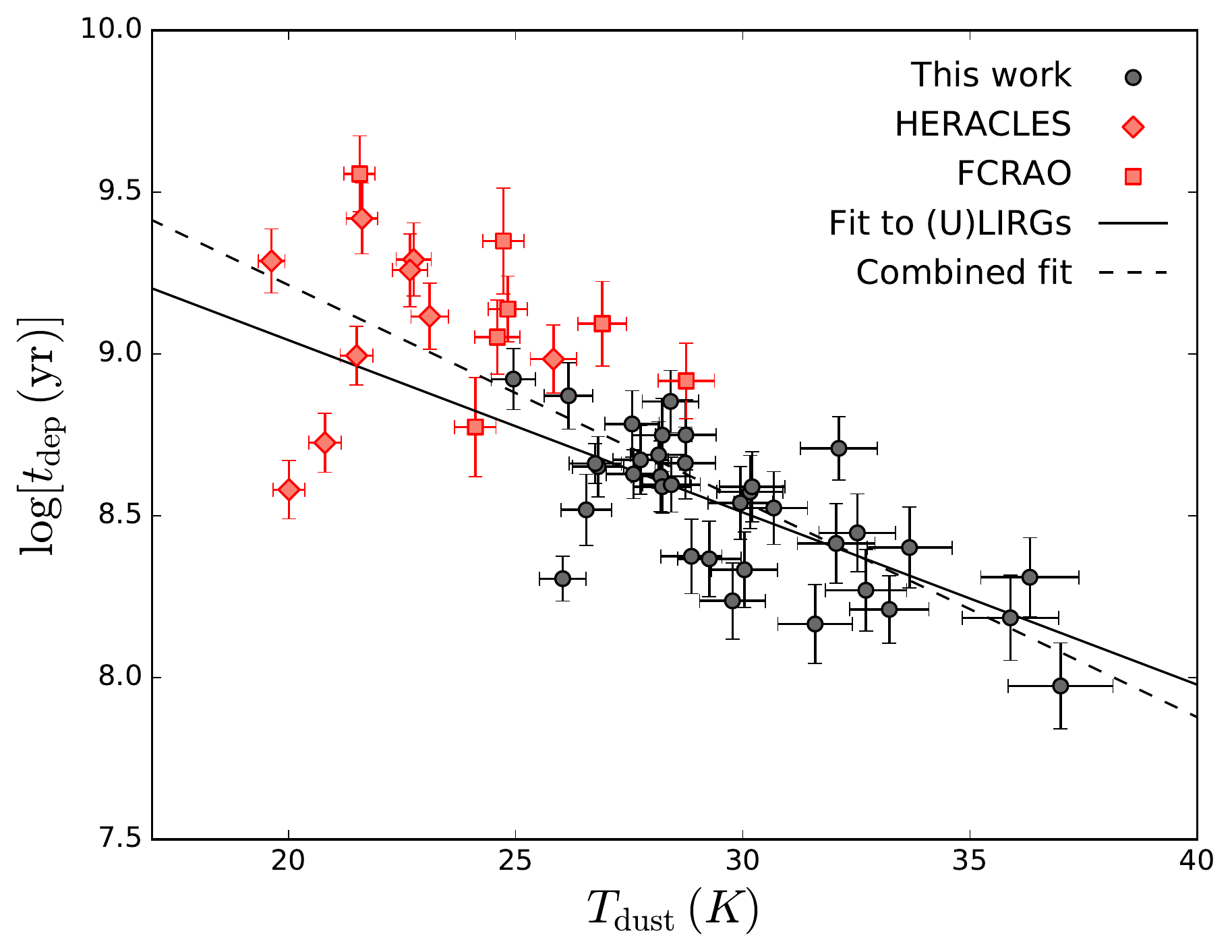}
\caption{Depletion time versus dust temperature, {using $T_\mathrm{dust}$ fitted values to obtain $M_\mathrm{dust}$}. 
Dust tmeperature is obtained from the SED model fit. The solid line represents the linear fit for the (U)LIRGs, yielding
$\log(t_{\rm dep}) = (-5.5\pm1.0)\times 10^{-2}\,T_\mathrm{dust} + (10.1\pm0.3)$. The dash lines represent the best fit considering also our local comparison sample; the fit yields 
$\log(t_{\rm dep}) = (-6.9\pm0.8)\times 10^{-2}\,T_\mathrm{dust} + (10.6\pm0.2)$,}
\label{fig:tdep_vs_tdust}
\end{figure}

Finally, {if fitted $T_\mathrm{dust}$ are used to estimate the dust masses}, our data also show a clear correlation between the depletion time and the dust temperature as shown in Fig.~\ref{fig:tdep_vs_tdust}. The fit of the combined data yields
\begin{equation}
\log(t_{\rm dep}) = (-6.9\pm0.8)\times 10^{-2}\,T_\mathrm{dust} + (10.6\pm0.2),
\end{equation}

with $\rho=-0.812$, {and a robust statistical significance (p-value of $1.1\times10^{-7}$).}
This trend shows a very sensitive variation, with depleting times being one order of magnitude lower with an increase of $\sim15$\,K. In other words, the star formation efficiency ($\propto 1/t_\mathrm{dep}$) increases by a factor of ten by such variation in the dust temperature.

\subsection{The \twCO / \thCO line intensity  ratio} \label{sec:12co13coratio}
\begin{figure}\centering
\includegraphics[width=0.9\columnwidth]{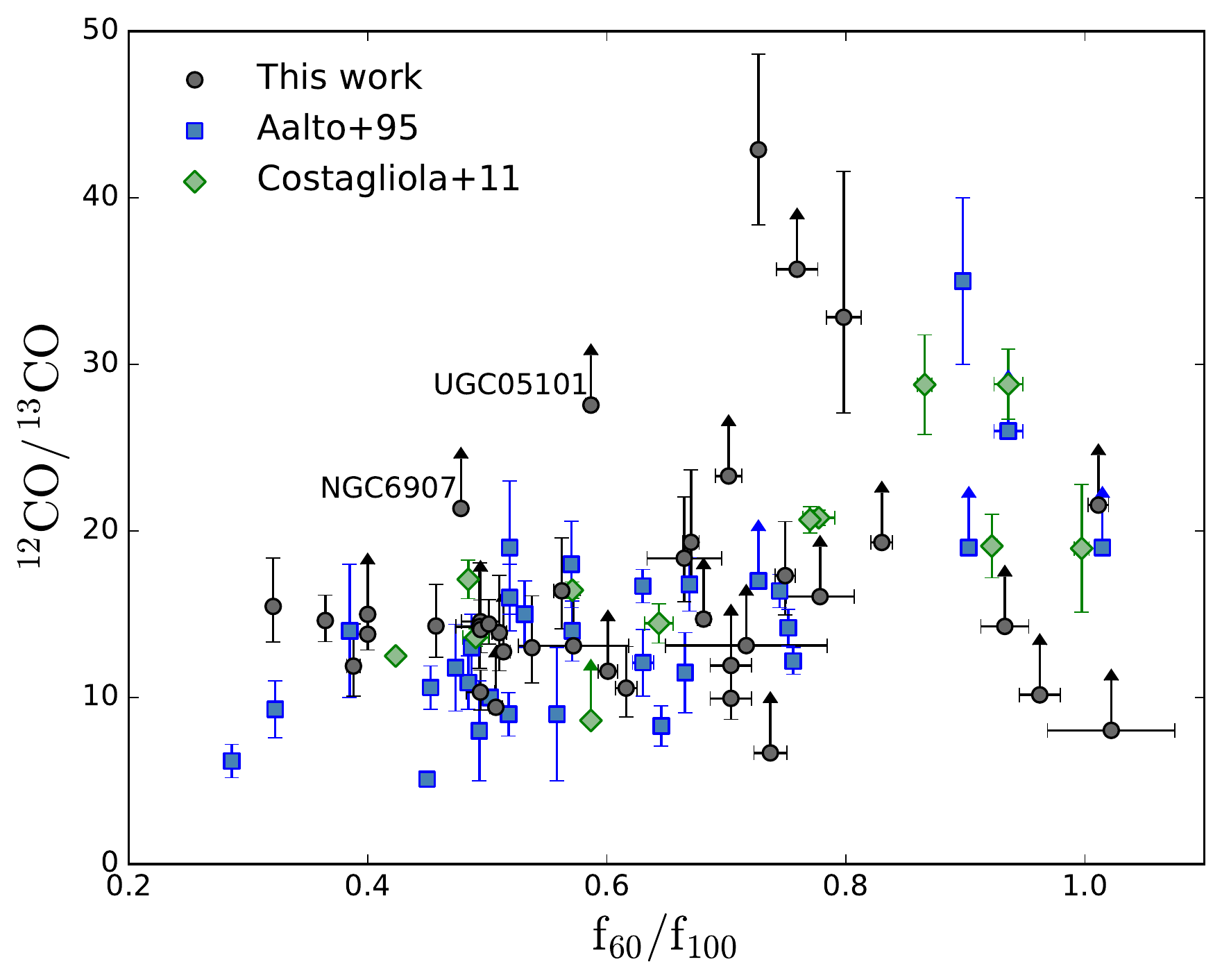}
\caption{\twCO over \thCO intensity ratio plotted vs. the \emph{IRAS} flux ratio $f_{60}/f_{100}$, used as a proxy for the dust temperature. For the line ratios, 
the uncertainties were derived using a Monte Carlo simulation using 10\,000 samples and following a similar prescription to \citet{crocker12}.
The sources are from different samples: this work (black circles), \citet[blue squares]{aalto95}, and \citet[green diamonds]{costagliola11}. {Errors are plotted at $1\sigma$ level, although x-axis errors are not appreciable for many sources.} The two labeled sources are at odds with the rest of the sample for their observed $f_{60}/f_{100}$ ratio.
}
\label{fig:12CO13CO_f60f100_intensity}
\end{figure}

In this sub-section, we examine the relationship between the \twCO/\thCO line intensity ratio and the dust temperature.
In Fig. \ref{fig:12CO13CO_f60f100_intensity} we present a plot of \twCO/\thCO for the (U)LIRGs in our sample versus the 60\,$\mu$m-to-100\,$\mu$m {\it IRAS} flux density ratio $f_{60}/f_{100}$, which is used as a proxy for the dust temperature. We do not use here our SED-fit-derived $T_\mathrm{dust}$ so as to be able to systematically compare with previous studies on LIRGs (for which \emph{Herschel} photometry is not available). We have confirmed nonetheless the very good correlation between $f_{60}/f_{100}$ and the derived $T_\mathrm{dust}$ for our sample, finding $T_\mathrm{dust}=(14.1\pm1.2)(f_{60}/f_{100}) + (20.6\pm0.8)$. {The data from the present study are compared with} published results from \cite{aalto95} and \cite{costagliola11}. {The former is formed of 31 local spiral galaxies, starbursts, interacting systems, and luminous mergers, while the latter includes a representation of 12 Seyferts, starbursts, and LIRGs.}
Our results are in agreement with those studies, finding lower values of $^{12}$CO/$^{13}$CO for $f_{60}/f_{100}<0.8$, and a relatively steep enhancement for $f_{60}/f_{100}>0.8$. This cutoff point corresponds to $\sim32$\,K. {This general trend is also in agreement with Milky Way measurements of molecular clouds \citep{barnes15}.} Only two of the sources in our sample present a lower-limit value that seems at odds with this behavior, showing too large $^{12}$CO/$^{13}$CO ratios ($>21.4$ and $>27.6$) for their FIR color ($0.48$ and $0.59$). These are NGC\,6907 and UGC\,05101, and they are marked in Fig.~\ref{fig:12CO13CO_f60f100_intensity} with their names.

The above relation can be explained as a result of the decreasing optical depth of the \twCO line with increasing ISM temperature \citep[e.g.,][]{young86a,aalto95}. Increased linewidths in diffuse or turbulent gas may also result in decreased line optical depths and elevated line ratios \citep[see, e.g.,][]{polk88, aalto95,aalto10,konig16}.

In addition to optical depth effects, the relative \twCO and \thCO abundances ratio will impact the intensity ratio. The less abundant isotopomer of \thCO will become selectively photodissociated in diffuse gas since it cannot self-shield like \twCO. Furthermore, stellar nucleosynthesis will alter the $^{12}$C/$^{13}$C abundance ratio and low-metallicity gas is expected to have high
$^{12}$C/$^{13}$C ratios \citep[see, e.g.,][]{casoli92,henkel14,tang19}. The ejecta of massive stars are also expected to have higher $^{12}$C/$^{13}$C abundance ratios, which means that in the early stages of a starburst \twCO/\thCO abundance ratios may be higher. Indeed, there seems to be an enrichment of the ISM due to a young starburst and/or a top-heavy initial mass function \citep{sliwa17}.

\section{Summary} \label{sec:summary}

We observed 55 (U)LIRGs with the IRAM-30\,m Telescope. 
In this work, we have focused on the \twCO$(1-0)$ transition as a tracer of molecular gas. We summarize our results as follows:

\begin{enumerate}
\item We observed \twCO and \thCO simultaneously, in the same band, obtaining detection rates of 96\% and 56\% for \twCO and \thCO, respectively, as expected from \thCO being much less abundant than \twCO.

\item We have used \emph{Herschel} data to construct the far infrared SED of the emitting region inside the IRAM-30\,m beam for each galaxy in our sample, and fitted the SED to a modified blackbody model, obtaining the dust masses and temperatures, as well as $L_\mathrm{FIR}$. We also have systematically fitted local galaxies from HERACLES and FCRAO, for which both CO and \emph{Herschel} data are available, to be used as a local comparison sample. Complementarily, we have used MIPS data to derive the total IR luminosities, $L_\mathrm{TIR}$, for both samples.

\item {We have determined an average CO-to-H$_2$ conversion factor for (U)LIRGs of $\langle\alpha\rangle=1.8^{+1.3}_{-0.8}\,M_\odot$\,(K km\,s$^{-1}$\,pc$^{2}$)$^{-1}$, $\sim2.2$ times smaller than the accepted conversion factor for local spiral galaxies, for an assumed constant gas-to-dust mass ratio between the local comparison sample and the (U)LIRGs sample. This value is obtained by assuming a fixed dust temperature of 25\,K for every source. If the SED fitted temperatures are used instead, the average value of $\alpha$ for (U)LIRGs would be $0.9^{+0.9}_{-0.5}$.}

\item We confirm the close linear correlation between the CO luminosity and $L_\mathrm{TIR,FIR}$. We have also found an enhancement in the star formation efficiency for more IR-luminous systems. Consequently, the depletion time for these systems is shorter, following a $\log-\log$ correlation with $L_\mathrm{IR,TIR}$. However, this effect with $t_\mathrm{dep}$ is not observed in the higher density phase of molecular gas.
{When $M_\mathrm{dust}$ are derived from fitted $T_\mathrm{dust}$}, the $t_\mathrm{dep}$ also shows a clear decreasing trend with dust temperature, implying an increase of one order of magnitude in the star formation efficiency per increase of $\sim 15$\,K.

\item We have re-examined the variation between the ratio of \twCO and its isotopologue \thCO with the dust temperature, as traced by the proxy $f_{60}/f_{100}$. Our data confirm previous results, with cooler galaxies having lower \twCO / \thCO ratios, with a steep increase above $f_{60}/f_{100}\sim0.8$.

\end{enumerate}

\begin{acknowledgements}
      We thank the anonymous referee for the useful comments. We also thank Nick Scoville for his discussion and comments on the paper. We are grateful to Manuel Gonz\'alez for his help in the preparation of the scripts and during the observations.
This work is based on observations carried out under project numbers 099-10, 092-11, 227-11, 076-12, 222-13, and D01-13 with the 30m telescope. IRAM is supported by INSU/CNRS (France), MPG (Germany), and IGN (Spain).
We thank the director of IRAM 30m for the approval of the discretionary time requested.
RHI, MAPT, and AA acknowledge support from the Spanish MINECO through grants AYA2012-38491-C02-02 and AYA2015-63939-C2-1-P.
G.C.P. was supported by a FONDECYT Postdoctoral Fellowship (No. 3150361).
{T.D.-S. acknowledges support from ALMA-CONICYT project 31130005 and FONDECYT regular project 1151239.}
This work was supported in part by National Science Foundation grant No. PHYS-1066293 and the hospitality of the Aspen Center for Physics.
This research has made use of the NASA/IPAC Extragalactic Database (NED), which is operated by the Jet Propulsion Laboratory, California Institute of Technology, under contract with the National Aeronautics and Space Administration.
\end{acknowledgements}

\bibliographystyle{aa}
\bibliography{/Users/rherrero/Dropbox/masterbibdesk}

\clearpage

\begin{appendix}
\section{Observed spectra and fitted Gaussian line components}\label{app:lineprof}
Figure~\ref{figapp:lineprof} presents the same results as Fig.~\ref{fig:lineprof} for our complete sample. We have also tabulated in Table~\ref{table:gaussian} the Gaussian components that were fitted to the \twCO spectra.

\begin{figure*}[htb]\centering
\includegraphics[width=.9\textwidth]{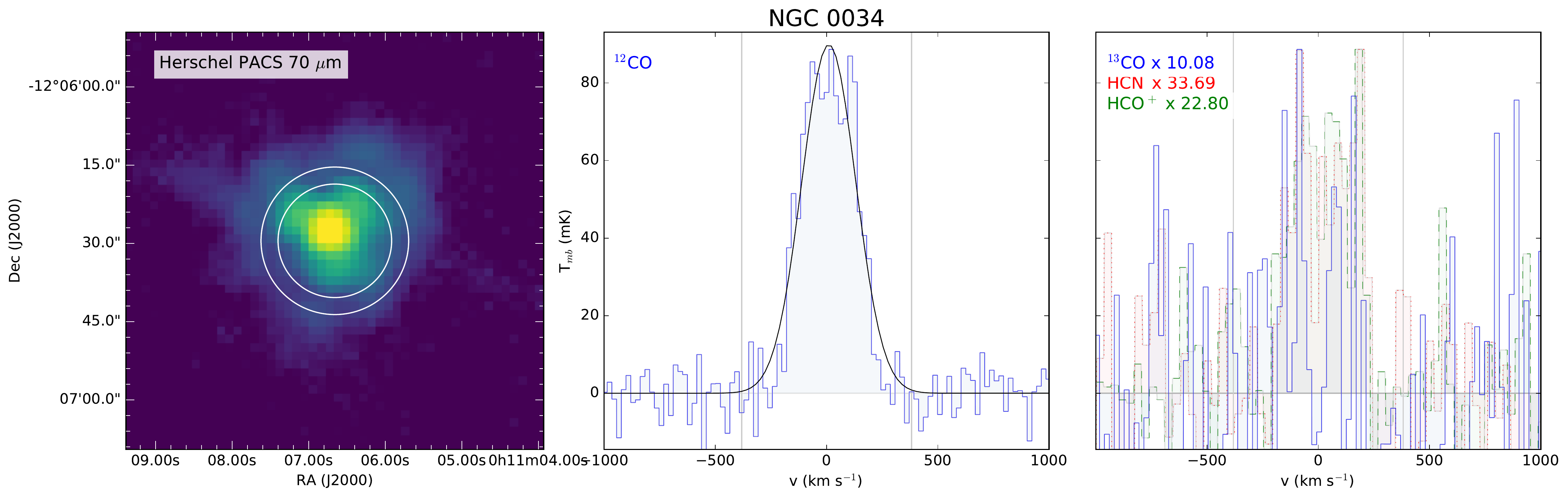}
\includegraphics[width=.9\textwidth]{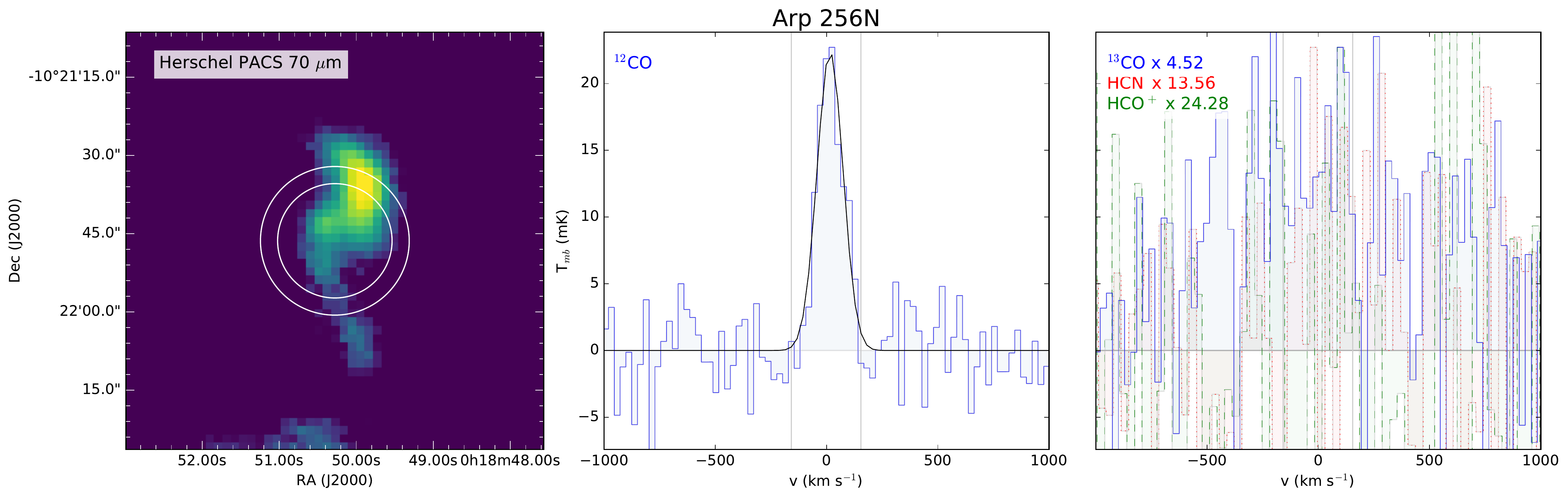}
\includegraphics[width=.9\textwidth]{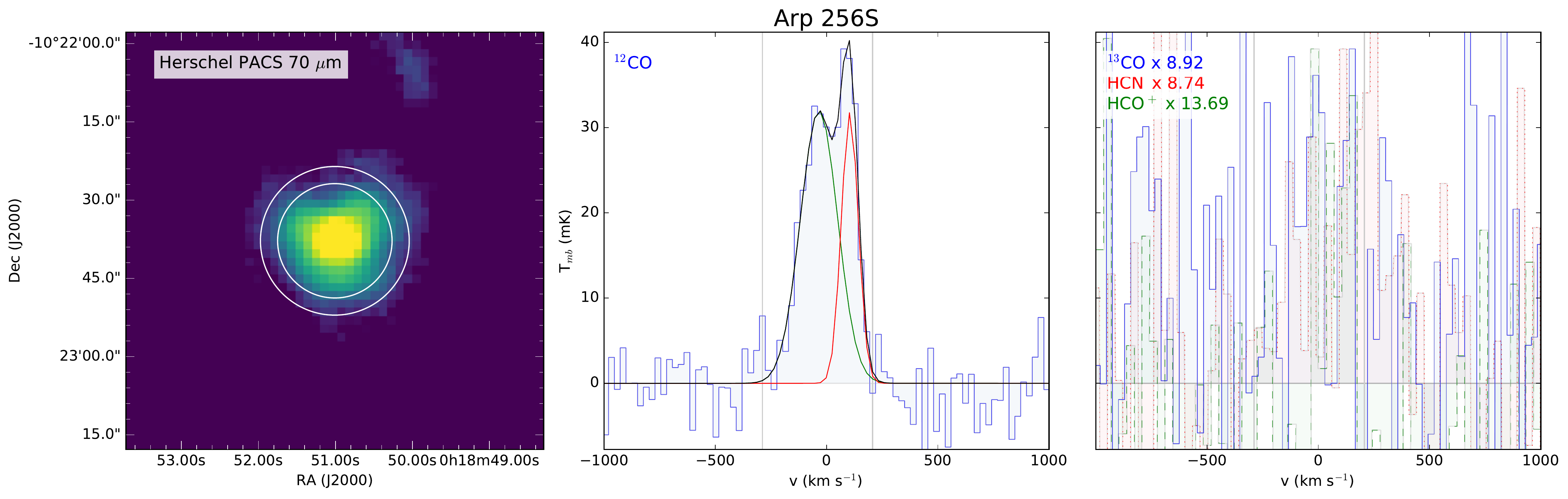}
\caption{Herschel images and IRAM-30\,m spectra. Figure~\ref{fig:lineprof} provides details.}
\label{figapp:lineprof}
\end{figure*}

\begin{figure*}[htb]\centering
\includegraphics[width=.9\textwidth]{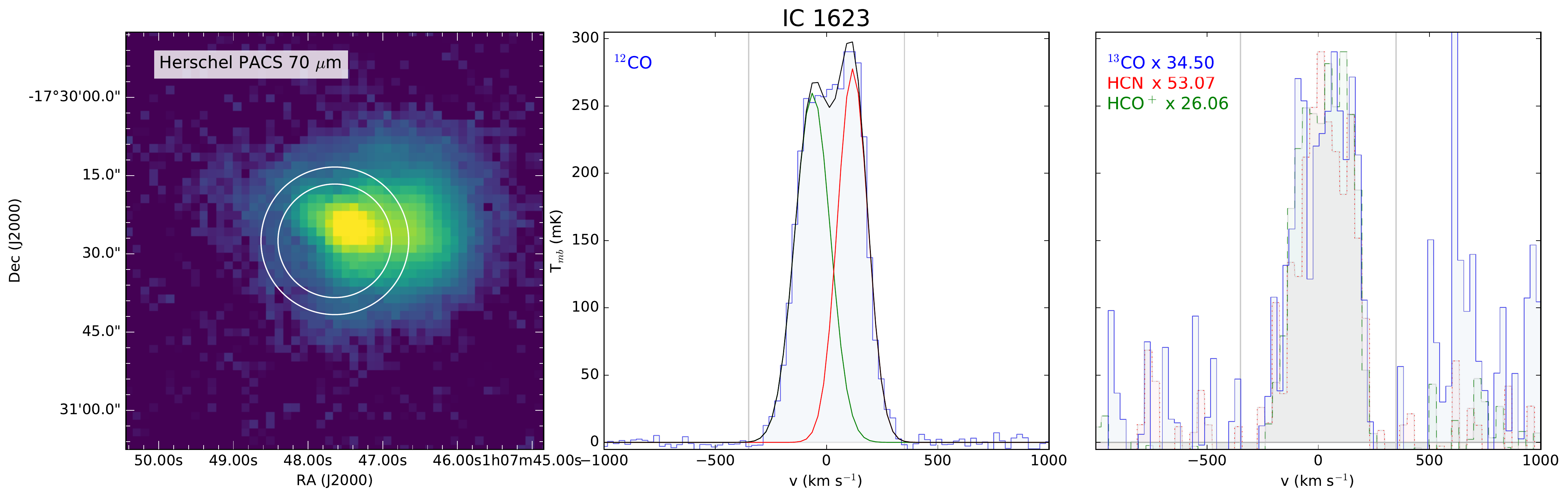}
\includegraphics[width=.9\textwidth]{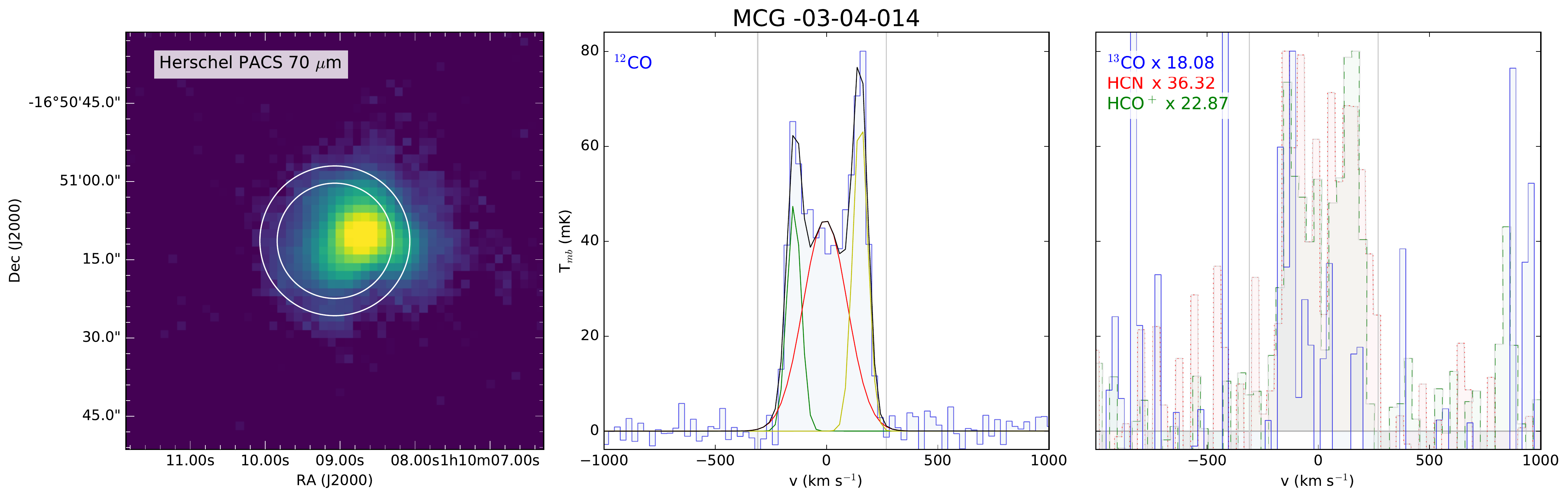}
\includegraphics[width=.9\textwidth]{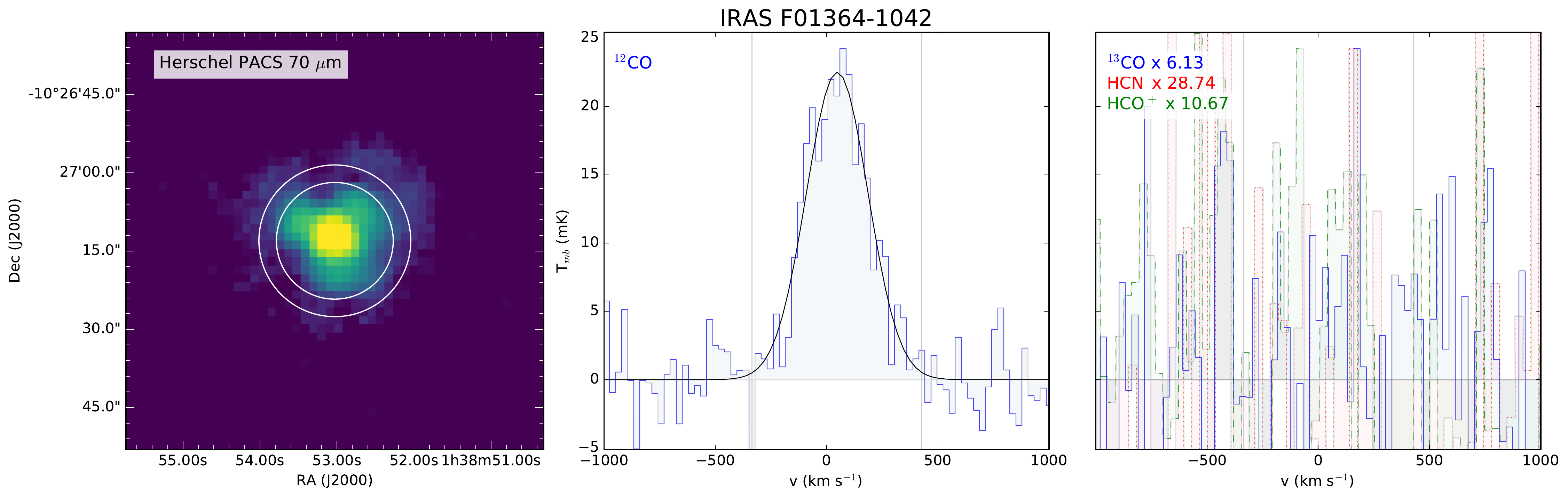}
\includegraphics[width=.9\textwidth]{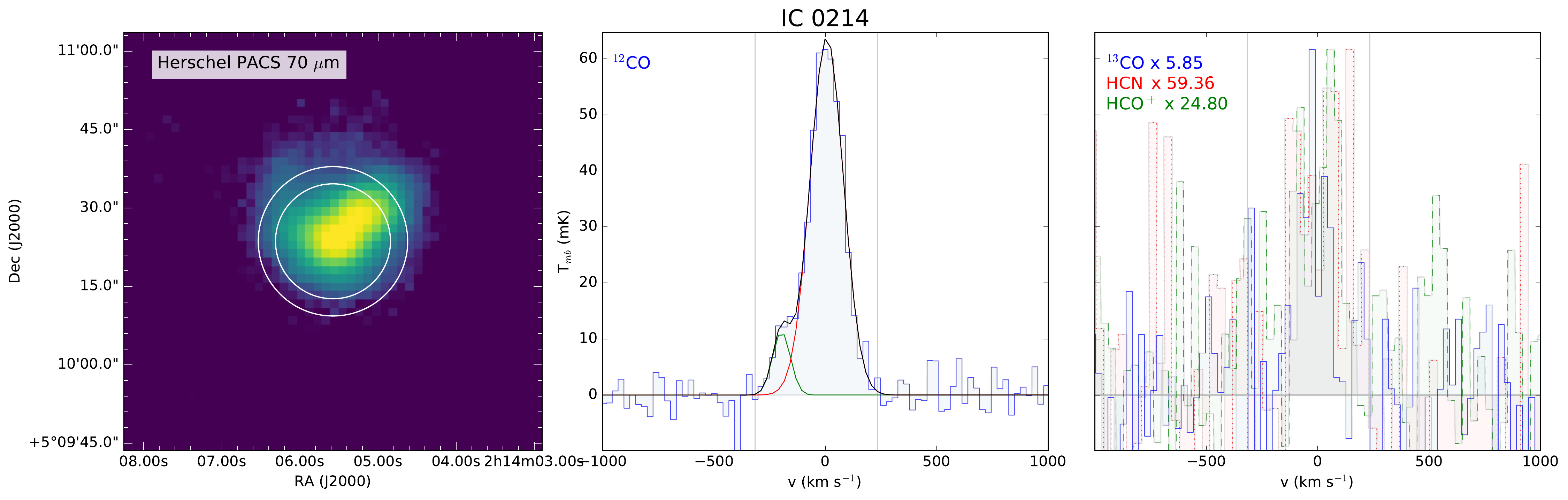}
\contcaption{\emph{continued}}
\end{figure*}

\begin{figure*}[htb]\centering
\includegraphics[width=.9\textwidth]{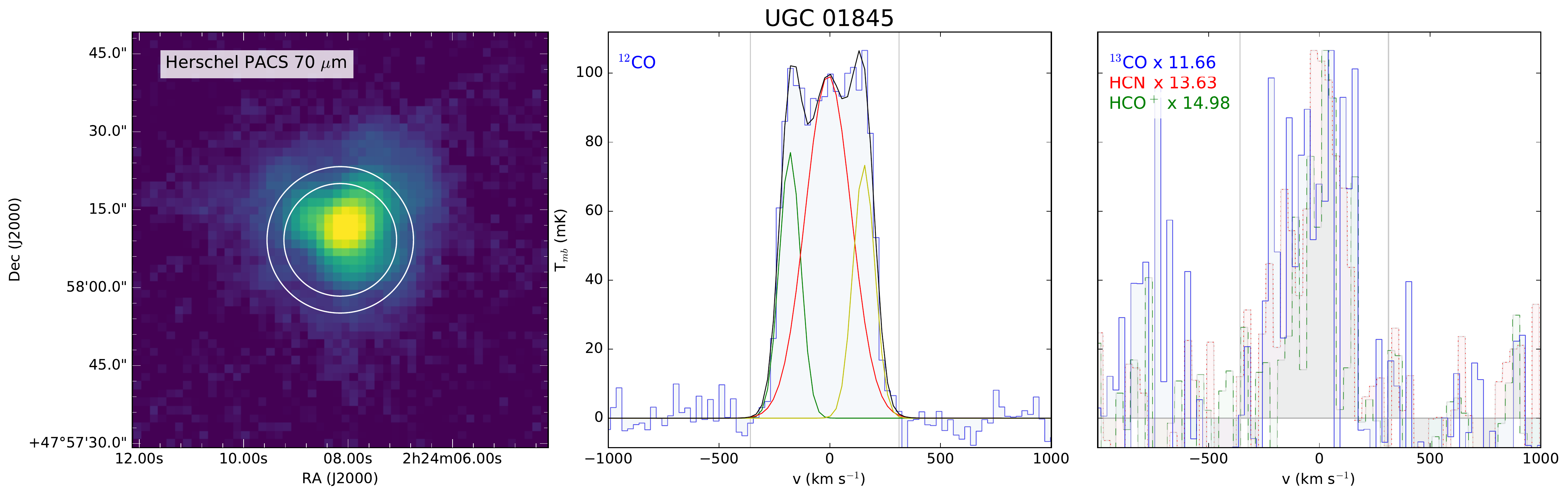}
\includegraphics[width=.9\textwidth]{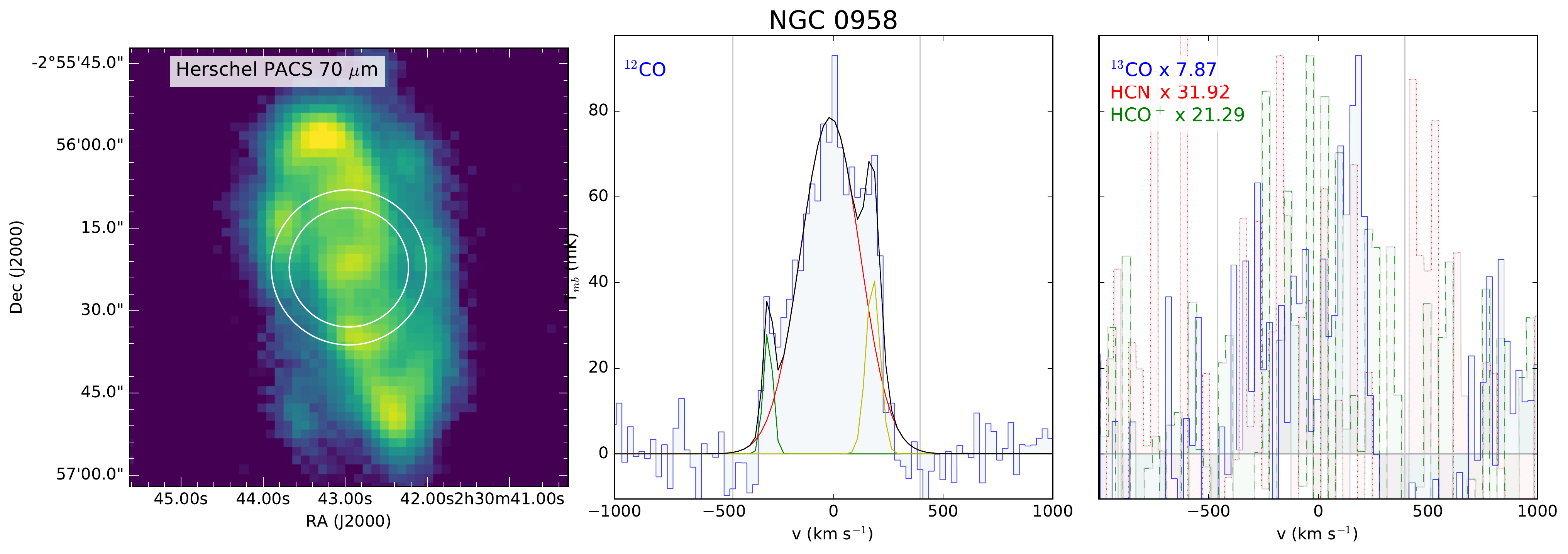}
\includegraphics[width=.9\textwidth]{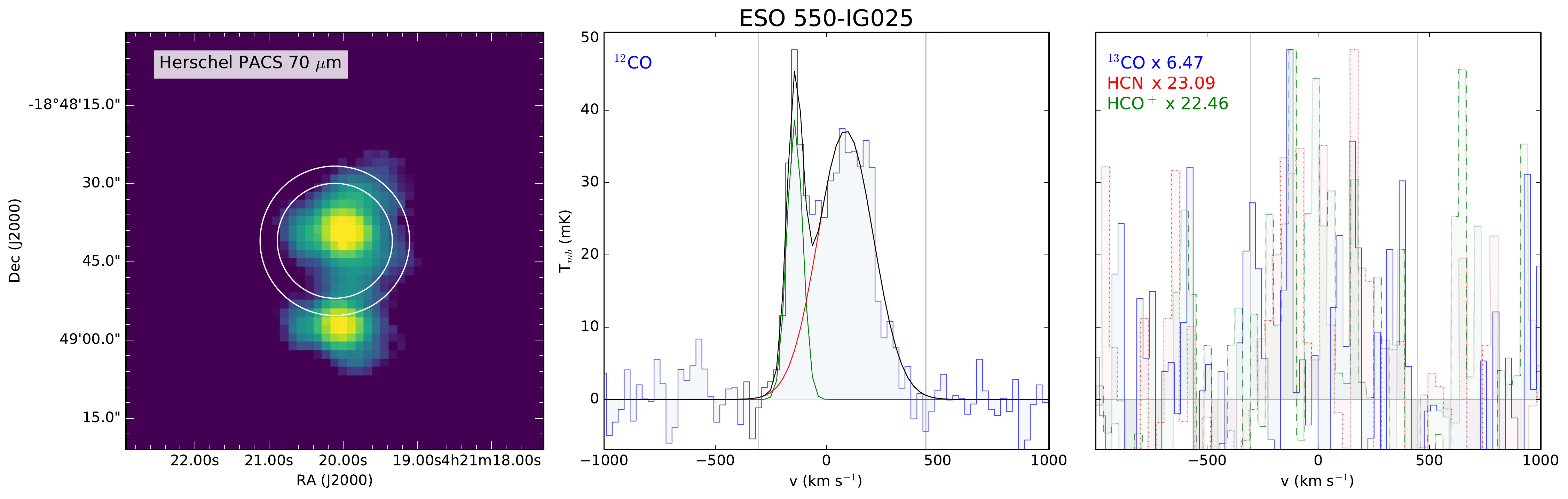}
\includegraphics[width=.9\textwidth]{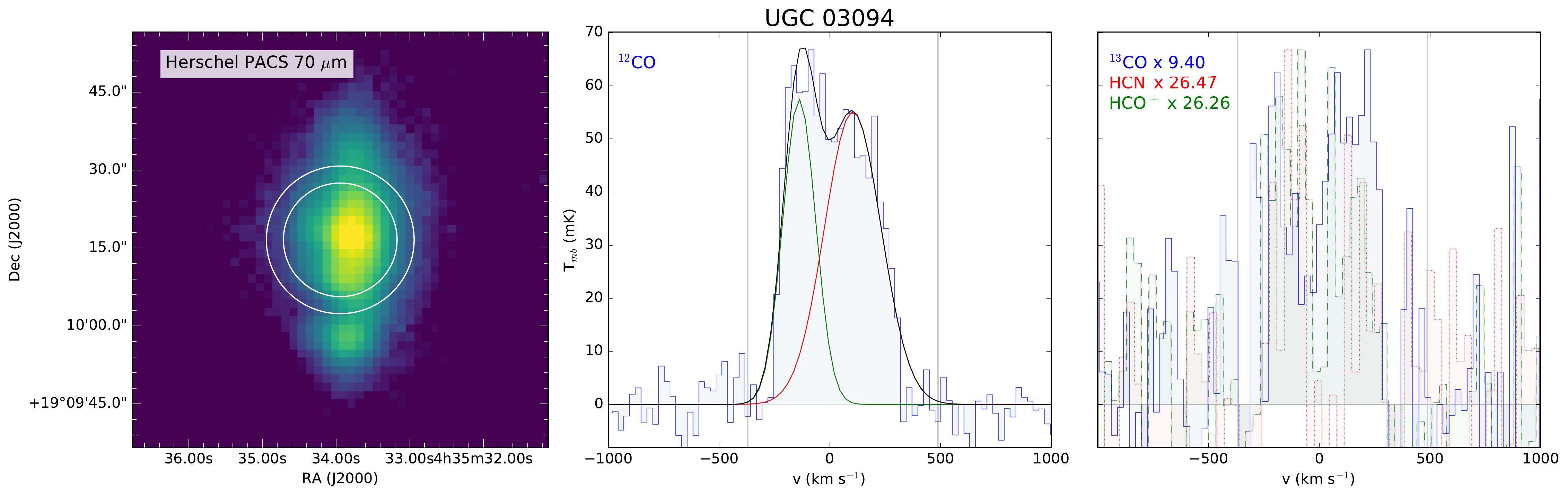}
\contcaption{\emph{continued}}
\end{figure*}

\begin{figure*}[htb]\centering
\includegraphics[width=.9\textwidth]{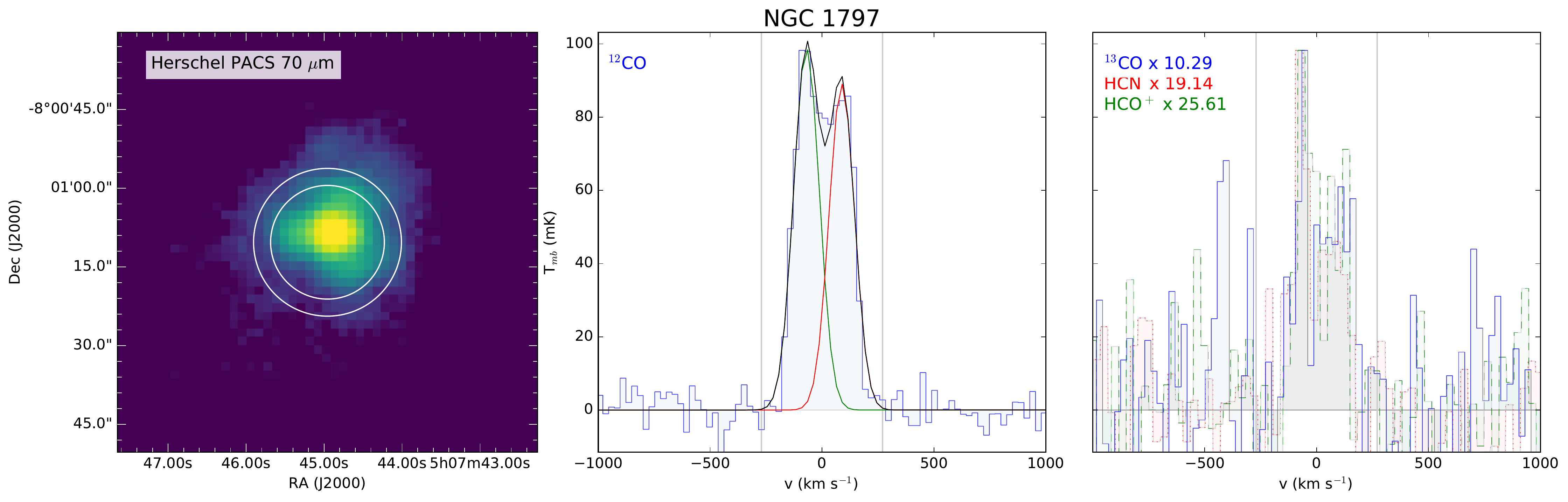}
\includegraphics[width=.9\textwidth]{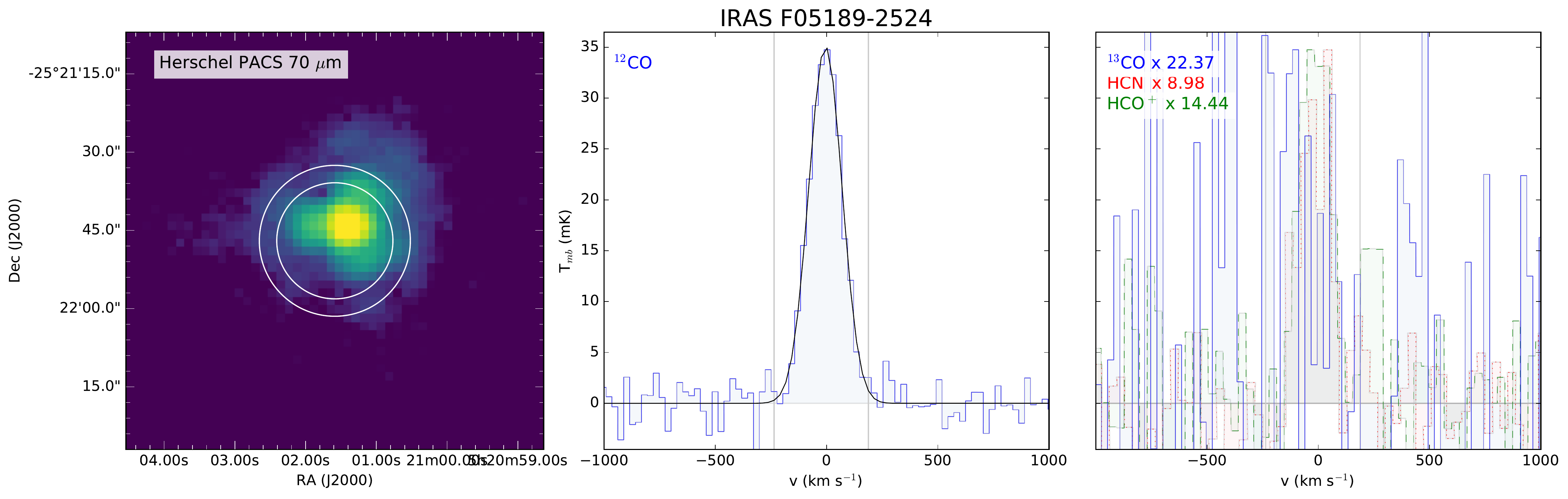}
\includegraphics[width=.9\textwidth]{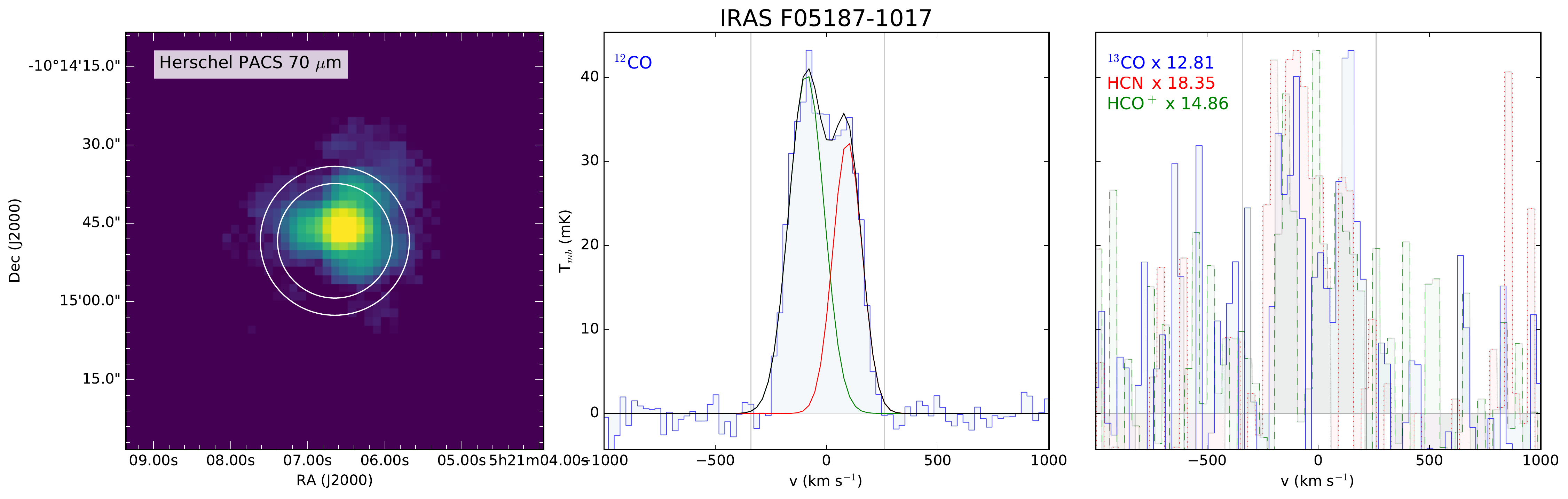}
\includegraphics[width=.9\textwidth]{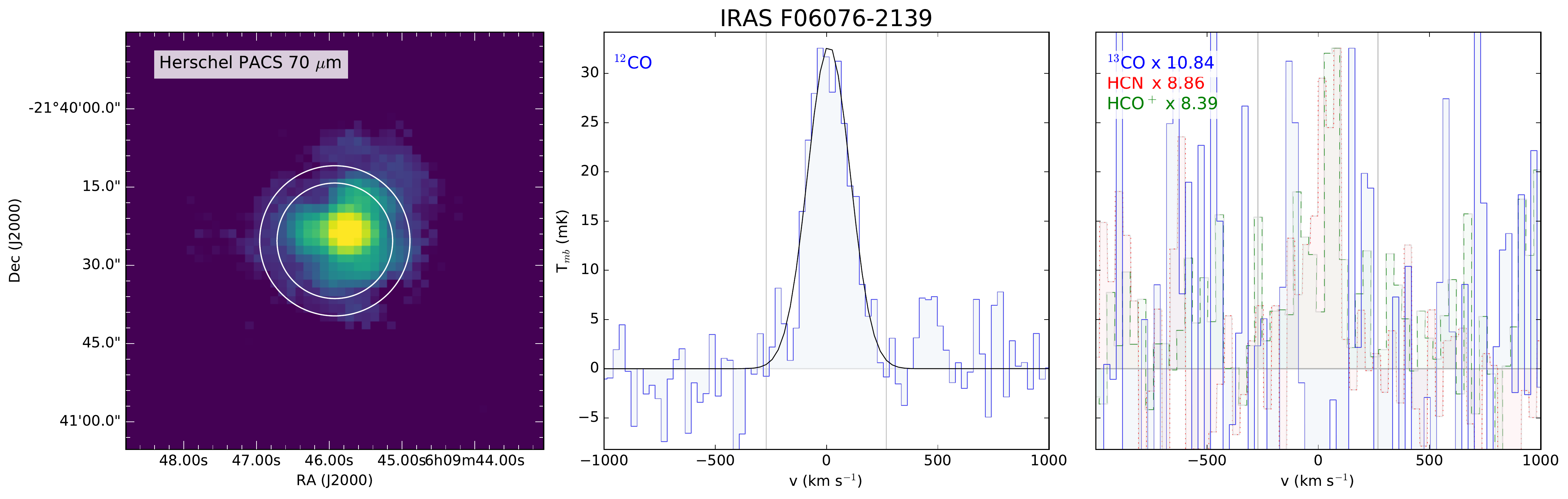}
\contcaption{\emph{continued}}
\end{figure*}

\begin{figure*}[htb]\centering
\includegraphics[width=.9\textwidth]{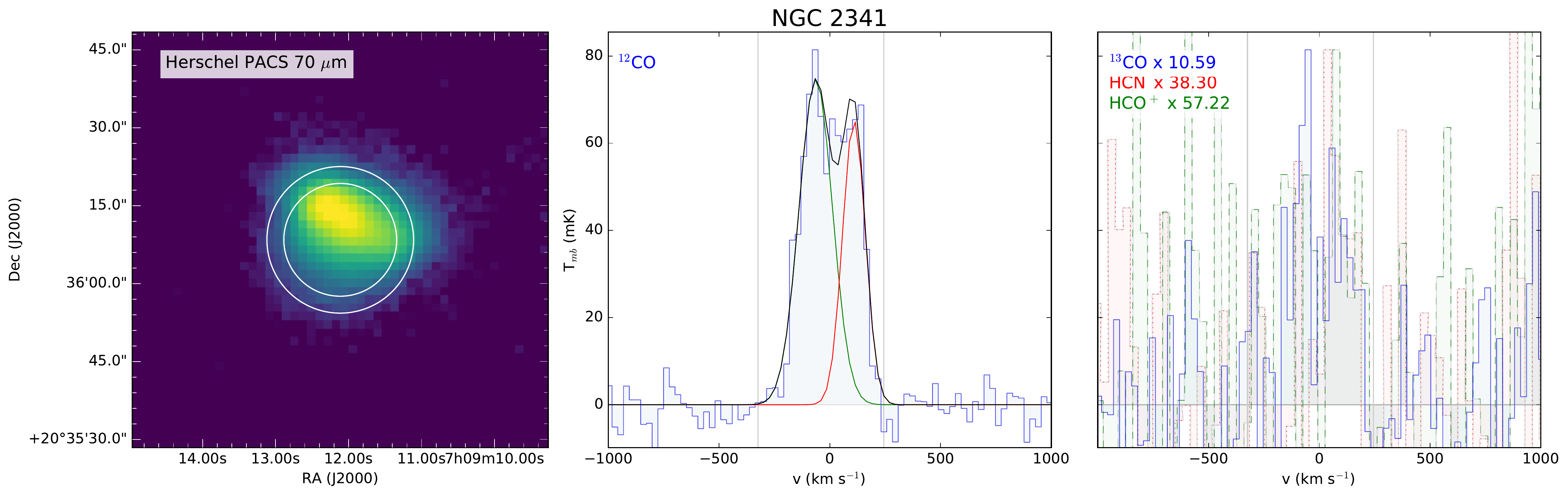}
\includegraphics[width=.9\textwidth]{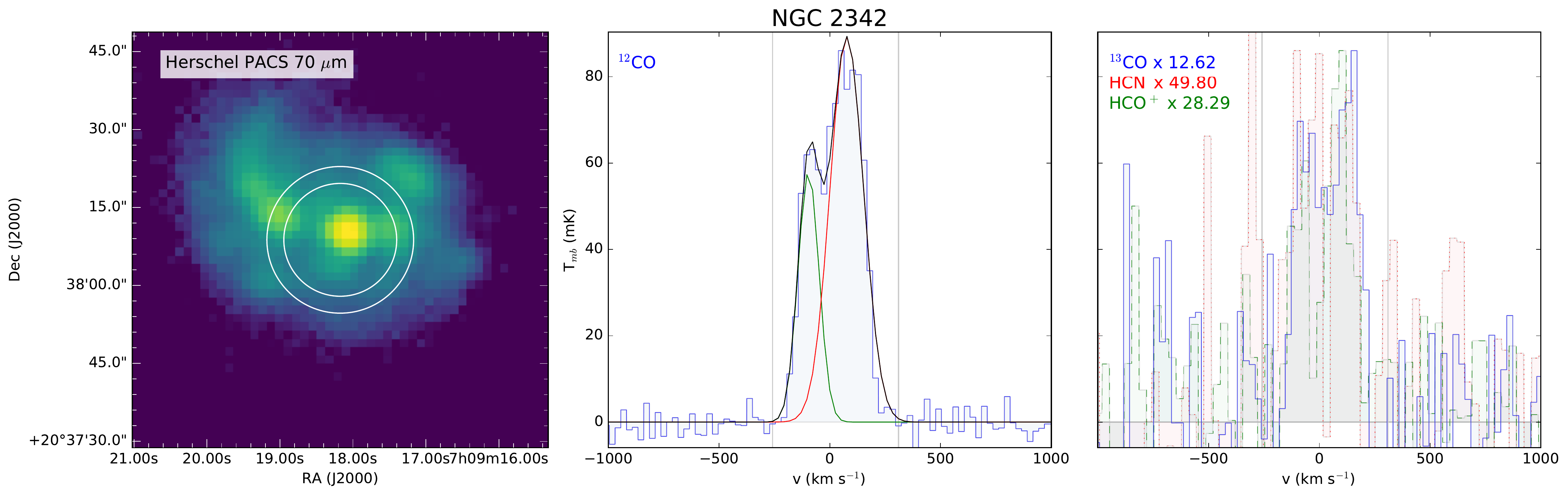}
\includegraphics[width=.9\textwidth]{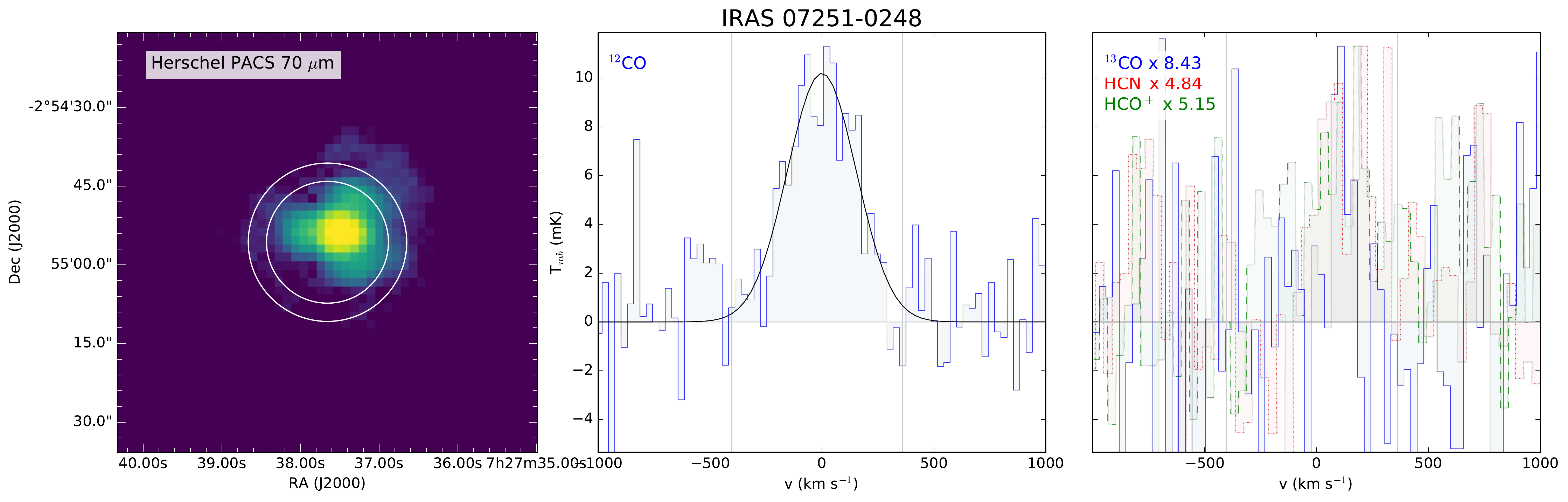}
\includegraphics[width=.9\textwidth]{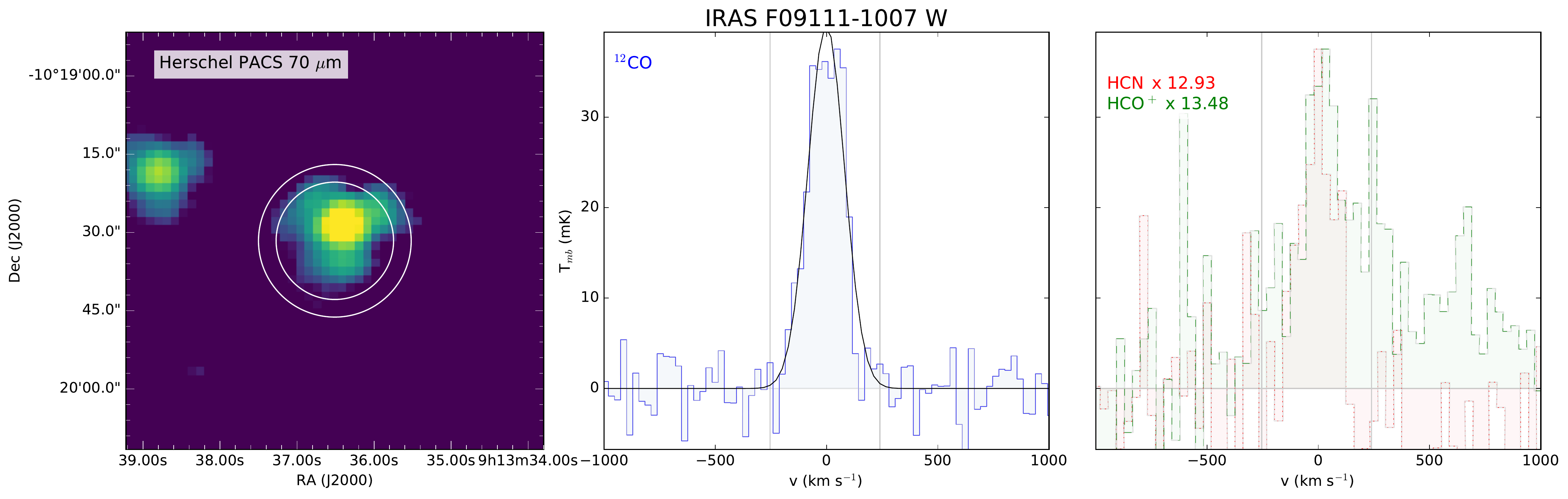}
\contcaption{\emph{continued}}
\end{figure*}

\begin{figure*}[htb]\centering
\includegraphics[width=.9\textwidth]{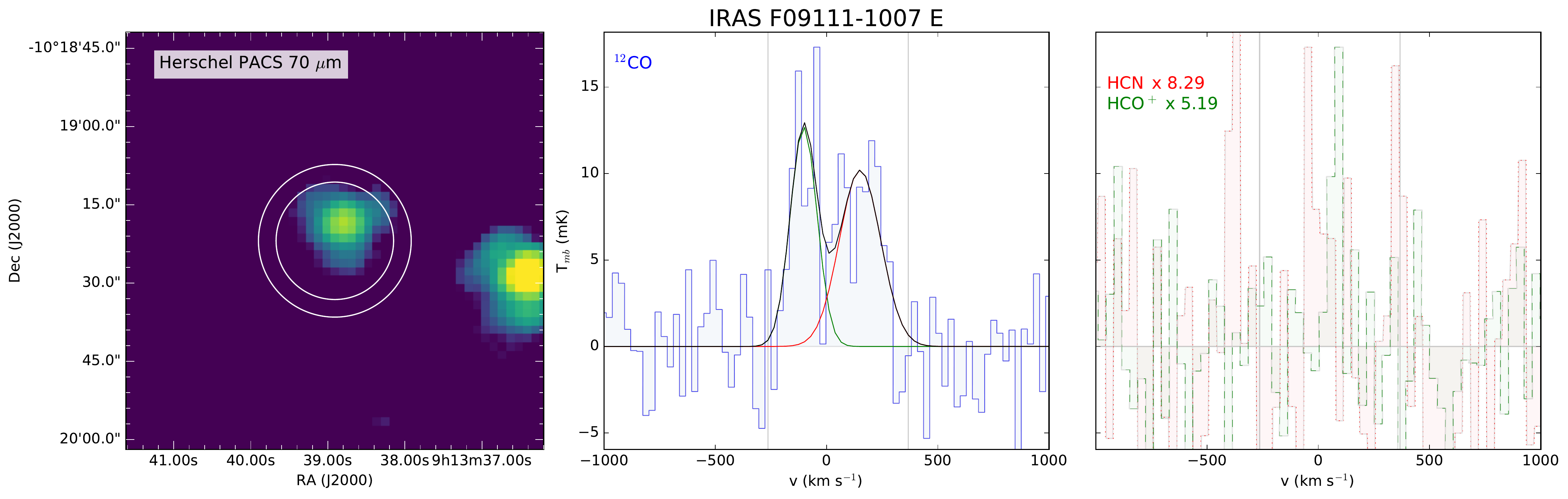}
\includegraphics[width=.9\textwidth]{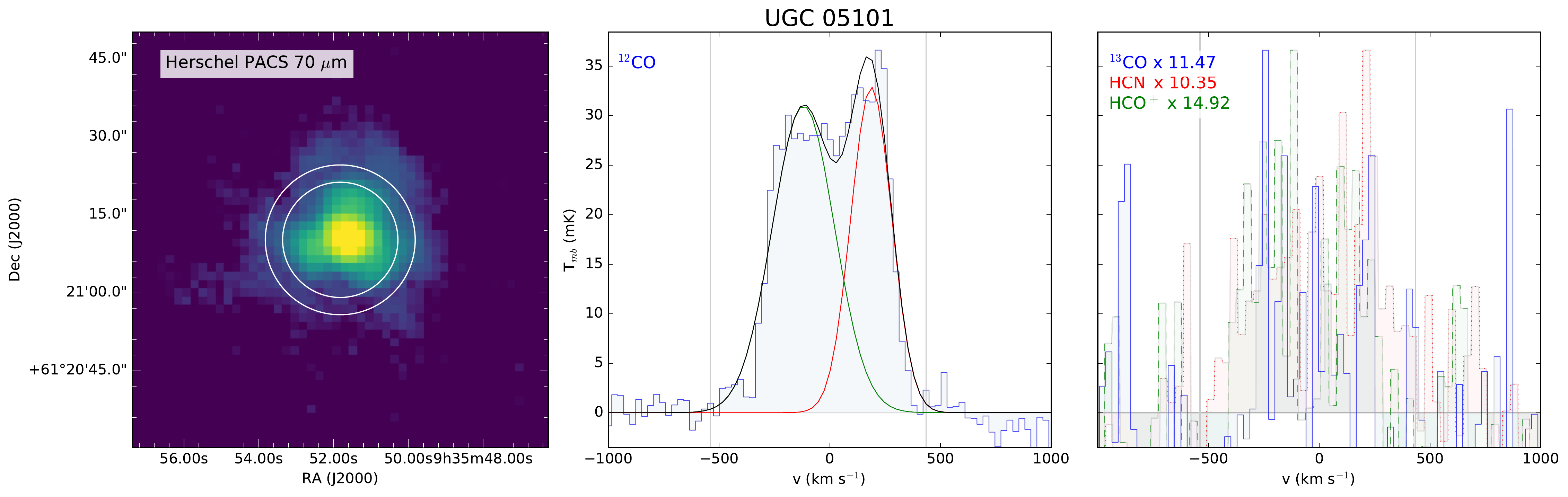}
\includegraphics[width=.9\textwidth]{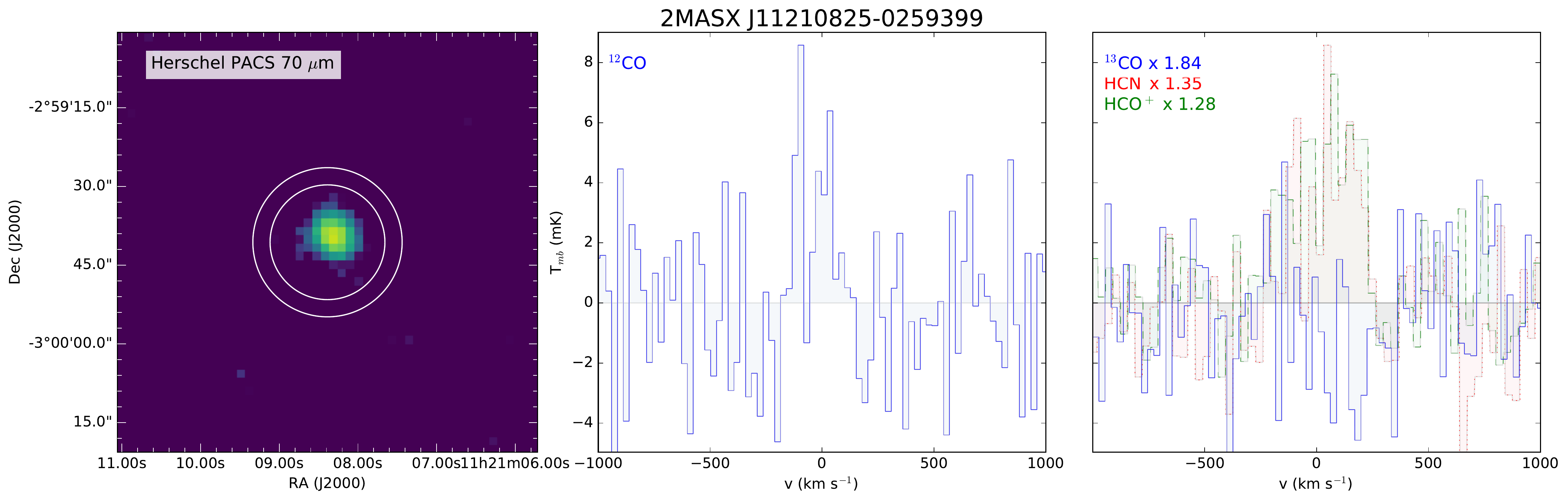}
\includegraphics[width=.9\textwidth]{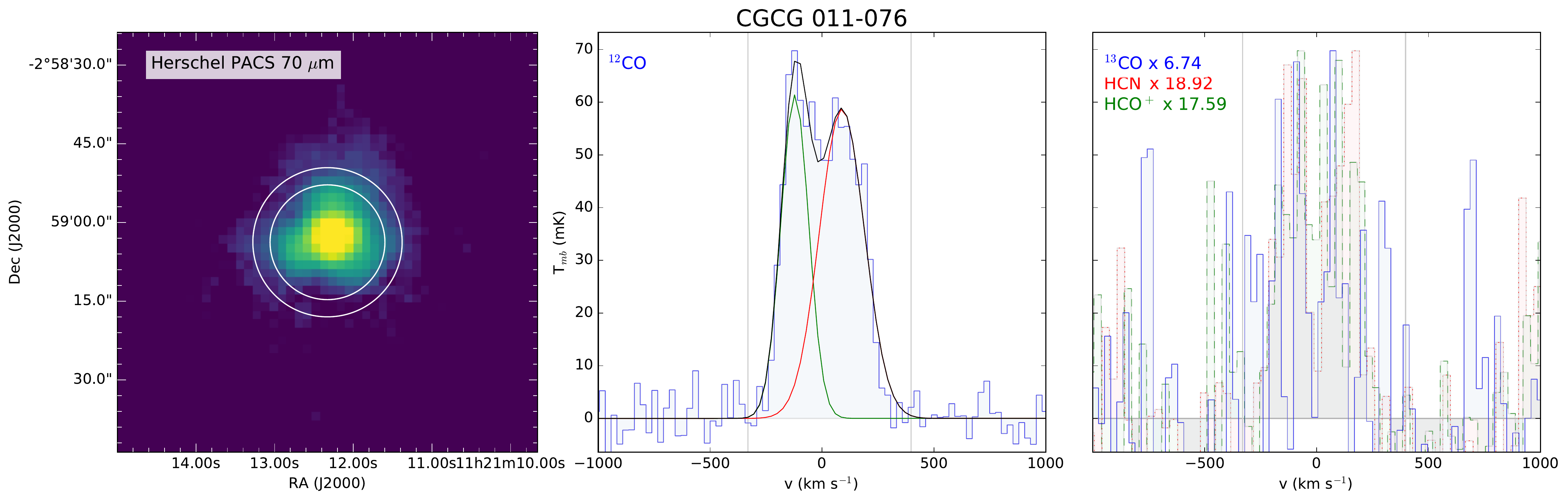}
\contcaption{\emph{continued}}
\end{figure*}

\begin{figure*}[htb]\centering
\includegraphics[width=.9\textwidth]{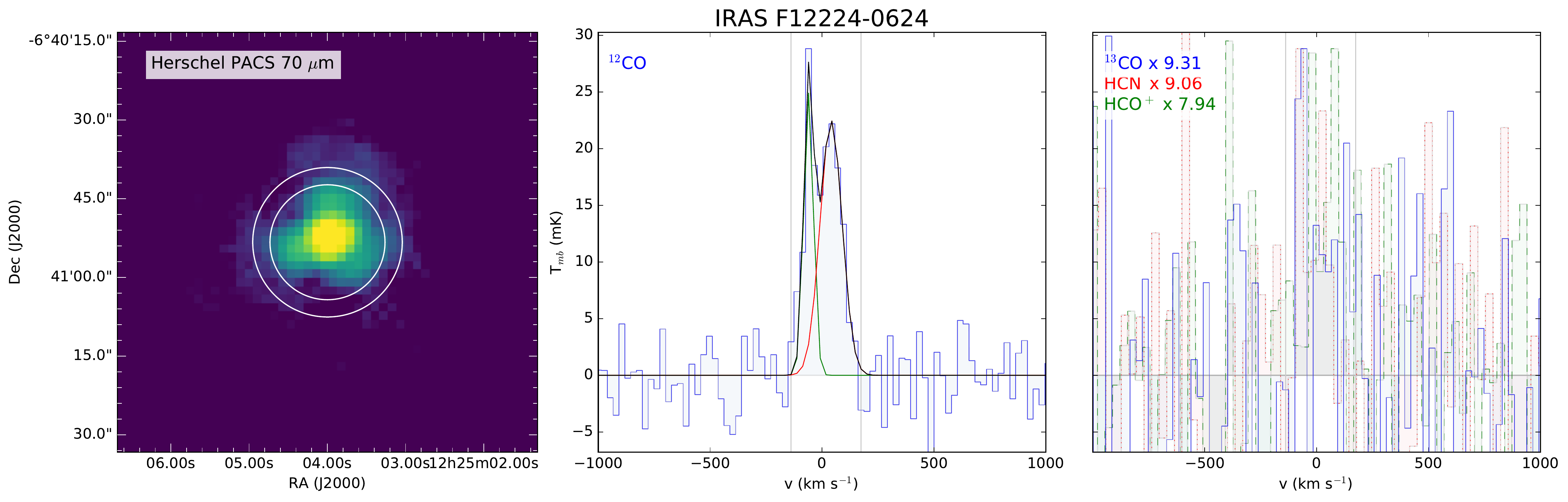}
\includegraphics[width=.9\textwidth]{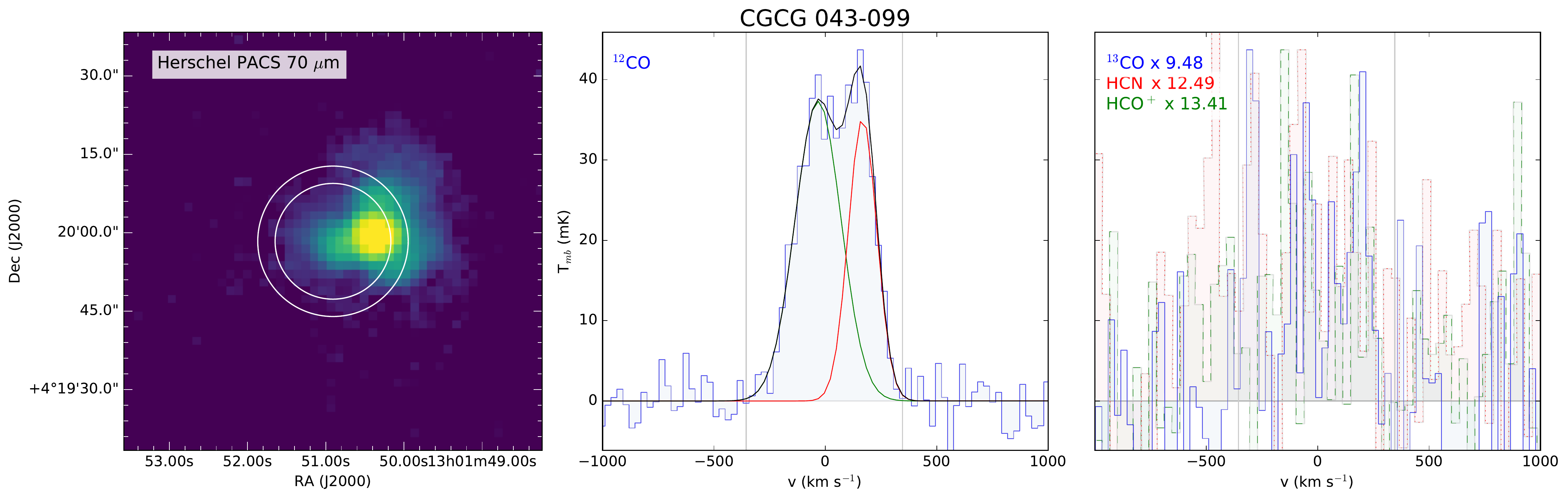}
\includegraphics[width=.9\textwidth]{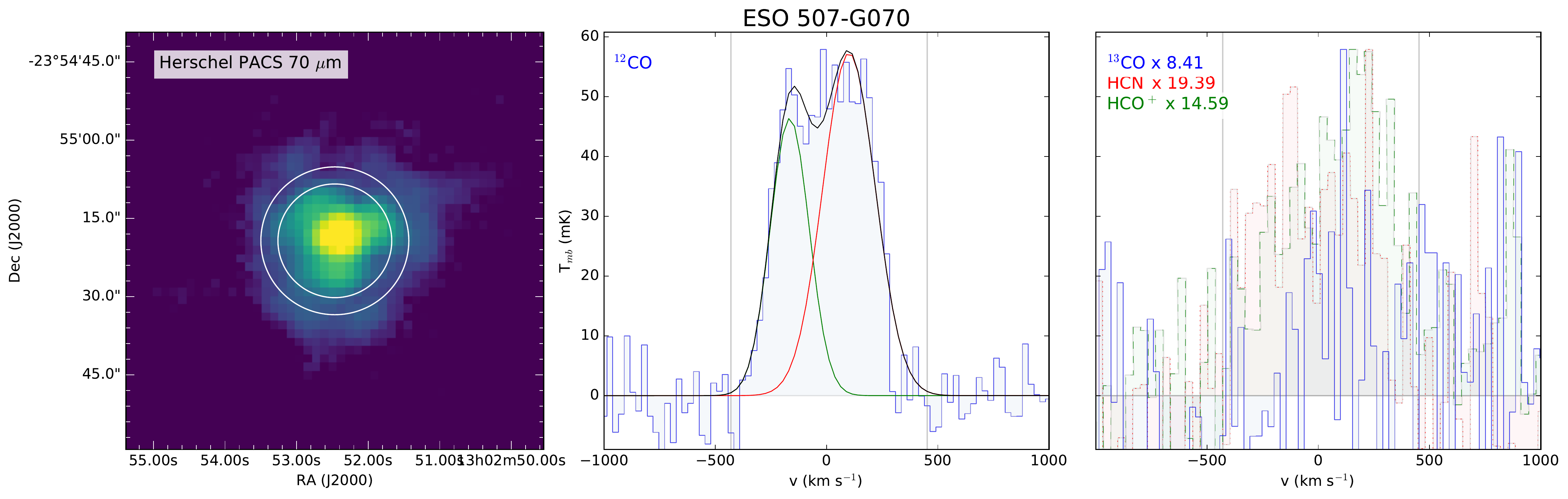}
\includegraphics[width=.9\textwidth]{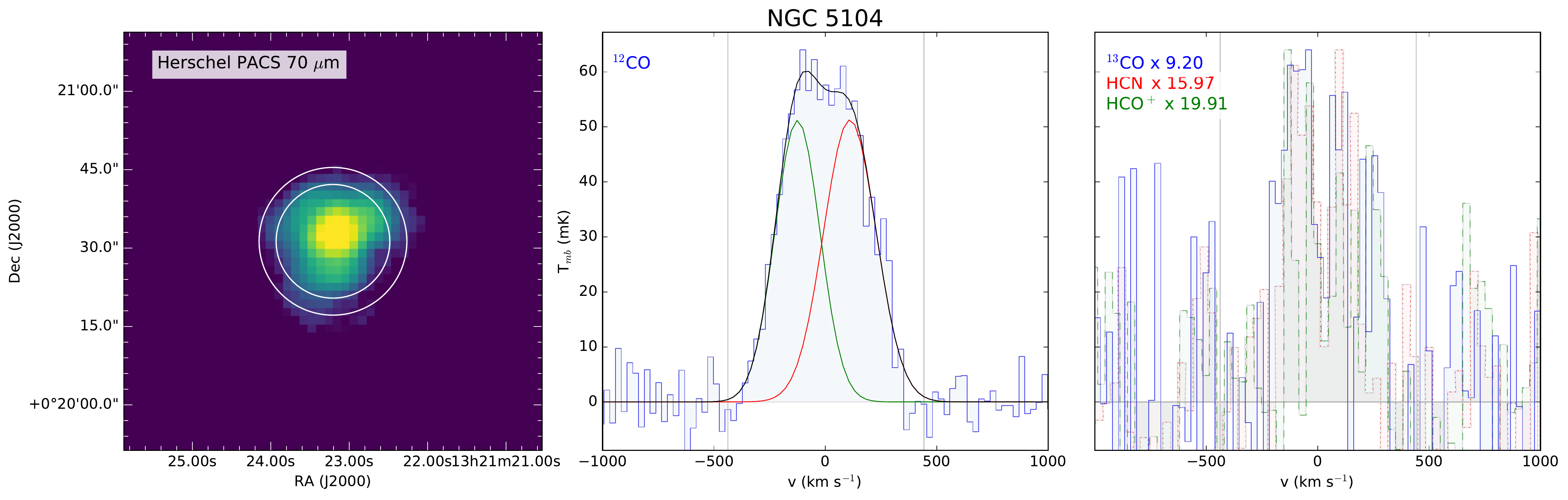}
\contcaption{\emph{continued}}
\end{figure*}

\begin{figure*}[htb]\centering
\includegraphics[width=.9\textwidth]{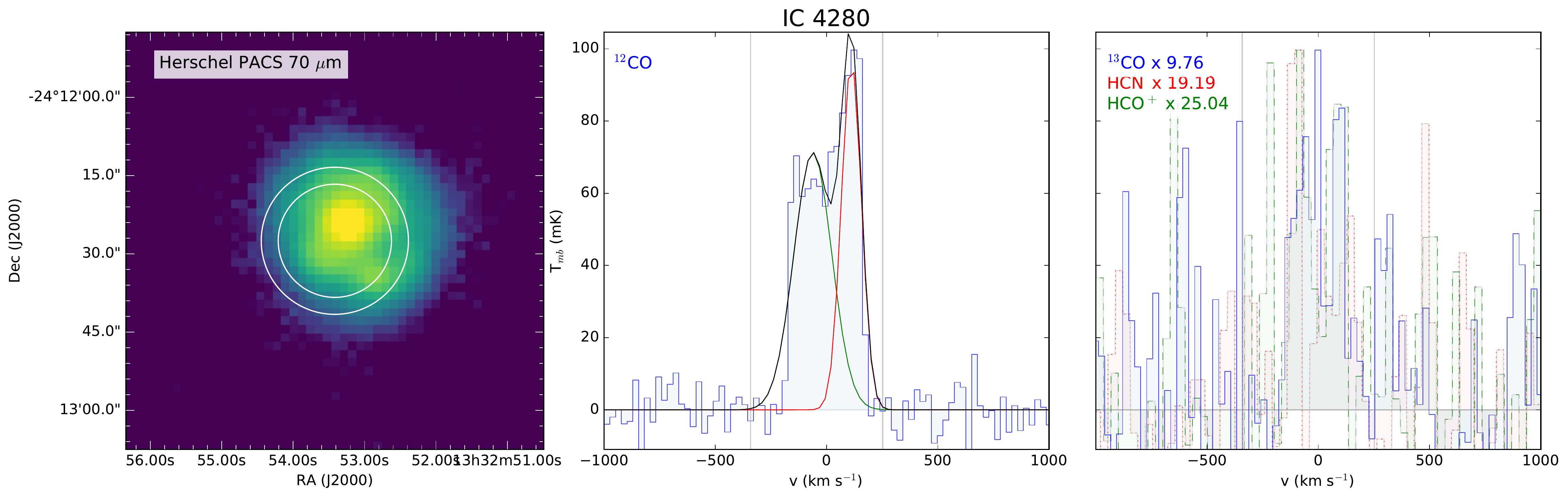}
\includegraphics[width=.9\textwidth]{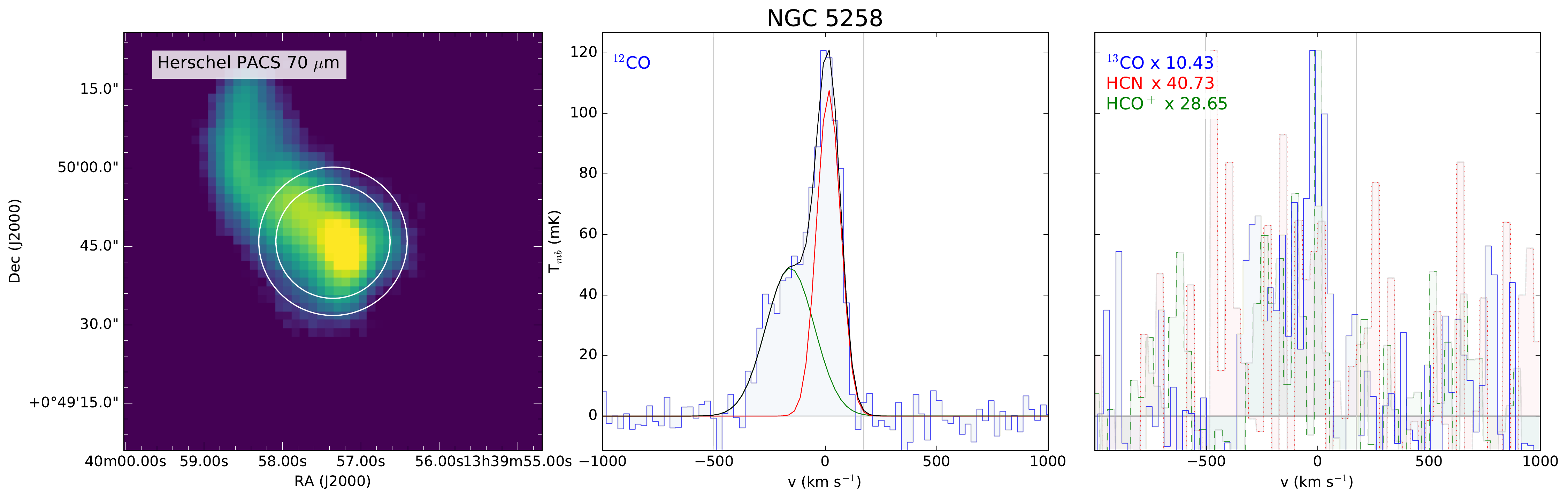}
\includegraphics[width=.9\textwidth]{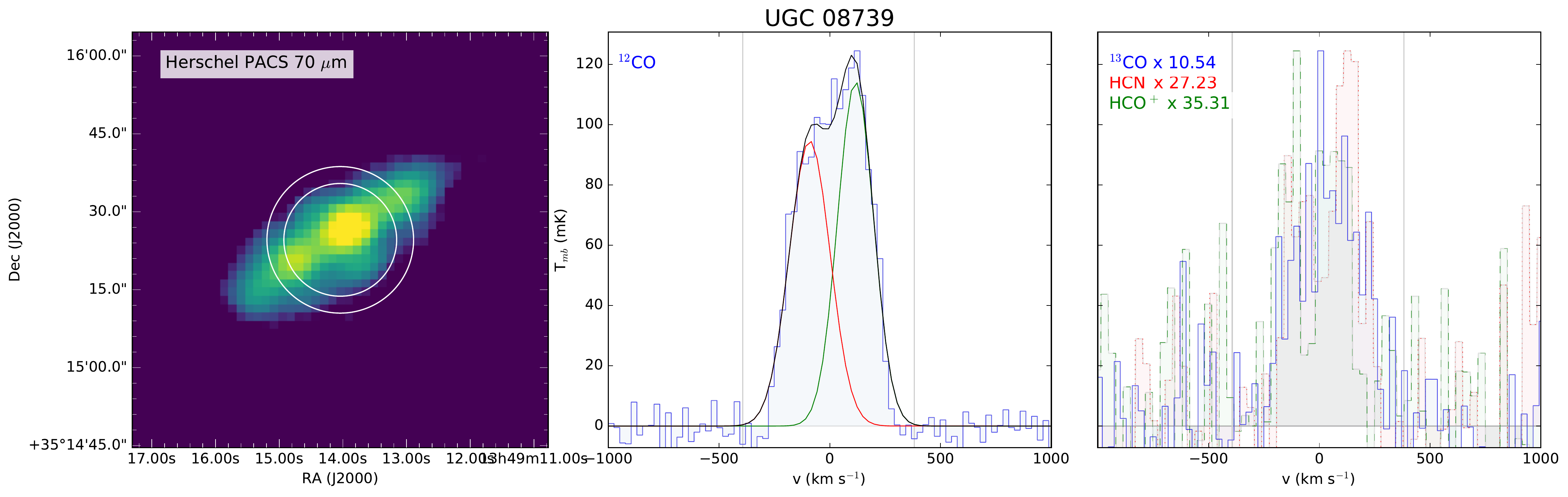}
\includegraphics[width=.9\textwidth]{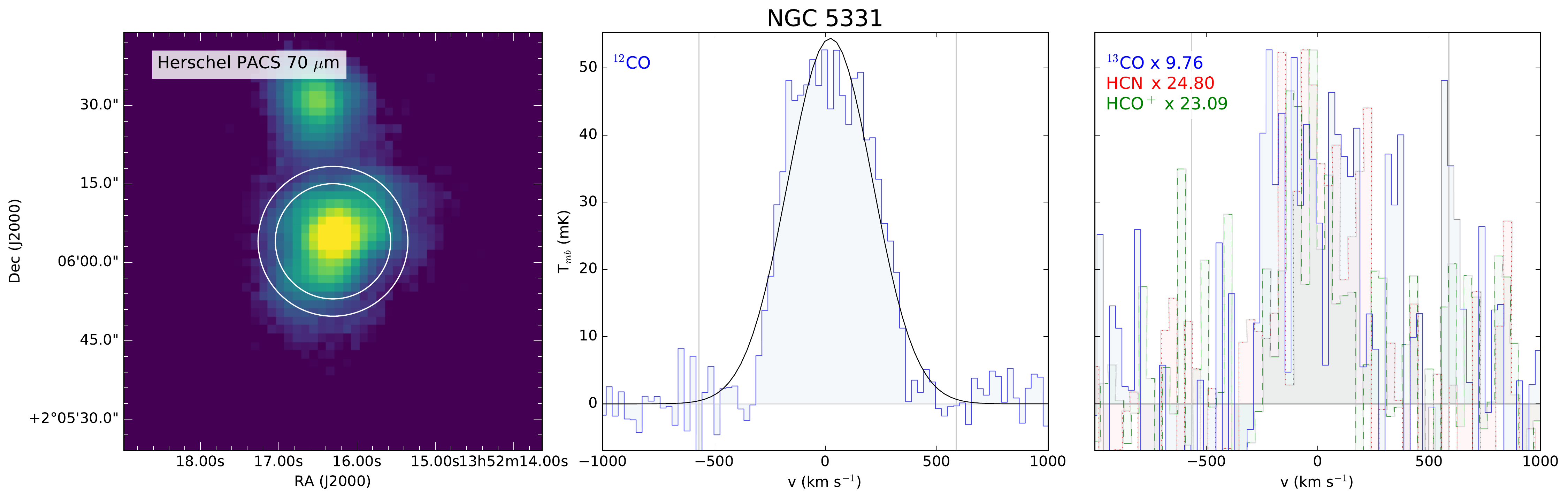}
\contcaption{\emph{continued}}
\end{figure*}

\begin{figure*}[htb]\centering
\includegraphics[width=.9\textwidth]{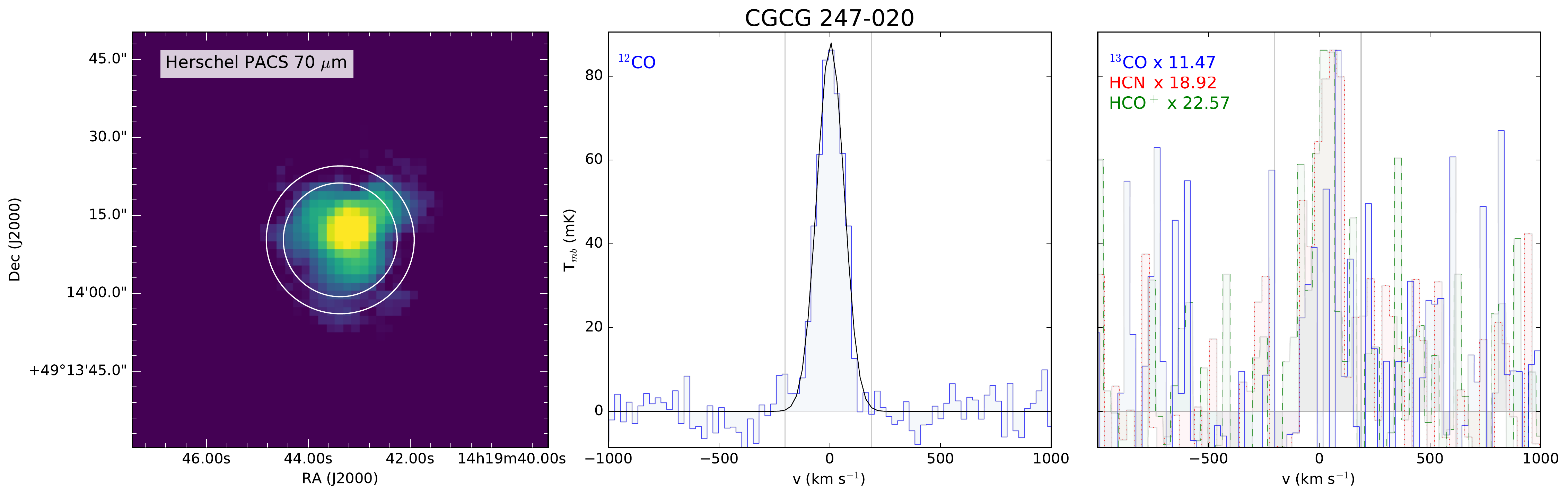}
\includegraphics[width=.9\textwidth]{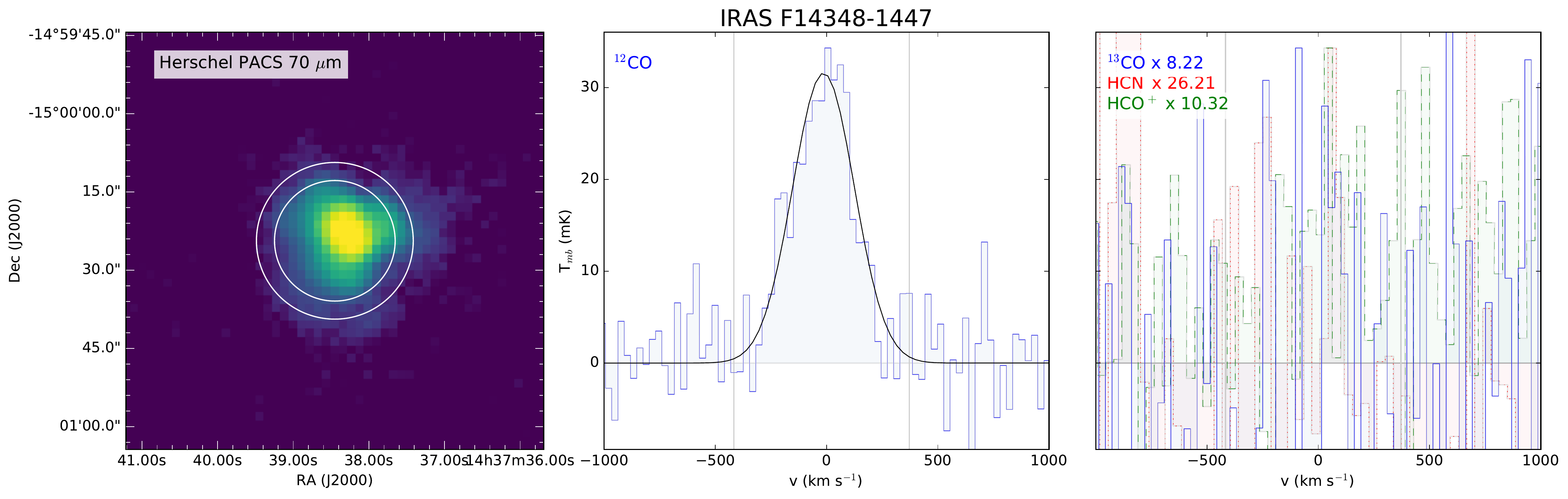}
\includegraphics[width=.9\textwidth]{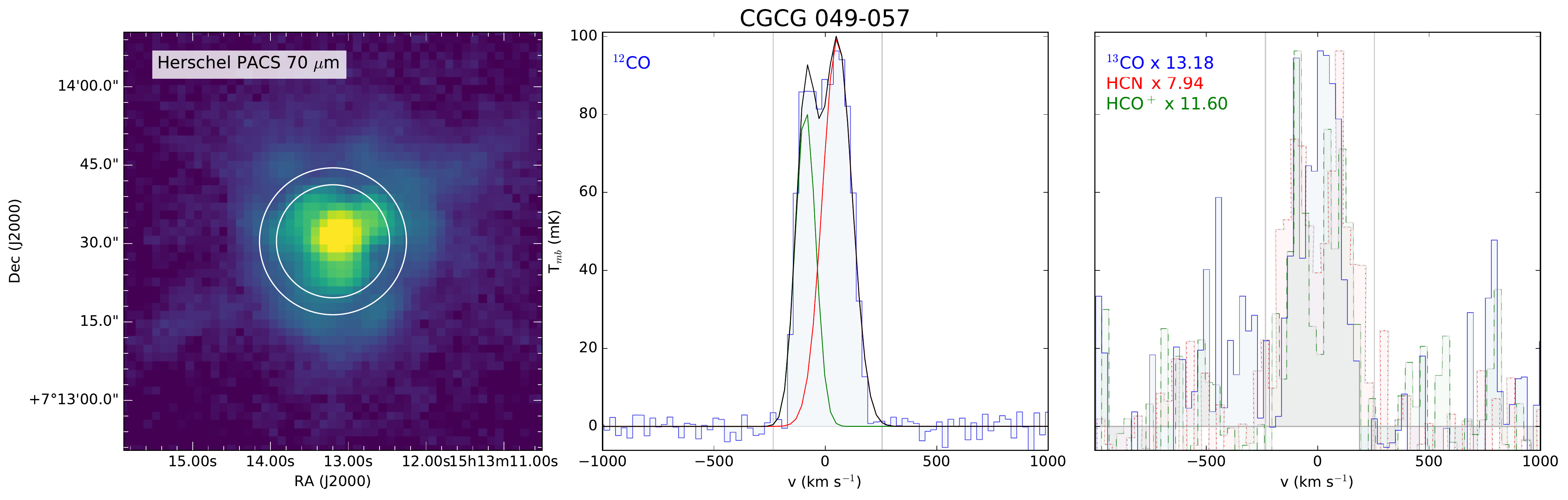}
\includegraphics[width=.9\textwidth]{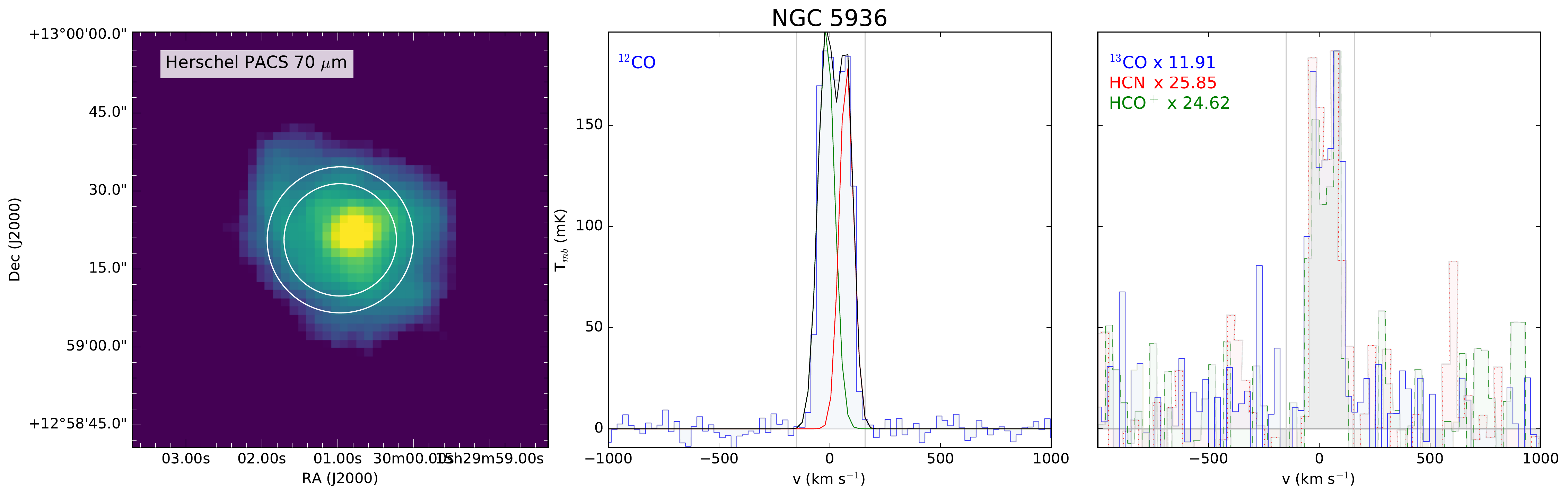}
\contcaption{\emph{continued}}
\end{figure*}

\begin{figure*}[htb]\centering
\includegraphics[width=.9\textwidth]{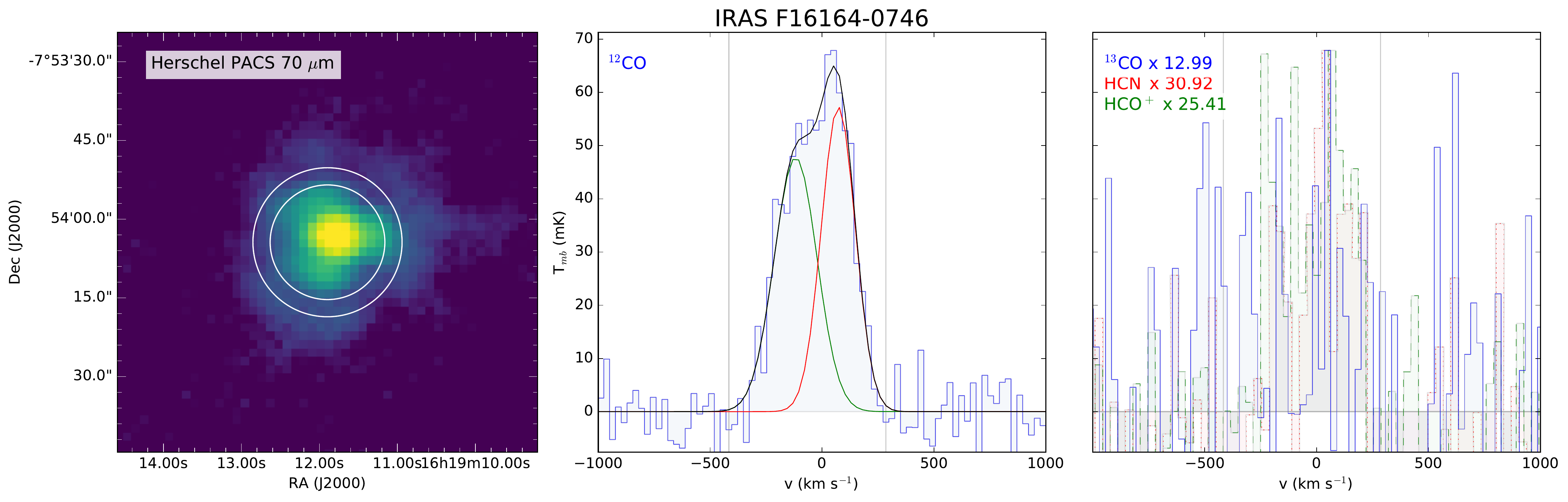}
\includegraphics[width=.9\textwidth]{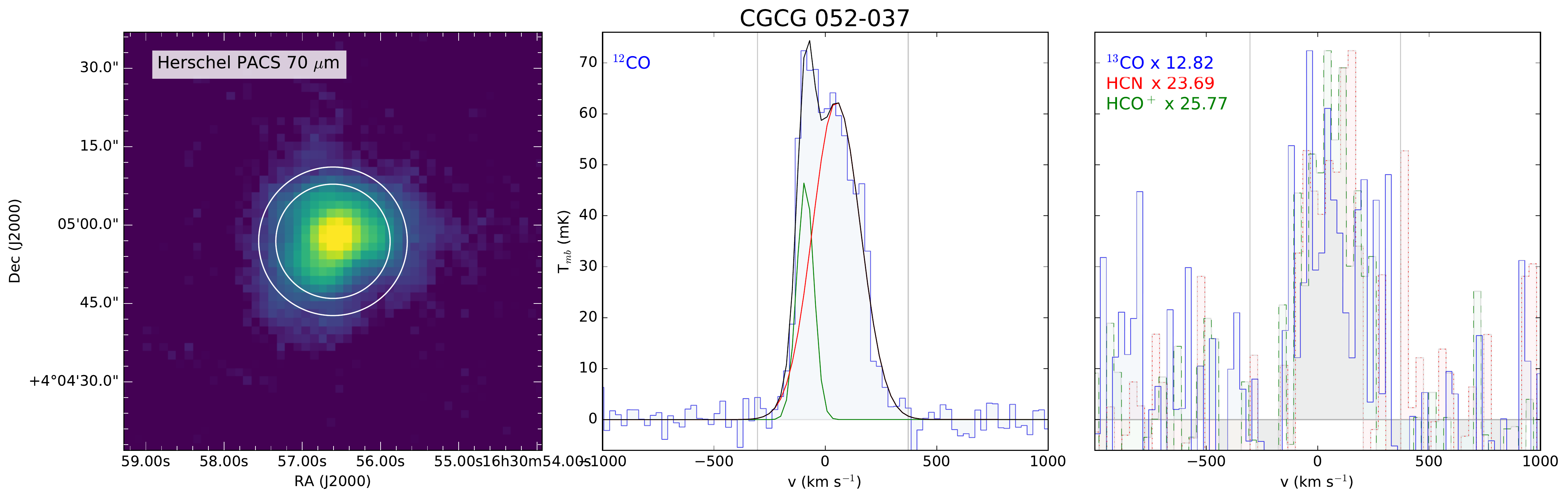}
\includegraphics[width=.9\textwidth]{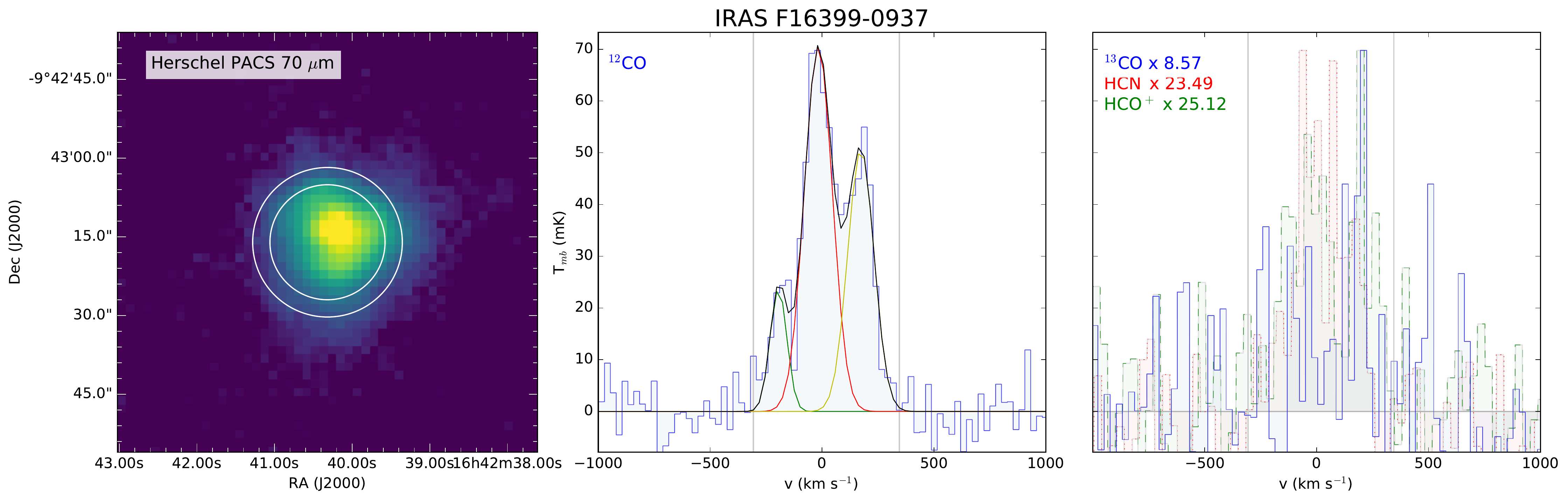}
\includegraphics[width=.9\textwidth]{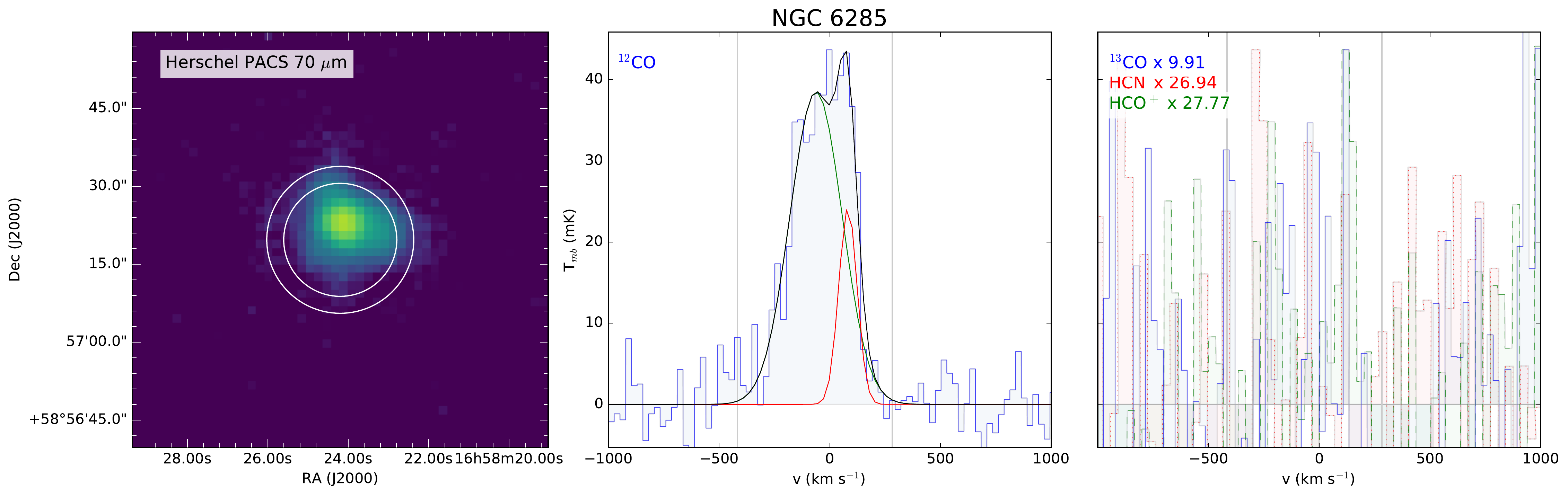}
\contcaption{\emph{continued}}
\end{figure*}

\begin{figure*}[htb]\centering
\includegraphics[width=.9\textwidth]{multipdf/NGC6286.pdf}
\includegraphics[width=.9\textwidth]{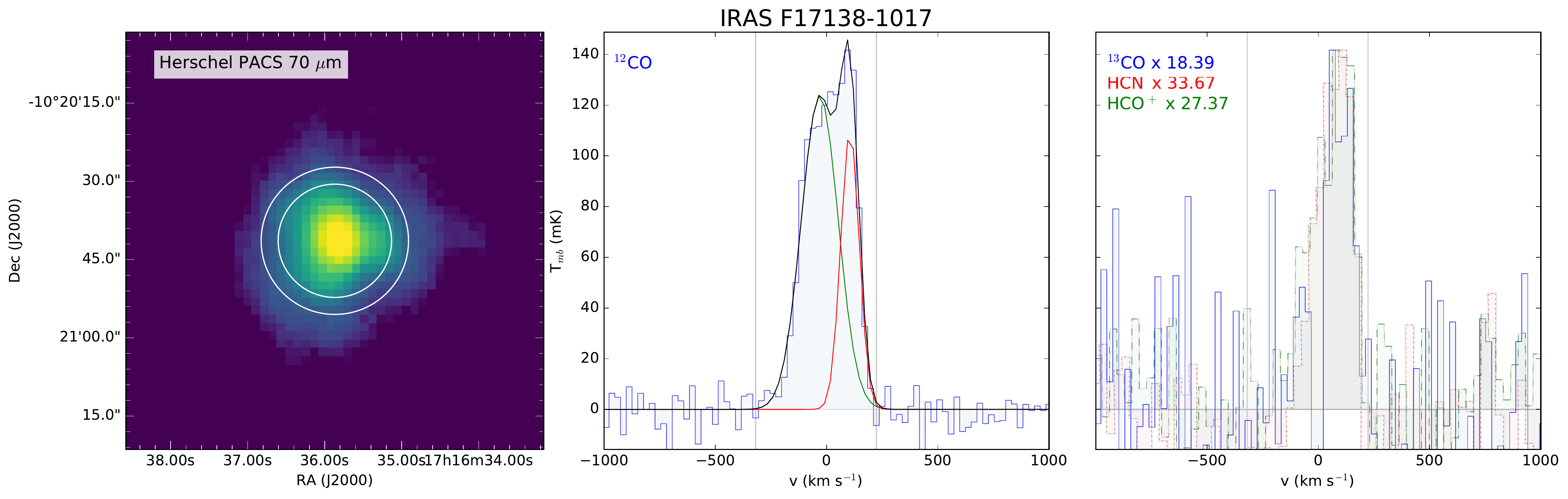}
\includegraphics[width=.9\textwidth]{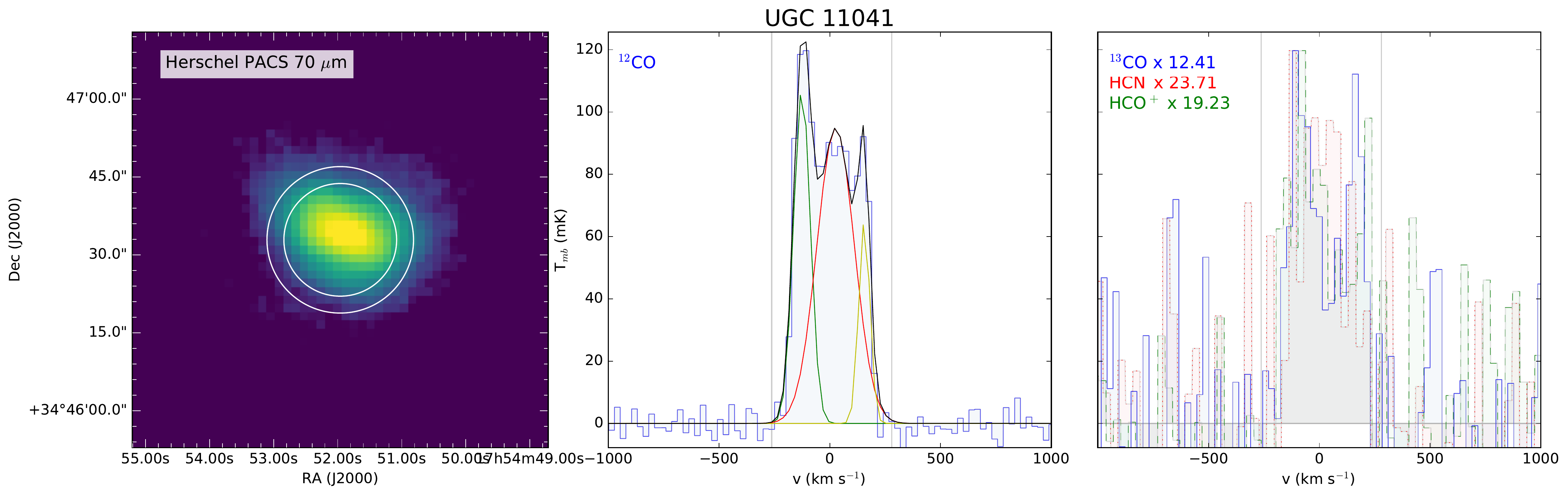}
\includegraphics[width=.9\textwidth]{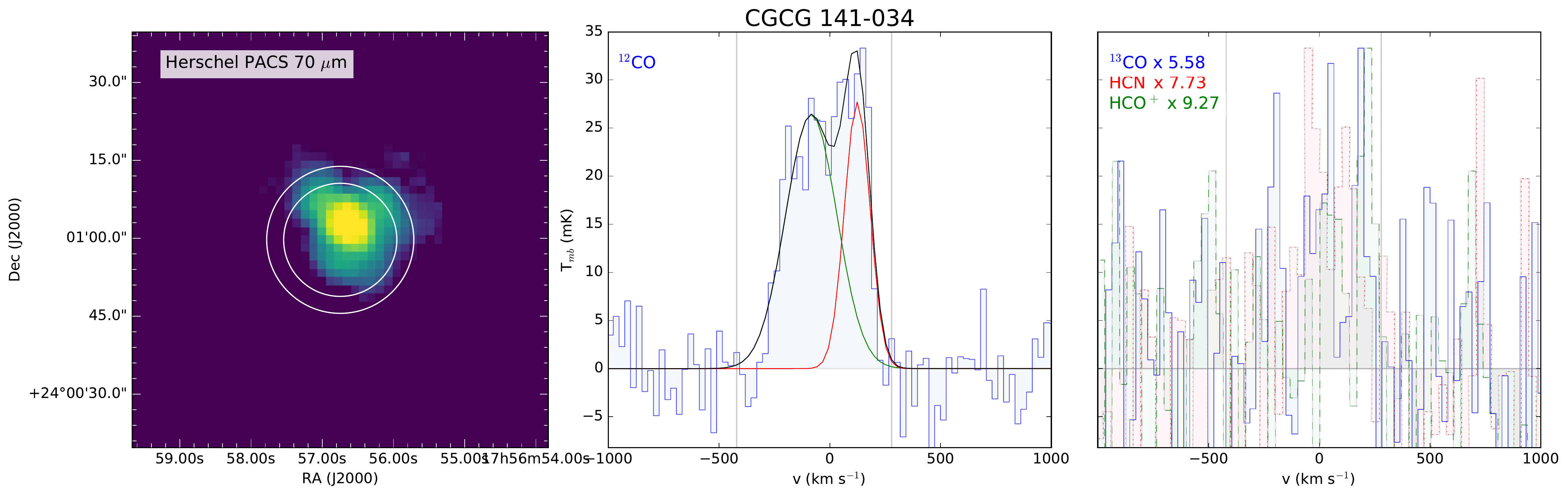}
\contcaption{\emph{continued}}
\end{figure*}

\begin{figure*}[htb]\centering
\includegraphics[width=.9\textwidth]{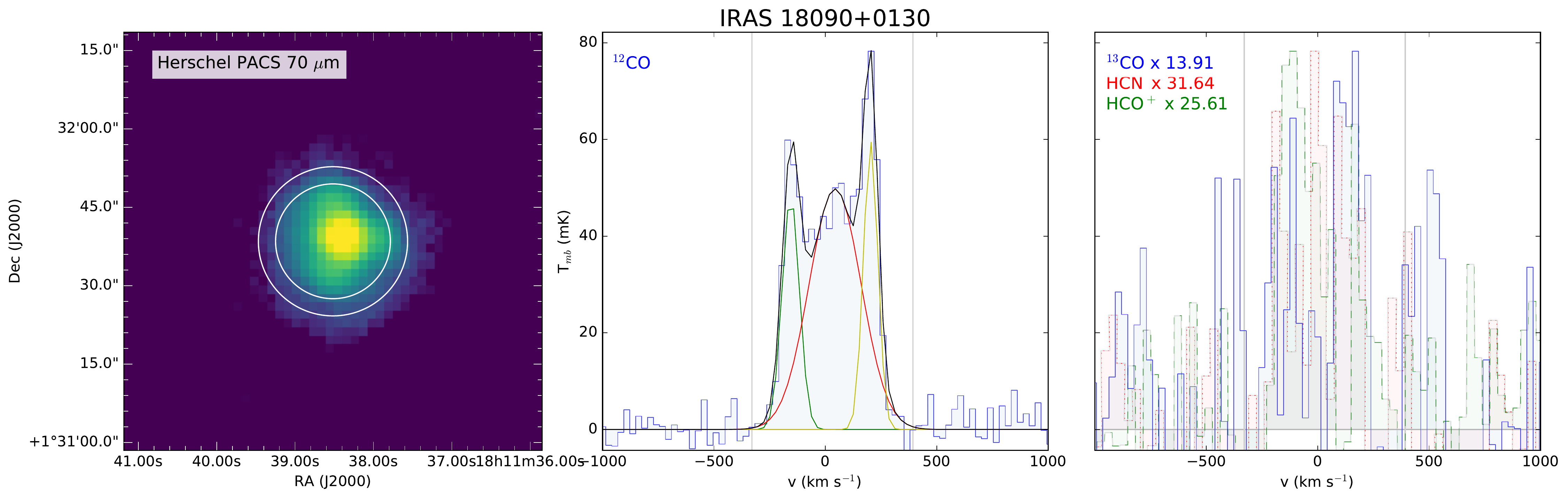}
\includegraphics[width=.9\textwidth]{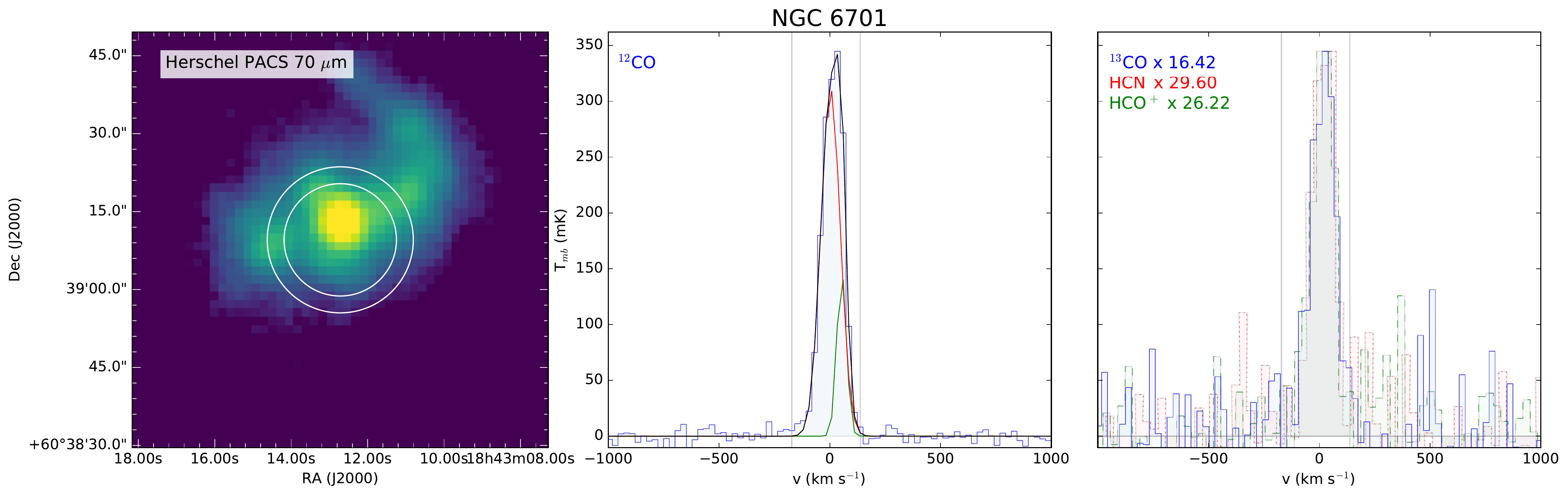}
\includegraphics[width=.9\textwidth]{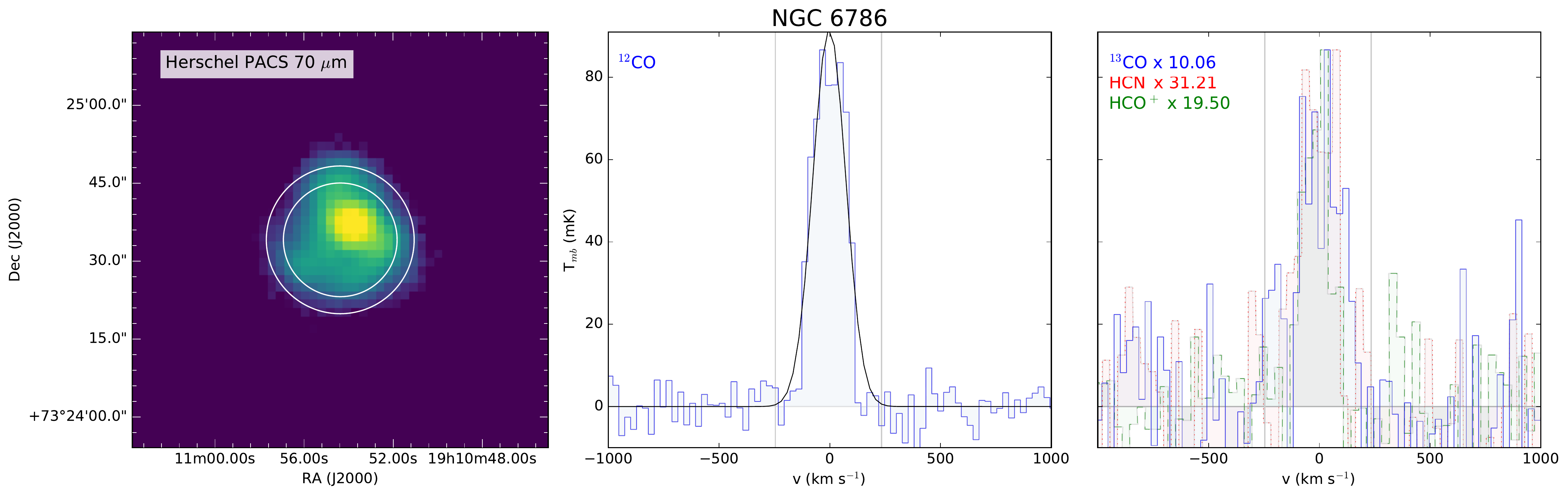}
\includegraphics[width=.9\textwidth]{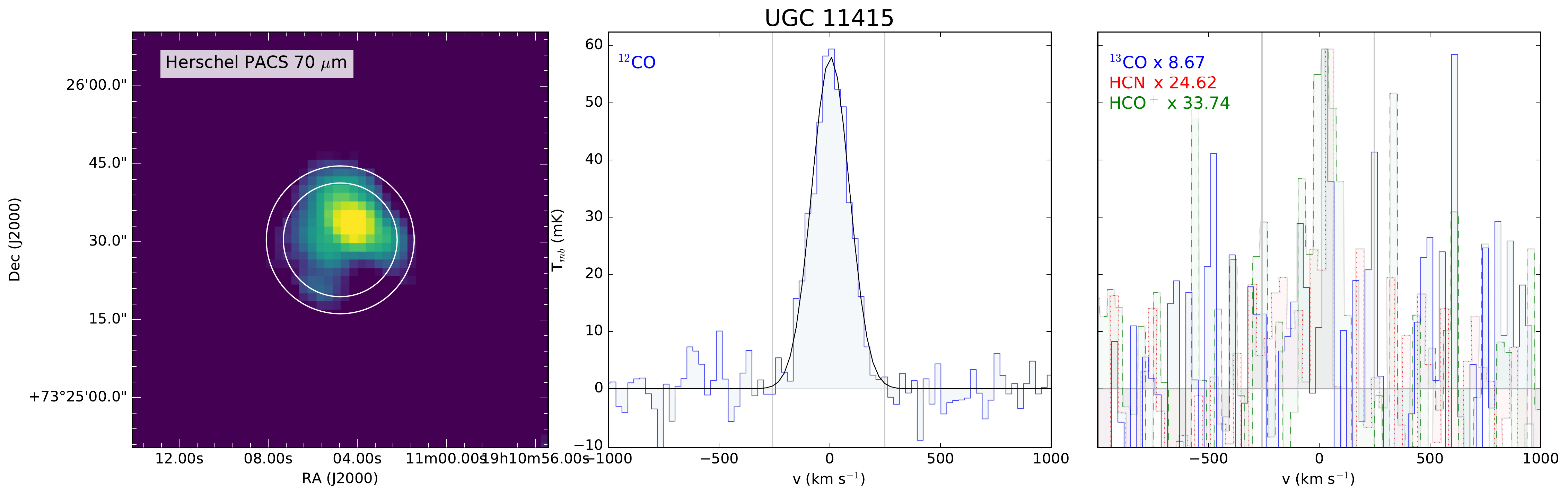}
\contcaption{\emph{continued}}
\end{figure*}

\begin{figure*}[htb]\centering
\includegraphics[width=.9\textwidth]{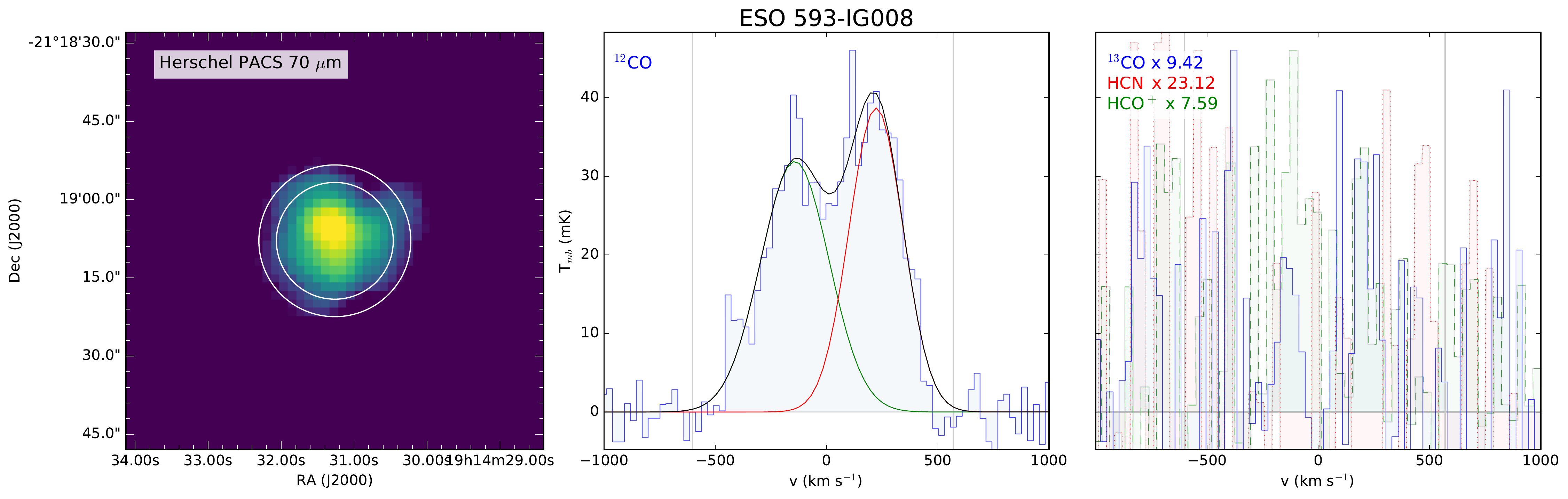}
\includegraphics[width=.9\textwidth]{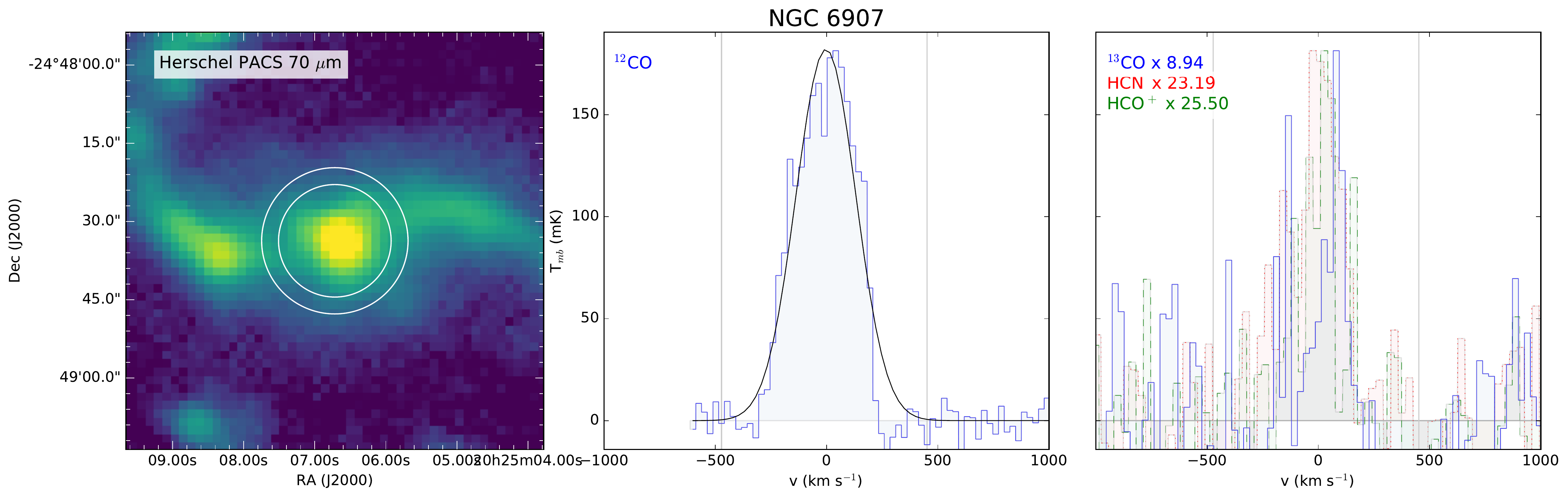}
\includegraphics[width=.9\textwidth]{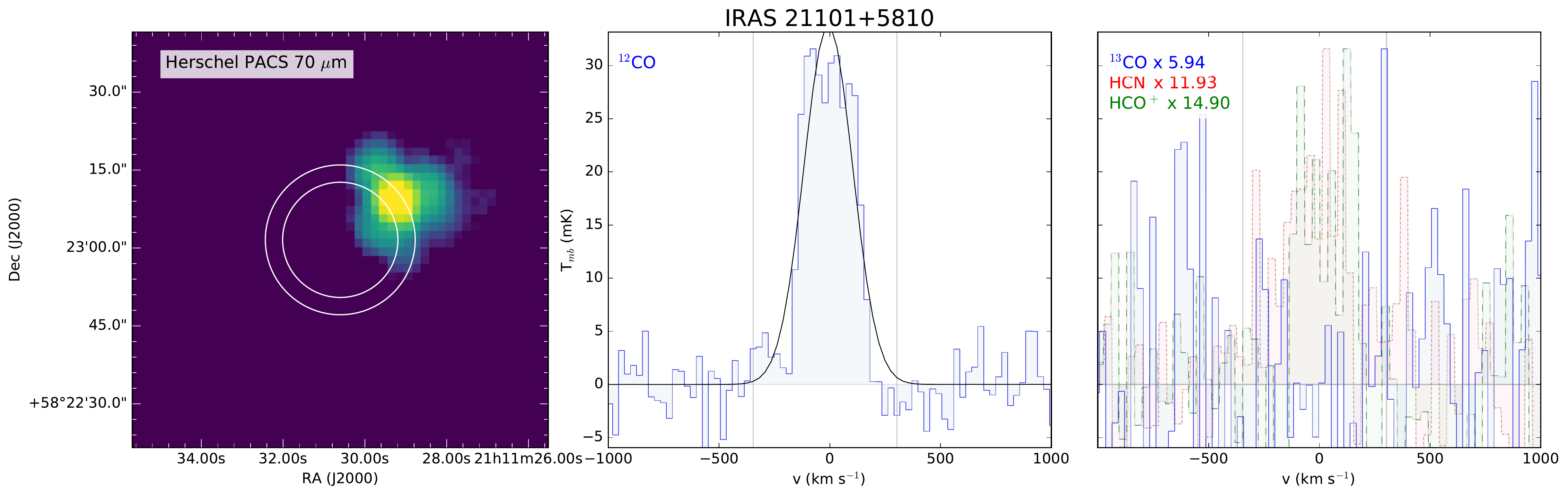}
\includegraphics[width=.9\textwidth]{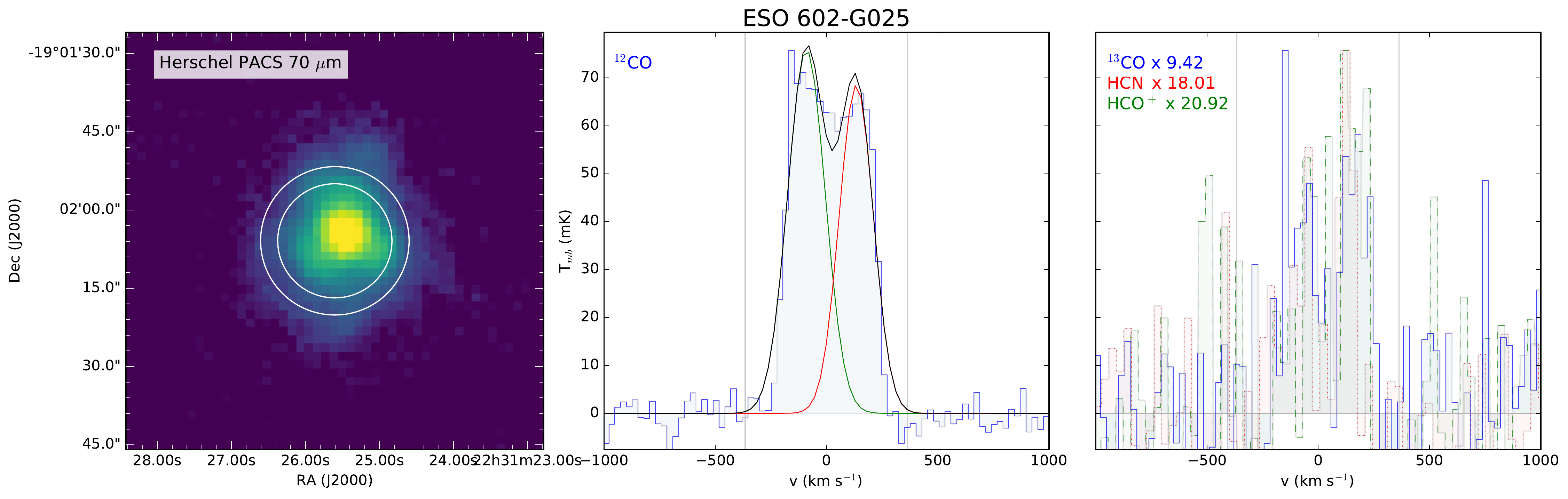}
\contcaption{\emph{continued}}
\end{figure*}

\begin{figure*}[htb]\centering
\includegraphics[width=.9\textwidth]{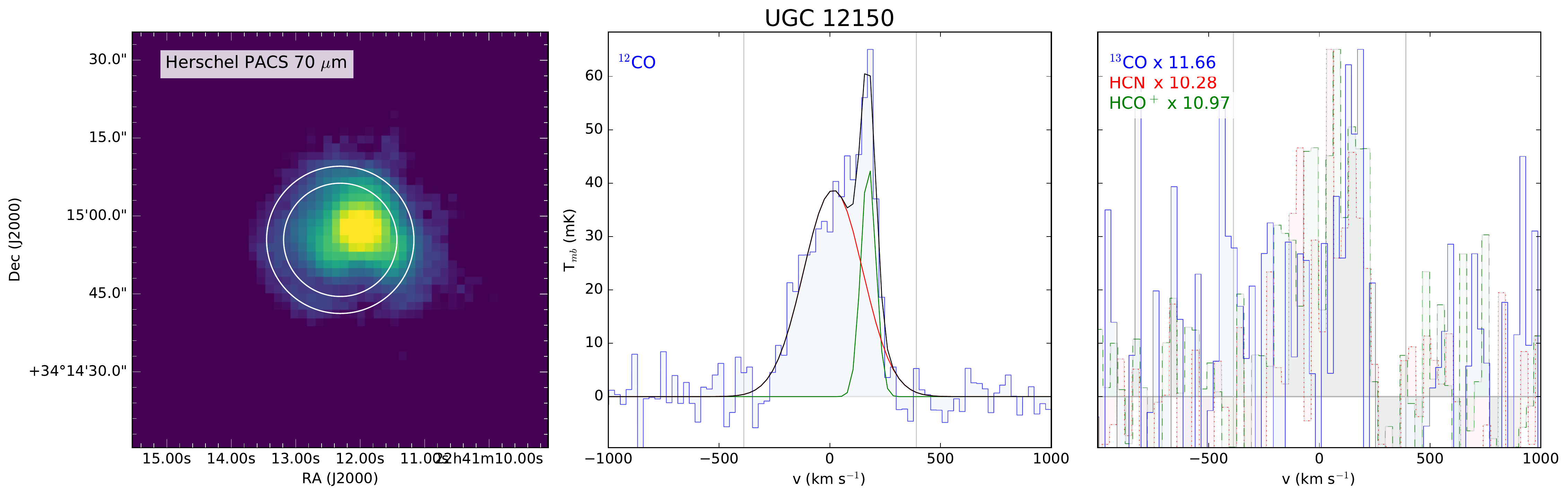}
\includegraphics[width=.9\textwidth]{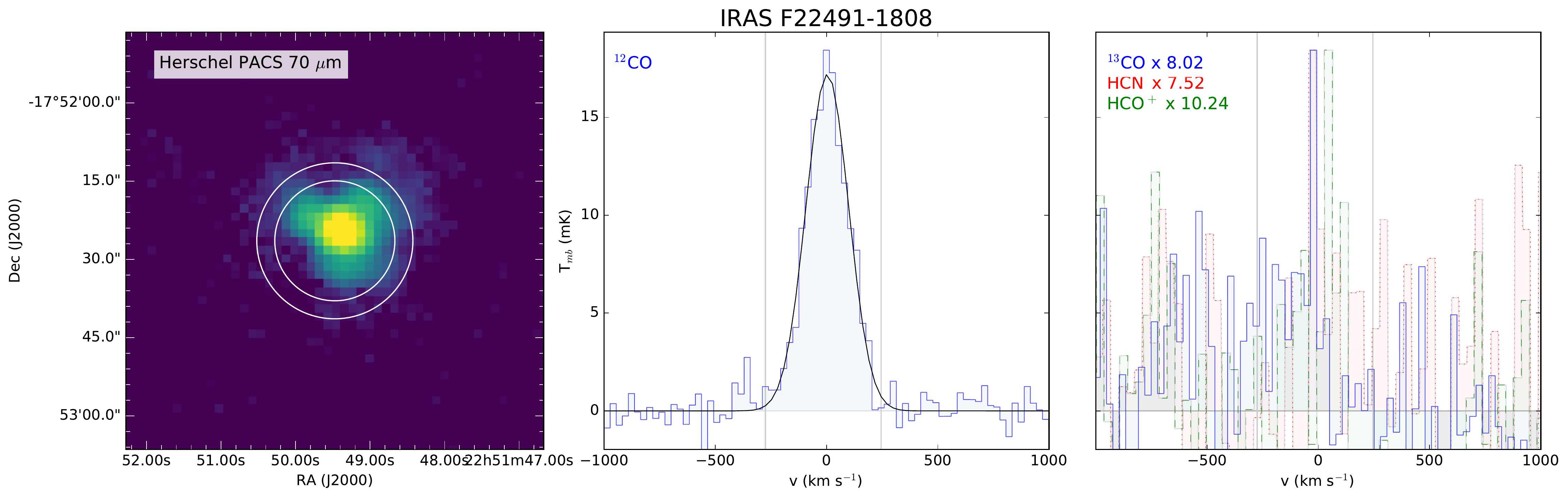}
\includegraphics[width=.9\textwidth]{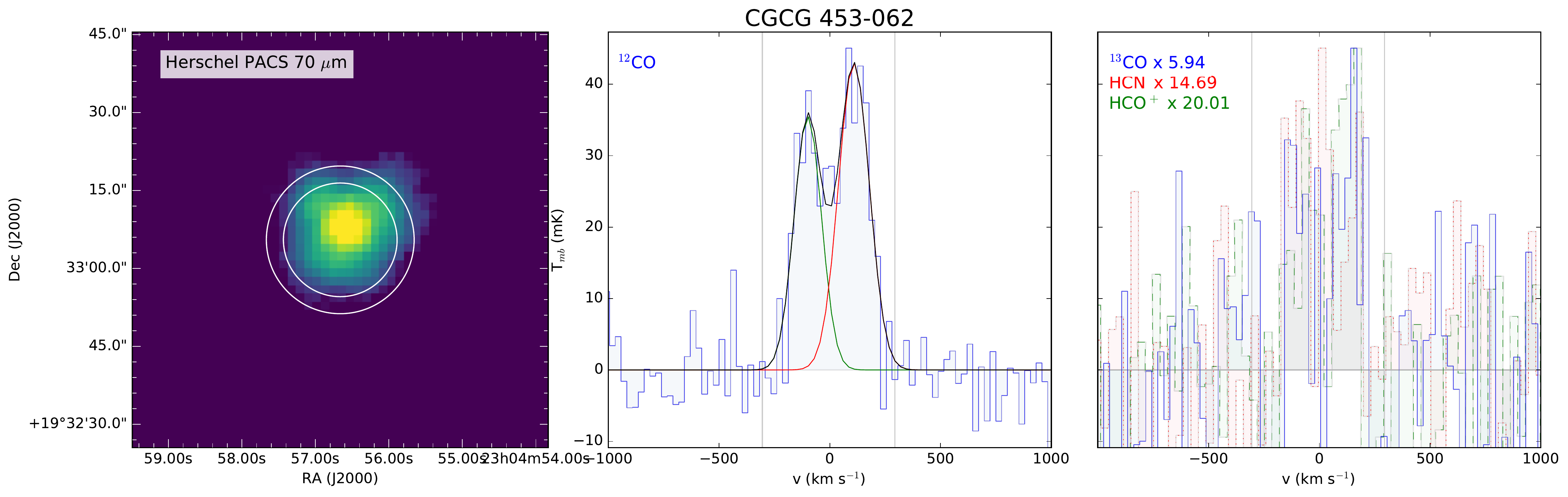}
\includegraphics[width=.9\textwidth]{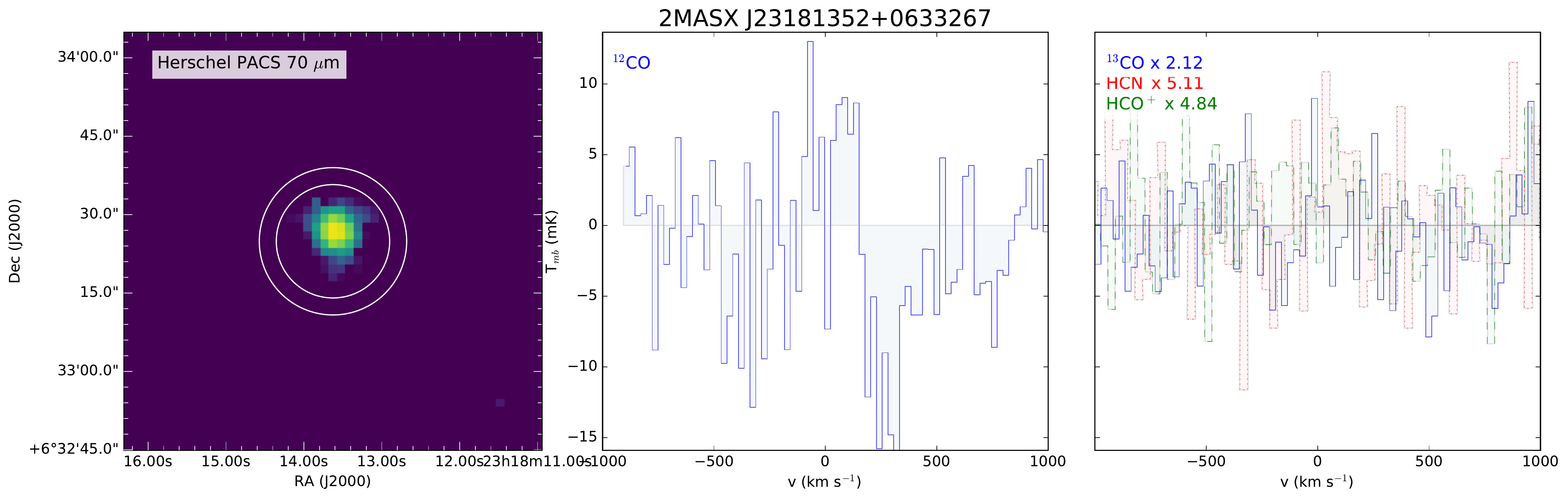}
\contcaption{\emph{continued}}
\end{figure*}

\begin{table*}
\caption{\label{table:gaussian}Fitted Gaussian line components to the \twCO spectra.}
\centering\small
\begin{tabular}{lccccccccccc}
\hline\hline
{} & \multicolumn{3}{c}{First component}  & {} &\multicolumn{3}{c}{Second component}  &  {} &\multicolumn{3}{c}{Third component}    \\
\cline{2-4}\cline{6-8}\cline{10-12} \\
{Source name}  & {Position} & {FWHM} & {Peak} & {} & {Position} & {FWHM} & {Peak}&  {} & {Position} & {FWHM} & {Peak} \\
{} & {(km\,s$^{-1}$)} & {(km\,s$^{-1}$)} & {mK} & {} & {(km\,s$^{-1}$)} & {(km\,s$^{-1}$)} & {mK}& {} & {(km\,s$^{-1}$)} & {(km\,s$^{-1}$)} & {mK} \\
\hline
NGC 0034 &                      $10.3$ & $285.9$ & $89.9$ &  & $\cdots$ & $\cdots$ & $\cdots$ &  & $\cdots$ & $\cdots$ & $\cdots$ \\
Arp 256N &                       $15.7$ & $137.2$ & $22.4$ &  & $\cdots$ & $\cdots$ & $\cdots$ &  & $\cdots$ & $\cdots$ & $\cdots$ \\
Arp 256S &                       $-33.8$ & $197.3$ & $32.0$ &  & $104.7$ & $89.6$ & $31.8$ &  & $\cdots$ & $\cdots$ & $\cdots$ \\
IC 1623 &                                $-62.8$ & $186.6$ & $259.6$ &  & $117.6$ & $159.7$ & $277.6$ &  & $\cdots$ & $\cdots$ & $\cdots$ \\
MCG -03-04-014 &                 $-145.9$ & $74.7$ & $48.2$ &  & $-5.9$ & $232.4$ & $44.5$ &  & $151.9$ & $79.1$ & $67.1$ \\
IRAS F01364-1042 &       $48.1$ & $326.8$ & $22.5$ &  & $\cdots$ & $\cdots$ & $\cdots$ &  & $\cdots$ & $\cdots$ & $\cdots$ \\
IC 0214 &                       $-197.0$ & $93.8$ & $11.3$ &  & $6.3$ & $176.1$ & $63.9$ &  & $\cdots$ & $\cdots$ & $\cdots$ \\
UGC 01845 &                      $-179.9$ & $112.8$ & $77.1$ &  & $-6.7$ & $242.7$ & $99.4$ &  & $154.2$ & $115.3$ & $73.5$ \\
NGC 0958 &                                $-298.6$ & $50.2$ & $29.1$ &  & $-15.5$ & $316.9$ & $78.5$ &  & $179.5$ & $74.7$ & $41.5$ \\
ESO 550-IG025 &                  $-142.7$ & $83.0$ & $38.7$ &  & $85.9$ & $293.5$ & $37.2$ &  & $\cdots$ & $\cdots$ & $\cdots$ \\
UGC 03094 &                      $-138.0$ & $174.3$ & $57.5$ &  & $103.9$ & $300.8$ & $55.1$ &  & $\cdots$ & $\cdots$ & $\cdots$ \\
NGC 1797 &                       $-69.6$ & $135.3$ & $98.8$ &  & $88.2$ & $133.5$ & $89.1$ &  & $\cdots$ & $\cdots$ & $\cdots$ \\
IRAS F05189-2524 &       $-5.1$ & $175.8$ & $35.1$ &  & $\cdots$ & $\cdots$ & $\cdots$ &  & $\cdots$ & $\cdots$ & $\cdots$ \\
IRAS F05187-1017 &       $-90.1$ & $185.8$ & $40.5$ &  & $94.0$ & $153.3$ & $32.5$ &  & $\cdots$ & $\cdots$ & $\cdots$ \\
IRAS F06076-2139 &       $10.5$ & $225.3$ & $32.8$ &  & $\cdots$ & $\cdots$ & $\cdots$ &  & $\cdots$ & $\cdots$ & $\cdots$ \\
NGC 2341 &                       $-63.3$ & $177.4$ & $74.7$ &  & $109.2$ & $120.6$ & $65.3$ &  & $\cdots$ & $\cdots$ & $\cdots$ \\
NGC 2342 &                       $-96.3$ & $111.7$ & $58.0$ &  & $76.6$ & $178.3$ & $89.3$ &  & $\cdots$ & $\cdots$ & $\cdots$ \\
IRAS 07251-0248 &                $-0.9$ & $363.1$ & $10.2$ &  & $\cdots$ & $\cdots$ & $\cdots$ &  & $\cdots$ & $\cdots$ & $\cdots$ \\
IRAS F09111-1007 W &     $-1.2$ & $193.2$ & $40.2$ &  & $\cdots$ & $\cdots$ & $\cdots$ &  & $\cdots$ & $\cdots$ & $\cdots$ \\
IRAS F09111-1007 E &     $-102.9$ & $141.5$ & $12.7$ &  & $150.6$ & $218.2$ & $10.2$ &  & $\cdots$ & $\cdots$ & $\cdots$ \\
UGC 05101 &                      $-118.9$ & $330.9$ & $31.0$ &  & $187.4$ & $216.1$ & $32.9$ &  & $\cdots$ & $\cdots$ & $\cdots$ \\
2MASX J11210825-0259399&  $\cdots$ & $\cdots$ & $\cdots$ &  & $\cdots$ & $\cdots$ & $\cdots$ &  & $\cdots$ & $\cdots$ & $\cdots$ \\
CGCG 011-076 &                   $-121.9$ & $146.7$ & $61.4$ &  & $89.0$ & $236.8$ & $58.7$ &  & $\cdots$ & $\cdots$ & $\cdots$ \\
IRAS F12224-0624 &       $-60.2$ & $51.8$ & $24.9$ &  & $40.5$ & $115.7$ & $22.5$ &  & $\cdots$ & $\cdots$ & $\cdots$ \\
CGCG 043-099 &                   $-32.5$ & $243.3$ & $37.3$ &  & $167.1$ & $151.0$ & $35.2$ &  & $\cdots$ & $\cdots$ & $\cdots$ \\
ESO 507-G070 &                    $-165.5$ & $205.9$ & $46.4$ &  & $101.1$ & $278.7$ & $57.3$ &  & $\cdots$ & $\cdots$ & $\cdots$ \\
NGC 5104 &                       $-125.2$ & $237.7$ & $51.2$ &  & $110.0$ & $276.8$ & $51.3$ &  & $\cdots$ & $\cdots$ & $\cdots$ \\
IC 4280 &                       $-61.7$ & $200.9$ & $71.3$ &  & $111.7$ & $105.7$ & $96.5$ &  & $\cdots$ & $\cdots$ & $\cdots$ \\
NGC 5258 &                       $-158.1$ & $256.1$ & $48.9$ &  & $15.0$ & $125.5$ & $107.7$ &  & $\cdots$ & $\cdots$ & $\cdots$ \\
UGC 08739 &                      $116.3$ & $190.6$ & $114.3$ &  & $-90.7$ & $216.9$ & $94.7$ &  & $\cdots$ & $\cdots$ & $\cdots$ \\
NGC 5331 &                        $22.3$ & $449.5$ & $54.4$ &  & $\cdots$ & $\cdots$ & $\cdots$ &  & $\cdots$ & $\cdots$ & $\cdots$ \\
CGCG 247-020 &                   $3.4$ & $143.8$ & $88.1$ &  & $\cdots$ & $\cdots$ & $\cdots$ &  & $\cdots$ & $\cdots$ & $\cdots$ \\
IRAS F14348-1447 &       $-13.5$ & $326.9$ & $31.6$ &  & $\cdots$ & $\cdots$ & $\cdots$ &  & $\cdots$ & $\cdots$ & $\cdots$ \\
CGCG 049-057 &                   $-89.1$ & $106.2$ & $81.6$ &  & $53.8$ & $155.4$ & $99.6$ &  & $\cdots$ & $\cdots$ & $\cdots$ \\
NGC 5936 &                       $-16.1$ & $88.6$ & $200.3$ &  & $75.0$ & $74.8$ & $182.3$ &  & $\cdots$ & $\cdots$ & $\cdots$ \\
IRAS F16164-0746 &       $-118.1$ & $225.0$ & $47.8$ &  & $72.7$ & $178.5$ & $57.3$ &  & $\cdots$ & $\cdots$ & $\cdots$ \\
CGCG 052-037 &                    $-89.9$ & $88.9$ & $47.1$ &  & $50.0$ & $252.4$ & $62.4$ &  & $\cdots$ & $\cdots$ & $\cdots$ \\
IRAS F16399-0937 &       $-194.7$ & $88.6$ & $23.7$ &  & $-16.7$ & $147.5$ & $70.6$ &  & $173.2$ & $138.6$ & $50.6$ \\
NGC 6285 &                       $-62.0$ & $277.1$ & $38.5$ &  & $81.8$ & $97.0$ & $24.3$ &  & $\cdots$ & $\cdots$ & $\cdots$ \\
NGC 6286 &                       $-141.0$ & $189.1$ & $138.2$ &  & $89.9$ & $338.7$ & $86.3$ &  & $\cdots$ & $\cdots$ & $\cdots$ \\
IRAS F17138-1017 &       $-30.6$ & $195.1$ & $123.7$ &  & $105.3$ & $97.1$ & $109.9$ &  & $\cdots$ & $\cdots$ & $\cdots$ \\
UGC 11041 &                      $-125.7$ & $88.9$ & $107.5$ &  & $26.3$ & $198.4$ & $94.9$ &  & $155.8$ & $59.2$ & $65.1$ \\
CGCG 141-034 &                   $-83.6$ & $272.6$ & $26.4$ &  & $124.0$ & $135.4$ & $27.7$ &  & $\cdots$ & $\cdots$ & $\cdots$ \\
IRAS 18090+0130 &                $-155.4$ & $91.6$ & $48.3$ &  & $43.8$ & $274.0$ & $49.8$ &  & $203.3$ & $75.8$ & $59.6$ \\
NGC 6701 &                       $53.2$ & $50.1$ & $147.9$ &  & $3.3$ & $102.8$ & $311.5$ &  & $\cdots$ & $\cdots$ & $\cdots$ \\
NGC 6786 &                       $-2.3$ & $174.8$ & $91.9$ &  & $\cdots$ & $\cdots$ & $\cdots$ &  & $\cdots$ & $\cdots$ & $\cdots$ \\
UGC 11415 &                      $4.0$ & $199.5$ & $58.0$ &  & $\cdots$ & $\cdots$ & $\cdots$ &  & $\cdots$ & $\cdots$ & $\cdots$ \\
ESO 593-IG008 &                   $-142.2$ & $361.1$ & $31.9$ &  & $222.7$ & $284.8$ & $38.7$ &  & $\cdots$ & $\cdots$ & $\cdots$ \\
NGC 6907 &                       $-2.8$ & $317.3$ & $182.1$ &  & $\cdots$ & $\cdots$ & $\cdots$ &  & $\cdots$ & $\cdots$ & $\cdots$ \\
IRAS 21101+5810 &                $-6.7$ & $259.0$ & $33.8$ &  & $\cdots$ & $\cdots$ & $\cdots$ &  & $\cdots$ & $\cdots$ & $\cdots$ \\
ESO 602-G025 &                   $-90.3$ & $198.5$ & $75.9$ &  & $133.6$ & $180.1$ & $68.5$ &  & $\cdots$ & $\cdots$ & $\cdots$ \\
UGC 12150 &                      $174.2$ & $78.6$ & $43.7$ &  & $15.6$ & $316.4$ & $38.7$ &  & $\cdots$ & $\cdots$ & $\cdots$ \\
IRAS F22491-1808 &       $3.5$ & $226.4$ & $17.2$ &  & $\cdots$ & $\cdots$ & $\cdots$ &  & $\cdots$ & $\cdots$ & $\cdots$ \\
CGCG 453-062 &                   $-99.3$ & $145.5$ & $35.5$ &  & $108.9$ & $164.3$ & $43.0$ &  & $\cdots$ & $\cdots$ & $\cdots$ \\
2MASX J23181352+0633267& $\cdots$ & $\cdots$ & $\cdots$ &  & $\cdots$ & $\cdots$ & $\cdots$ &  & $\cdots$ & $\cdots$ & $\cdots$ \\
\hline
\end{tabular}
\end{table*}

\clearpage
\section{Modified blackbody model SED fitting}\label{app:mbbsedfit}

\begin{figure*}[htb]\centering
\includegraphics[width=.8\textwidth]{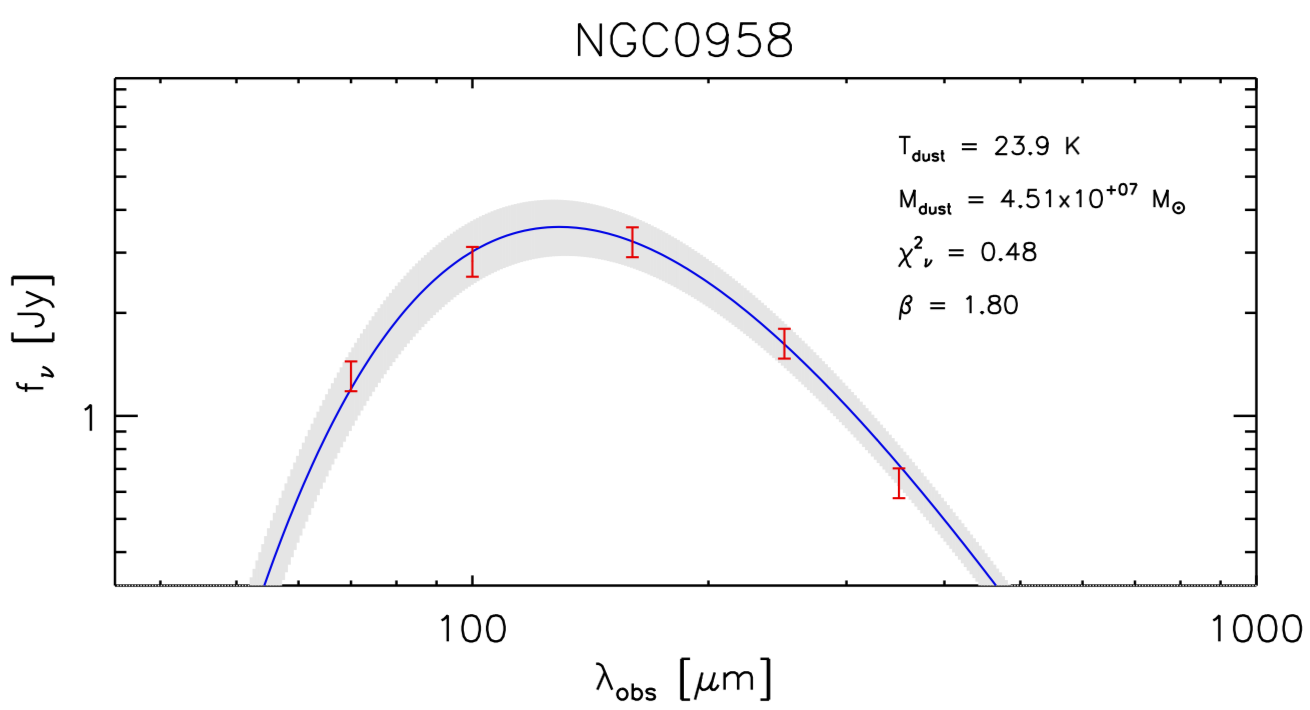}
\caption{SED fitting for NGC\,0958. {Red lines show the \emph{Herschel} photometry with its uncertainty}. The best model is shown as a solid blue line and the shaded area represents the uncertainty in the fit.}
\label{fig:exampleSEDfit}
\end{figure*}
Assuming that all dust grains have the same size and and composition, their emission can be modeled by an optically thin modified blackbody,
\begin{equation}\label{eq:mbb}
S(\nu)=\frac{Q(\nu)B(\nu,T)N\sigma}{D^2},
\end{equation}
where $S(\nu)$ is the flux density, $Q(\nu)$ is the emissivity, $B(\nu,T)$ is the blackbody intensity, $N$ and $\sigma$ are the number of identical dust particles and their cross section, respectively, and $D$ is the distance to the dust cloud.

We have used \emph{Herschel} data from both PACS (70, 100, and 170\,$\mu$m) and SPIRE (250 and 350\,$\mu$m, {excluding 500\,$\mu$m for having a too coarse resolution}) to construct the SED of each source. To do so, we smoothed the images to the FWHM of the IRAM-30\,m observations, {except for the 350\,$\mu$m images, which already have an angular resolution comparable with the IRAM-30\,m beam}. We then used a $\chi^2$ minimization method to obtain the best model. The emissivity index $\beta$ (where $\kappa(\nu)\propto\nu^\beta$) was fixed to a value of $\beta=1.8$, to avoid degeneracies between $\beta$ and $T_\mathrm{dust}$ in $\chi^2$ minimization fits \citep[][]{blain03}.

We can rewrite Eq.~\ref{eq:mbb} in terms of the dust mass, $M_d$, as
\begin{equation}
M_d=\frac{S(\nu_\mathrm{obs})D^2}{\kappa(\nu_\mathrm{rest})B(\nu_\mathrm{rest},T)(1+z)},
\end{equation}
where $\kappa(\nu)$ is the so-called grain absorption cross section per unit mass or mass absorption coefficient. This parameter is highly uncertain \citep[see, e.g.,][]{draine84, 
  davies12b, 
   bianchi13}. For our model we have adopted an intermediate value of $\kappa(250\,\mu\mathrm{m})=0.48\,\mathrm{m}^2\,\mathrm{kg}^{-1}$. Complementary to the dust masses and temperatures, we have derived $L_\mathrm{FIR}$ from direct integration below the fitted curve between 42.5 and 122.5\,$\mu$m. 
Figure~\ref{fig:exampleSEDfit} shows an example for the fit in the galaxy NGC\,0958.

\end{appendix}

\end{document}